\PassOptionsToPackage{unicode}{hyperref}
\PassOptionsToPackage{hyphens}{url}
\documentclass[
  10pt,
]{article}
\usepackage{xcolor}
\usepackage[a4paper,margin=17mm]{geometry}
\usepackage{amsmath,amssymb}
\usepackage[T1]{fontenc}
\usepackage[utf8]{inputenc}
\usepackage{textcomp}
\usepackage{mathptmx}
\IfFileExists{upquote.sty}{\usepackage{upquote}}{}
\IfFileExists{microtype.sty}{
  \usepackage[]{microtype}
  \UseMicrotypeSet[protrusion]{basicmath} 
}{}
\makeatletter
\@ifundefined{KOMAClassName}{
  \IfFileExists{parskip.sty}{%
    \usepackage{parskip}
  }{
    \setlength{\parindent}{0pt}
    \setlength{\parskip}{6pt plus 2pt minus 1pt}}
}{
  \KOMAoptions{parskip=half}}
\makeatother
\usepackage{amsmath,amssymb,booktabs,longtable,array,graphicx}
\usepackage{caption}
\usepackage{xurl}
\usepackage{needspace}
\AtBeginDocument{\hypersetup{colorlinks=true,urlcolor=black,linkcolor=black}}
\usepackage{bookmark}
\IfFileExists{xurl.sty}{\usepackage{xurl}}{} 
\hypersetup{
  pdftitle={Reference Coordinates{,} Output Perturbations{,} and Information Loss in EEG Decoding},
  pdfauthor={Runni Zhou, Rongrong Fu, Xinghao Yan, Xiaoyang Li},
  hidelinks,
  pdfcreator={LaTeX via pandoc}}

\title{Reference Coordinates, Output Perturbations, and Information Loss
in EEG Decoding}
\author{Runni Zhou, Rongrong Fu, Xinghao Yan, Xiaoyang Li}
\date{}

\usepackage{titlesec}
\usepackage{flushend}
\titleformat{\section}{\normalfont\bfseries\fontsize{11}{13}\selectfont}{}{0pt}{}
\titleformat{\subsection}{\normalfont\bfseries\fontsize{10}{12}\selectfont}{}{0pt}{}
\titlespacing*{\section}{0pt}{9pt}{4pt}
\titlespacing*{\subsection}{0pt}{7pt}{3pt}

\begin{document}
\twocolumn[{\begin{minipage}{\textwidth}\begin{center}{\fontsize{17}{20}\selectfont\bfseries Reference Coordinates, Output Perturbations, and Information Loss
in EEG Decoding\par}\vspace{8pt}{\large Runni Zhou, Rongrong Fu, Xinghao Yan, Xiaoyang Li\textsuperscript{*}}\end{center}

\begin{center}
College of Medicine and Biological Information Engineering, Northeastern University\\
Life Science Building B, No. 195 Chuangxin Road, Hunnan District, Shenyang, Liaoning 110016, China\\
Corresponding author: Xiaoyang Li, \href{mailto:20246389@stu.neu.edu.cn}{20246389@stu.neu.edu.cn}\\
{\small ORCID: Runni Zhou, \href{https://orcid.org/0009-0006-8365-4852}{0009-0006-8365-4852}; Rongrong Fu, \href{https://orcid.org/0009-0008-8879-3370}{0009-0008-8879-3370}\\
Xinghao Yan, \href{https://orcid.org/0009-0002-4437-957X}{0009-0002-4437-957X}; Xiaoyang Li, \href{https://orcid.org/0009-0002-4863-5761}{0009-0002-4863-5761}}
\end{center}

\vspace{3pt}\noindent\textbf{Abstract}\par

Electroencephalography (EEG) decoders can respond to reference changes
that preserve all relative-potential information. We use established
measurement-operator theory to distinguish coordinate sensitivity,
additional output-channel perturbations, and electrode loss. Exact
pseudoinverse alignment supplies a consistency control;
measurement-operator canonicalization (MOC) denotes a graph-Tikhonov
estimator. The broader evaluation covers three public motor-imagery
datasets and four decoders. The primary EEGNet comparison on Cho2017 and
PhysioNet uses three paired training seeds, three shared noise
realizations, and six front ends, with identical CAR training inputs
sharing checkpoints. At 10 dB output perturbation, MOC-minus-raw
disagreement was -0.144 and -0.107, while balanced-accuracy differences
were 0.044 and 0.050, respectively; these four contrasts passed the
retained Holm correction. MOC also reduced disagreement relative to
pseudoinversion across all three seeds in both cohorts. However,
comparison with fixed graph smoothing after pseudoinversion did not
establish an additional MOC benefit. Preserving zero reference rows
changed raw disagreement by less than 0.001 in magnitude; inverse-method
predictions were unchanged. For canonical front ends, cross-reference
disagreement was close to the reference obtained from independent noise
within the same operator. Controlled weak-mode reconstruction and
equal-rank electrode masks exposed separate stability and
information-loss effects. The results locate the complete-reference
gains within spatial processing under the specified stress model, while
requiring decoding accuracy alongside consistency to assess practical
utility.

\textbf{Keywords:} electroencephalography; measurement operator;
reference; canonicalization; motor imagery; robustness

\end{minipage}\vspace{14pt}}]
\section{1. Introduction}\label{introduction}

An EEG channel is defined by both its electrode location and its
measurement rule. A waveform labelled C3 may represent the potential
difference between C3 and Cz, a common average, or another reference.
These signals can have different numerical values while encoding the
same relative-potential field. A decoder trained on one representation
may therefore change its decision when the reference changes, even
though the underlying trial remains recoverable. Reference choice is
well established as a determinant of EEG analysis {[}1, 2{]}. For
decoding, it raises a precise question: how much prediction variation
arises from measurement coordinates alone?

Answering this question requires separating reference changes from
electrode loss. An information-equivalent reference transformation
preserves the relative field while changing its coordinates. Removing
electrodes can instead make spatial components unobservable. Both
changes affect the input tensor, but they call for different remedies:
coordinate alignment in the first case and estimation from incomplete
measurements in the second. An explicit measurement model makes this
distinction accessible to a downstream learning system.

EEG reference theory already establishes rank-one reference
transformations, their composition rules, and pseudoinverse recovery of
the average-reference representation {[}1, 2{]}. Hu et al.~explicitly
derive the projector identity used here {[}2{]}. REST and regularized
REST (rREST) estimate an infinity-referenced field using
volume-conduction priors and statistical regularization {[}3, 4{]}.
RESIT connects reference standardization with missing-channel
reconstruction through a lead-field model and evaluates signal
reconstruction in resting-state and event-related EEG {[}5{]}; spherical
splines instead use spatial smoothness {[}6{]}. Reference
standardization has also been evaluated directly in motor-imagery
classification: Zhao et al.~compared classification before and after
REST {[}7{]}, and REI-Net combines reference-standardization
interpolation, a three-dimensional CNN, and transfer learning {[}8{]}.
Those studies motivate downstream evaluation rather than a claim that
reference-aware decoding is new. Our target is the recorded mean-zero
field on a specified finite electrode domain, with a graph smoothness
penalty rather than a source/lead-field prior. We use these established
estimation ideas to ask how preserved information, added output
perturbations, and lost spatial observations affect downstream
predictions.

Learned interfaces provide complementary ways to accommodate
heterogeneous EEG inputs. BIOT and CBraMod learn channel or
spatial-temporal representations {[}9, 10{]}; LUNA and EEGMirror address
varying electrode configurations {[}11, 12{]}. E2EG2G uses latent nodes
and relation grids, and MGF-Net models bipolar derivation relationships
{[}13, 14{]}. Here, known transformations of a fixed recorded trial
provide a controlled setting in which coordinate sensitivity can be
separated from information loss before interpreting a decoder's
response.

The contribution is an empirical separation of coordinate alignment,
perturbation sensitivity, and task performance. The known projector
identity supplies a consistency control. A paired EEGNet reanalysis of
previously evaluated held-out subjects in two cohorts uses shared
observations, three fixed training seeds, and three noise realizations
to compare six front ends. Reusing a checkpoint when CAR training inputs
coincide separates operator-dependent estimation from fixed graph
smoothing at test time. Fixed-model zero-row and same-operator noise
controls locate the sensitivity of these comparisons without treating
disagreement as a causal decomposition. Historical architecture and
equal-rank electrode-loss analyses delimit the scope of the evidence.

\section{2. Methods}\label{methods}

\subsection{2.1. Measurements and information
equivalence}\label{measurements-and-information-equivalence}

Let \(v(t)\in\mathbb{R}^{K}\) denote electrode potentials on a specified
domain. A recording with \(C\) channels is

\begin{equation}
y(t)=Mv(t)+\epsilon(t),\qquad M\in\mathbb{R}^{C\times K}.
\label{eq:measurement}
\end{equation}

Each row of \(M\) defines a signed measurement. A single-reference row
is \(e_i^\top-e_r^\top\); a bipolar row is \(e_i^\top-e_j^\top\). A
weighted reference has \(M=I-\mathbf{1}w^\top\) with
\(w^\top\mathbf{1}=1\). Uniform weights give the common-average
reference (CAR). For the missing-electrode experiments,
\(S\in\mathbb{R}^{m\times K}\) selects the retained sites and
\(M=P_mS\), where \(P_m=I_m-\mathbf{1}_m\mathbf{1}_m^\top/m\).
Measurements are re-centered over the retained sites, giving rank
\(m-1\); simply selecting rows of a fixed full-domain reference would
define a different operator. Figure 1 connects these objects to the
implemented inverse and its canonical decoder input.

All potential-difference operators satisfy \(M\mathbf{1}=0\). Thus \(v\)
and \(v+c\mathbf{1}\) yield the same noiseless measurements: the
identifiable object is an equivalence class modulo an additive constant.
We choose its mean-zero representative,

\begin{equation}
Pv,\qquad P=I-\frac{1}{K}\mathbf{1}\mathbf{1}^{\top}.
\label{eq:projection}
\end{equation}

Two operators on the same electrode domain are information-equivalent if
both have rank \(K-1\) and annihilate \(\mathbf{1}\). Each retains the
entire relative field, even if their output dimensions differ, and their
noiseless measurements are related by \(y_2=M_2M_1^\dagger y_1\).
Information equivalence is defined for the noiseless field; noise
covariance remains operator-dependent. The electrode domain must
include, or explicitly account for, any external physical reference.

When \(\operatorname{rank}(M)<K-1\), the null space contains nonconstant
potential patterns. Distinct mean-zero fields then give identical
observations. Estimating an unobserved component requires assumptions
beyond the measurements, regardless of whether the reconstruction
outputs a value at every electrode.

\begin{figure*}[!t]\centering
\includegraphics[width=\textwidth]{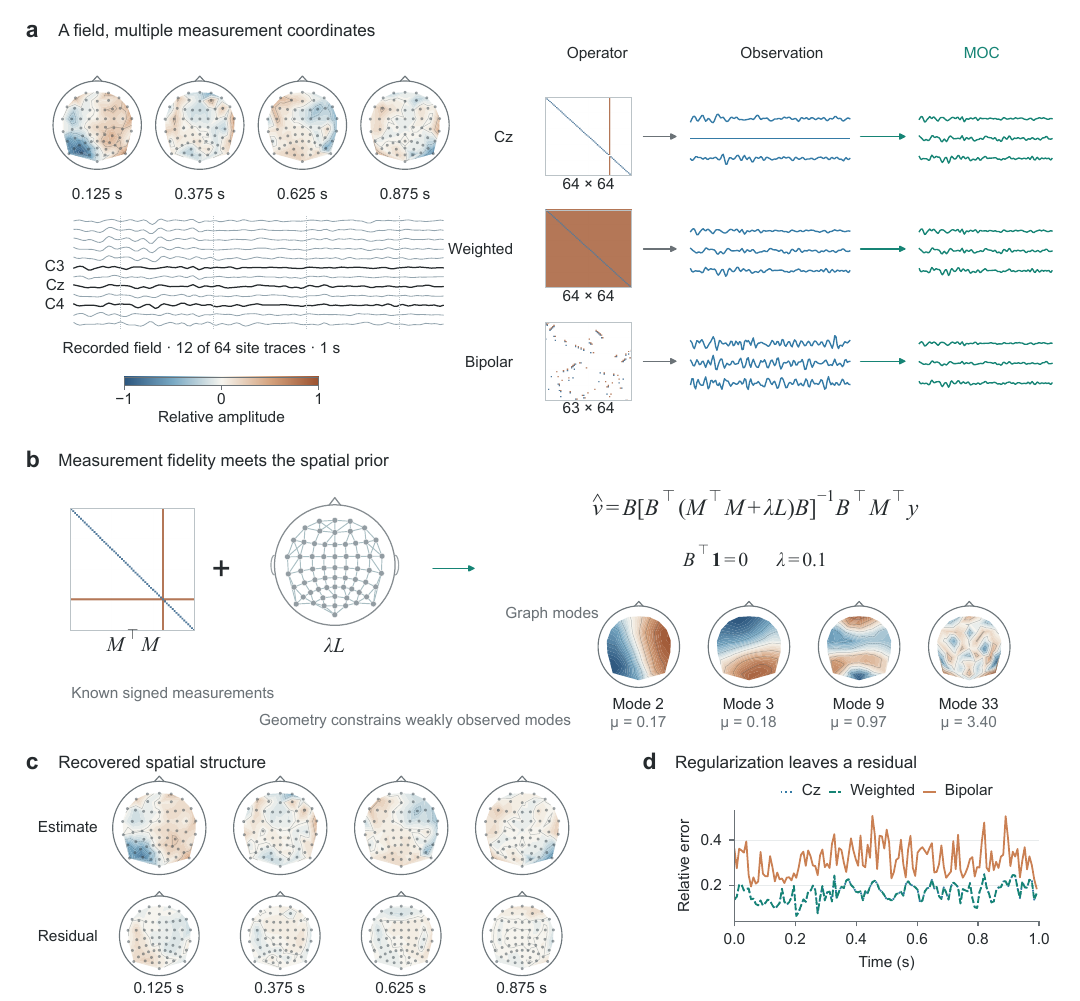}
\caption{Measurement coordinates and spatially regularized recovery. (a) Four snapshots and 12 traces from the first second of the first cached Cho2017 trial. Operator matrices show signed support (blue positive, rust negative). Branches display C3/Cz/C4 for reference observations and reconstructions, and the first three bipolar observations, on a shared trace scale. (b) The measurement Gram matrix and electrode graph define the constrained inverse. Columns of B span the mean-zero subspace. Graph modes are numbered including the constant mode; mu denotes the eigenvalue. Each mode has unit maximum absolute coefficient. (c) Bipolar MOC estimates and estimate-minus-recorded residuals at the same times, sharing the recorded-snapshot amplitude scale in a. (d) Field-error norm divided by recorded-field norm. Topographies linearly interpolate electrode values within their convex hull. The fixed-trial illustration uses lambda = 0.1 without added output-channel noise.}\label{fig:1}\end{figure*}

\subsection{2.2. Exact and regularized
canonicalization}\label{exact-and-regularized-canonicalization}

For a full-quotient-rank operator, the standard row-space projector
identity, given explicitly for unipolar references in {[}2{]}, gives

\begin{equation}
M^\dagger M=P,\qquad \widehat v=M^\dagger y=Pv+M^\dagger\epsilon.
\label{eq:identity}
\end{equation}

Rank-nullity gives \(\ker M=\operatorname{span}(\mathbf{1})\), so
\(M^\dagger M\) projects orthogonally onto \(\ker(M)^\perp\). Every
information-equivalent operator therefore yields the same noiseless
canonical input and, for a fixed deterministic decoder, the same
prediction. This identity provides the exact consistency baseline; task
accuracy is evaluated separately.

With additional output perturbations, inversion sensitivity depends on
the positive spectrum. If \(\sigma_{\min}^{+}(M)\) is the smallest
positive singular value, then

\begin{equation}
\|\widehat v-Pv\|_2\leq\frac{\|\epsilon\|_2}{\sigma_{\min}^{+}(M)}.
\label{eq:noise}
\end{equation}

We compare exact inversion with graph-Tikhonov estimation, denoted MOC:

\begin{equation}
\widehat v=\arg\min_{u\perp\mathbf{1}}
\left\{\|Mu-y\|_{R^{-1}}^2+\lambda u^\top Lu\right\}.
\label{eq:moc}
\end{equation}

The first term enforces measurement fidelity; the Laplacian \(L\)
penalizes differences between nearby electrode potentials. We used
Euclidean distances between standard three-dimensional electrode
coordinates, the symmetric union of four nearest nonself neighbors, and
affinities \(W_{ij}=\exp[-d_{ij}^2/(2\sigma^2)]\) on that union. The
unnormalized Laplacian is \(L=D-W\). The implemented bandwidth is the
median across sites of the fifth ordered nonself distance (Section S2).
The neighborhood size and bandwidth rule are fixed geometry
hyperparameters chosen independently of test decoding performance. For
an orthonormal basis \(B\) of \(\mathbf{1}^{\perp}\), the solution is

\begin{equation}
\begin{split}
H&=B^\top M^\top R^{-1}MB+\lambda B^\top LB,\\
\widehat v&=BH^{-1}B^\top M^\top R^{-1}y.
\end{split}
\label{eq:solution}
\end{equation}

The reduced system is formed in float64; the implementation uses NumPy
inversion and falls back to the Moore--Penrose pseudoinverse only on a
singular-matrix exception. SVD pseudoinverses use a relative cutoff of
\(10^{-15}\) (Section S3). We used identity noise precision,
\(R^{-1}=I\), in every experiment. MOC is a linear graph-Tikhonov
estimator whose spatial penalty trades reconstruction bias for reduced
sensitivity to weakly observed components. Under rank loss, the same
penalty also constrains unobserved directions.

For complete references \(M=I-\mathbf{1}w^\top\), the reduced Gram
matrix is

\begin{equation}
B^\top M^\top MB=I_{K-1}+K(B^\top w)(B^\top w)^\top.
\label{eq:reference_spectrum}
\end{equation}

The positive singular values are \(1\) with multiplicity \(K-2\) and
\(\sqrt{K}\|w\|_2\). Thus, for the unscaled complete references studied
here, \(\sigma_{\min}^{+}=1\) and \(\|M^\dagger\|_2=1\). Their condition
numbers reflect an enlarged maximum singular value, not a weak positive
mode. At CAR, MOC reduces to \(\widehat v=(I+\lambda L)^{-1}Py\), a
graph smoother. The weak-mode mechanism applies to the bipolar and
controlled ill-conditioned operators. To isolate the operator-dependent
part of complete-reference estimation, we add a fixed smoother after
Pinv:

\begin{equation}
\widehat v_{\mathrm{smooth}}=(I+\lambda L)^{-1}M^\dagger y.
\label{eq:fixed_smooth}
\end{equation}

At CAR its linear map coincides with MOC, including for arbitrary added
CAR output noise. The same trained MOC checkpoint therefore evaluates
both test front ends. Away from CAR, the maps generally differ: one
applies the same graph smoother to the pseudoinverse field, whereas MOC
uses the operator in its fidelity term. No performance ordering follows
from this distinction.

\subsection{2.3. Data and experimental
regimes}\label{data-and-experimental-regimes}

We analyzed left-hand versus right-hand motor imagery from BNCI2014-001,
Cho2017, and PhysioNet {[}15, 16, 17{]}. The MOABB pipeline {[}18{]}
applied a 4--38 Hz bandpass and resampling to 128 Hz. Epoch intervals
were 2--6 s for BNCI and 0--3 s for the other datasets. Trials were
mean-centered across electrodes before generating measurement views.
This processed recorded field, including its existing recording noise,
served as the reconstruction target. Table 1 summarizes the cohorts.

\begingroup\small\setlength{\tabcolsep}{4pt}\renewcommand{\arraystretch}{1.15}
\begin{table*}[t]\centering\caption{Processed datasets and subject partitions. All signals were resampled to 128 Hz; partition counts are train/validation/test.}
\begin{tabular}{l l l l l l}
\toprule
Dataset & Subjects & Trials & Sites & Samples & Partition \\
\midrule
BNCI2014-001 & 9 & 2592 & 22 & 513 & 5/2/2 \\
Cho2017 & 52 & 10520 & 64 & 384 & 32/10/10 \\
PhysioNet & 104 & 4680 & 64 & 385 & 62/21/21 \\
\bottomrule\end{tabular}\end{table*}\endgroup

Subject partitions were fixed, and all measurement views of a trial
inherited its partition. The test sets contained 2, 10, and 21 subjects.
The fixed PhysioNet exclusion list comprises subjects 88, 89, 92, 100,
and 104. The available records establish the list but not individual
exclusion reasons (Section S1). Preprocessing used loader event and
epoch selection without additional artifact rejection. Secondary
subject-grouped analyses assessed sensitivity to the partition within
the same cohorts.

The suite without added output-channel noise comprised CAR, three single
references, two weighted references, and four connected bipolar trees.
Raw fixed-channel decoders accepted six site-like operators: CAR, Cz,
Pz, Oz (C3 on BNCI), and the two weighted references. Inverse methods
also accepted the \(K-1\) bipolar channels. Their broader input
compatibility was evaluated in that suite; matched method comparisons
used the common six-reference set and its fifteen unordered pairs.

The primary EEGNet comparison on Cho2017 and PhysioNet was an
output-channel perturbation stress test: \(y_j=M_jv+\epsilon_j\), with
independent Gaussian entries added after each operator at 10 dB. Every
method shared the corresponding observation arrays. For each consecutive
block of at most 256 trials within a split and operator,
\(\sigma_\epsilon^2=\operatorname{mean}[(M_jv)^2]10^{-\mathrm{SNR}/10}\),
averaging all block entries. This nominal variance-based calibration is
unchanged from the original analysis. The first test realization
reproduces its observation hashes; two additional independent
realizations use the same rule. Training and validation arrays are fixed
across seeds. All measurement rows initially receive noise, including
single-reference zero rows. At each fixed checkpoint, a paired control
sets only those nominally zero rows exactly to zero, without rescaling
other rows. A second reference measure compares independent noise
realizations within the same operator. It describes noise sensitivity
alongside cross-reference disagreement; their difference is not a causal
partition. The earlier one-seed check is retained separately in Section
S12.

Digital re-referencing of a shared noisy electrode field instead gives
\(y_j=M_j(v+\eta)\) and \(M_j^\dagger y_j=P(v+\eta)\) for every
information-equivalent operator. Electrode noise is transformed together
with the field; its within-view covariance becomes
\(M_j\Sigma_\eta M_j^\top\) {[}2{]}. Existing acquisition noise
therefore does not invalidate exact alignment of the same field. Our
added output perturbations test a different sensitivity and do not
reproduce that acquisition-noise model.

Separate response analyses used 5, 10, 20, 30 dB and no added output
noise. Synthetic zero-sum matrices had positive-spectrum condition
numbers 1, 3, 10, 30, 100, 300, and 1000. These controlled matrices vary
inversion difficulty independently of native geometry. Reconstruction
used blocks of at most 256 trials; separate architecture and response
decoding used entire split arrays for each operator, and augmentation
used individual trials. Section S3 specifies the processing unit and
noise sharing for each analysis. Primary comparisons shared observations
across methods; response decoding and grouped analyses used
method-specific draws.

Information-loss analyses retained nominal electrode fractions 1, 0.875,
0.75, 0.625, and 0.5, with rounded site counts and recorded matrix
ranks. Each non-full level included five random masks and one mask from
each of two directed families that preferentially preserved or removed
predefined motor-region sites. Identical stored copies of directed masks
were collapsed, and disagreement across those copies was omitted.
Decoders were trained at full coverage and evaluated after electrode
removal without added output-channel noise.

\begin{figure*}[!t]\centering
\includegraphics[width=\textwidth]{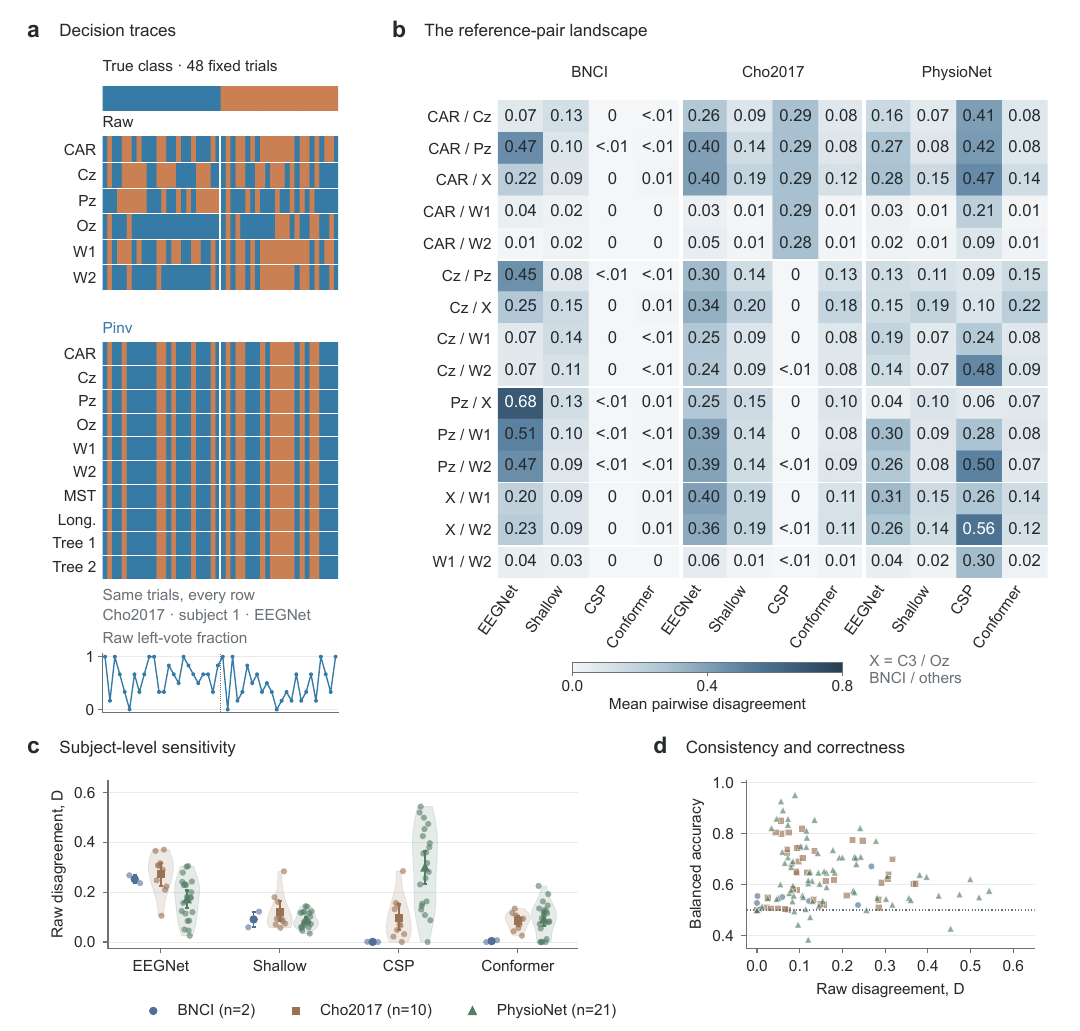}
\caption{Prediction sensitivity across trials, reference pairs and subjects, without added output-channel noise; existing recording noise is retained. (a) The first 24 trial IDs per class for Cho2017 subject 1, selected independently of predictions. Blue/rust denote left/right. Both rasters use the same trials; Pinv additionally accepts four bipolar operators. The trace shows the fraction of six raw predictions voting left in display order. (b) Mean disagreement for 15 unique reference pairs, four decoders and three cohorts, derived from 132 subject matrices. X denotes C3 for BNCI and Oz otherwise. Each column mean reproduces its cohort-model D; small nonzero values are distinguished from zero. (c) Individual raw D values, means and 95\% subject bootstrap intervals. Violin envelopes are kernel density estimates bounded by the observed range, shown for cohorts with at least ten subjects. (d) Raw D and balanced accuracy (BAC) for 132 subject-model combinations; subjects recur across decoders. Test n = 2, 10 and 21. Canonical endpoints are in Section S4. Decoder outputs in this figure retain historical fits; the exact-alignment identity itself is model independent.}\label{fig:2}\end{figure*}

\subsection{2.4. Decoders, controls, and
inference}\label{decoders-controls-and-inference}

We used EEGNet {[}19{]}, ShallowConvNet {[}20{]}, common spatial
patterns (CSP) followed by linear discriminant analysis, and EEG
Conformer {[}21{]}. Primary repeated training is restricted to EEGNet on
Cho2017 and PhysioNet; BNCI and other architectures retain their
historical fits. Models are trained separately for each dataset. The two
primary cohorts do not constitute external cross-dataset validation of a
single trained model. CSP uses MOABB microvolt-scaled inputs, averages
temporally centered trial covariances in float64 without trace
normalization, and adds \(10^{-6}I\) to each class covariance. SciPy's
symmetric generalized eigensolver then supplies the two smallest and two
largest eigenvectors. This diagonal loading handles rank-deficient CAR
inputs without a separate quotient projection or adaptive rank cutoff.
Clipped log-variance features (\(10^{-10}\) floor) pass through
StandardScaler and SVD-based linear discriminant analysis. Section S3
gives the full numerical rules and architecture settings.

The six front ends are Raw, Pinv, interpolation + centering,
augmentation + centering, MOC, and fixed smoothing after Pinv. For each
cohort and each seed fixed for this repeated analysis (20260919,
20260920, 20260921), the seed is set before network construction, and
every training group starts from the same parameter and buffer state.
Data order, dropout, validation iteration, and observation generation
have independently recorded random streams. CUDA benchmarking and TF32
are disabled; deterministic algorithms are enabled. All fits use 15
epochs, batch size 128, Adam (learning rate 0.001, weight decay 0.0001),
cross-entropy, and the first checkpoint achieving the lowest validation
loss. Float64 map identities and actual float32 training/validation
arrays are checked before sharing a model: Pinv shares its checkpoint
with complete-site interpolation, and MOC shares with fixed smoothing.
These aliases use identical training tensors, losses, and optimization
settings. Their non-CAR noisy test inputs are calculated separately.
Thus four training groups per cohort/seed suffice; one full repeated Raw
fit per cohort checks reproducibility. Section S13 records all checks
and the training ledger.

Augmentation selects CAR or an available Cz/Pz/Oz/C3 reference once per
training or validation trial, adds output noise, then interpolates and
centers. Its fixed arrays are shared across the three training seeds,
with separate streams for reference choice and noise. At complete
coverage, interpolation copies sites and centers them. Since \(PM_j=P\),
this preprocessing removes noiseless reference offsets; augmentation
varies output-noise scale and realization. It is not exposure to
different noiseless reference coordinates. Its test front end is
interpolation + centering.

The main MOC coefficient was fixed at \(\lambda=0.1\), following
validation reconstruction selection on BNCI and Cho2017, and transferred
to downstream comparisons. Selection emphasized noisy ill-conditioned
reconstruction rather than test accuracy. An earlier exploratory
reconstruction analysis used 0.01 on PhysioNet and remains separately
identified. Spline regularization was \(10^{-5}\). Studies without and
with added output noise had different fixed training budgets; their
scores were not pooled.

Prediction disagreement for subject \(s\), with \(N_s\) trials and \(J\)
operators, was

\begin{equation}
D_s=\frac{\sum_{n=1}^{N_s}\sum_{i<j}\mathbb{I}\{\widehat y_{sni}\ne\widehat y_{snj}\}}
{N_s\binom{J}{2}}.
\label{eq:disagreement}
\end{equation}

Balanced accuracy (BAC) and macro-F1 were calculated within
subject/operator and averaged across operators. Normalized mean squared
error (NMSE) measured reconstruction error relative to each trial's
processed field energy and was averaged within subject and condition.
For the primary comparison, seed and noise-realization values are
averaged within subject before cohort inference. Same-operator noise
disagreement averages the three unordered pairs of noise realizations
over six operators. Subjects receive equal weight; seeds and
realizations do not enlarge the independent sample size. Per-seed
effects and training variability are reported separately (Section S13).
For MOC-minus-control contrasts, negative \(\Delta D\) favors
consistency and positive \(\Delta\mathrm{BAC}\) favors accuracy.

We calculated percentile 95\% confidence intervals (CIs) from 10,000
subject bootstrap resamples, using paired differences for method
contrasts. Two-sided sign-flip tests enumerate assignments for up to
sixteen subjects and use 10,000 random assignments with a plus-one
correction otherwise. Before training in this repeated-analysis phase,
we fixed the contrasts and retained the original Holm family of twelve
dataset/control tests per endpoint: four historical controls across
three datasets. Its two larger cohorts now use repeated subject means;
the frozen two-subject BNCI cells remain descriptive and retain their
family membership. MOC versus fixed smoothing forms a separate
two-cohort Holm family per endpoint. D and BAC are the primary
endpoints; macro-F1 is secondary. CIs are not multiplicity-adjusted.
Noise controls are descriptive paired sensitivity estimates. Three
training seeds characterize this fixed training procedure, not the full
distribution of possible fits. Historical architecture and grouped
analyses retain their original families and implementation limits.

The paired repeated analysis reuses held-out subjects whose results had
already been evaluated in earlier analyses. It is a reanalysis after
correcting the fixed training protocol, not validation on a newly
recruited or previously unevaluated test cohort. Statements that
settings were fixed before training or evaluation refer only to this
repeated-analysis phase; they do not imply study-wide preregistration or
that test results had never been examined. Section S13 documents the
analysis history and retained records.

\subsection{2.5. Reproducible analysis and
visualization}\label{reproducible-analysis-and-visualization}

Subject-level predictions, measurement matrices, and fixed
configurations underlie the reported statistics. Python scripts generate
figures directly from these data and mathematical operators. Spatial
illustrations use a fixed recorded trial and linear interpolation within
the electrode convex hull; cohort summaries use equal subject weights.
The versioned Supplementary\_data\_and\_code.tar.xz archive (release FS2,
21 September 2026) provides protocols, checkpoints, observation hashes,
training histories, paired predictions, source hashes, and numerical
checks.

\section{3. Results}\label{results}

\begin{figure*}[!t]\centering
\includegraphics[width=\textwidth]{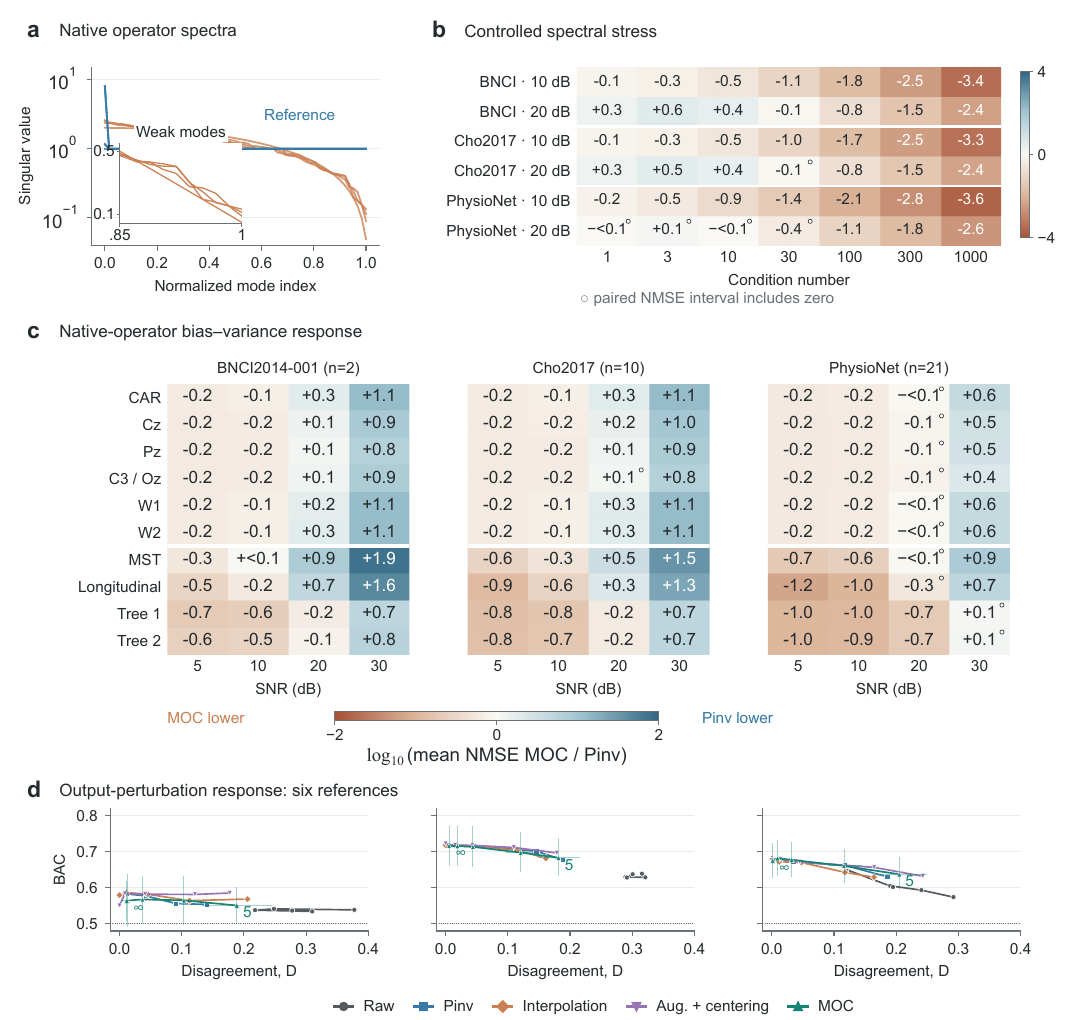}
\caption{Output-channel perturbations: weak-mode reconstruction and complete-reference decoding. (a) Positive Cho2017 singular spectra and an inset of weak bipolar modes. All complete references have minimum positive singular value one; the controlled matrices in b have minimum positive singular value 1 / condition number. (b) Controlled mixtures at seven condition numbers and two SNRs, with operator seeds averaged within subject. (c) Ten native operators at four SNRs. Matrix values are log10(mean NMSE MOC / mean NMSE Pinv), using equal-subject means. Rust/negative favours MOC; blue/positive favours Pinv. Symmetric colour limits are 4 in b and 2 in c. Open circles mark paired 95\% subject-bootstrap intervals for the NMSE difference that include zero; these intervals are unadjusted. (d) Separate six-reference EEGNet mean D/BAC trajectories at 5, 10, 20, 30 dB and no added noise, in the cohort order of c. Labels identify MOC endpoints; pale crosses are marginal 95\% subject intervals. The decoding response study uses method-specific output-noise draws; it does not identify weak-mode stabilization as the cause of classification changes. Aug. + centering denotes augmentation followed by interpolation and centering. Reconstruction without added output noise, absolute NMSE and remaining intervals are in Section S7 and the digital tables. These SNR decoding fits retain historical initialization and are descriptive; the paired primary evidence is in Figure 4.}\label{fig:3}\end{figure*}

\subsection{3.1. Predictions changed when the relative field did
not}\label{predictions-changed-when-the-relative-field-did-not}

Changing the reference altered raw EEGNet decisions while preserving the
recoverable relative-potential field (Figure 2a). The trial illustration
uses the first 24 trial IDs from each class of a fixed held-out Cho2017
subject, selected independently of predictions. Across the cohorts, raw
disagreement was 0.252 on BNCI, 0.274 on Cho2017, and 0.172 on
PhysioNet. The pairwise atlas shows which reference changes contributed
to this sensitivity, with clear variation among decoder families and
subjects (Figure 2b,c).

Pseudoinverse canonicalization reduced disagreement without added output
noise to zero for every tested subject and decoder. Canonical inputs
agreed across all six references and four bipolar trees, as illustrated
by the lower raster in Figure 2a. Numerical projector checks reproduced
this equality. The reference-pair atlas summarizes the fifteen unique
raw pairs for each cohort and decoder, connecting the trial-level
patterns to the aggregate disagreement statistic.

Consistency alone did not establish useful decoding (Figure 2d). The raw
BNCI Conformer had disagreement 0.0037, while some noisy canonical fits
operated at chance-level BAC. Their stable predictions are not evidence
of useful robustness. Historical regularized EEGNet SNR-response fits
retained disagreement without added output noise of 0.012, 0.006, and
0.001. These residuals are compatible with operator-dependent
regularization bias, but their magnitudes also condition on those fitted
models. Full classification endpoints are in Section S4.

\begin{figure*}[!t]\centering
\includegraphics[width=\textwidth]{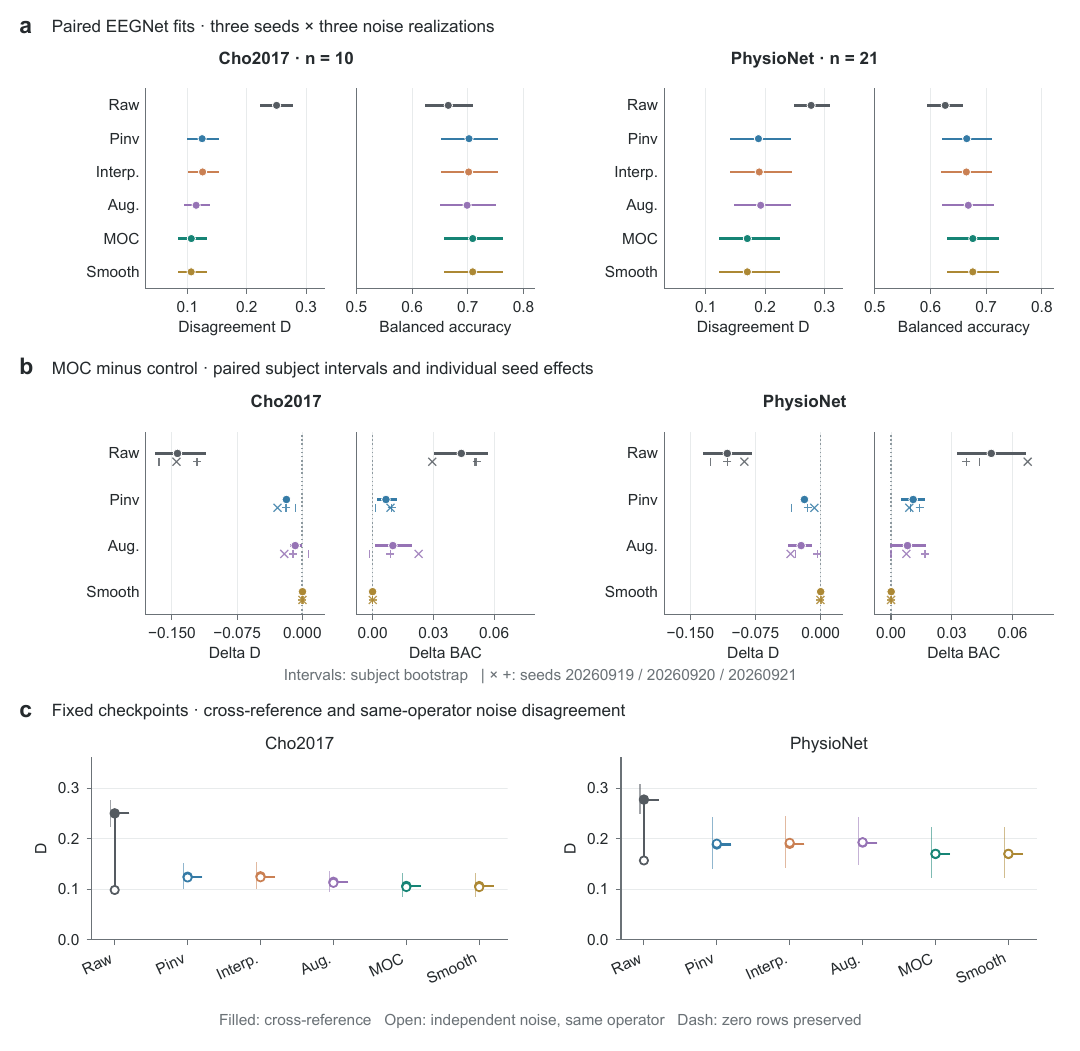}
\caption{Repeated EEGNet comparison and fixed-checkpoint perturbation controls. (a) Equal-subject means and unadjusted 95\% bootstrap intervals after averaging three training seeds and three shared noise realizations within subject. Six front ends use four training groups: Pinv/interpolation and MOC/fixed smoothing share CAR-trained checkpoints, but apply distinct non-CAR test maps. (b) MOC-minus-control subject-mean differences and paired intervals. The three small marks show cohort effects at the training seeds fixed for the repeated-analysis phase; they are not independent subjects. Tables 2 and 3 give absolute values and adjusted inference; intervals alone do not indicate Holm significance. Fixed smoothing has a separate two-cohort test family. (c) Cross-reference D (filled circles and subject intervals), same-operator independent-noise D (open circles), and cross-reference D with zero rows preserved (short dashes). All use unchanged checkpoints. Connecting segments compare descriptive levels, not a causal decomposition. Aug. denotes augmentation + centering; Smooth denotes fixed graph smoothing after Pinv. The budget, front ends and seeds were fixed for this repeated-analysis phase, which reused previously evaluated held-out subjects.}\label{fig:4}\end{figure*}

\subsection{3.2. Weak-mode stabilization and reference smoothing are
distinct}\label{weak-mode-stabilization-and-reference-smoothing-are-distinct}

Reference and bipolar operators shared full quotient rank but differed
in their positive spectra (Figure 3a). All six complete references had
minimum positive singular value one; their pseudoinverses did not
amplify the Euclidean norm of arbitrary output perturbations. Bipolar
trees and the controlled matrices included weak modes. In the controlled
experiment at 10 dB, pseudoinverse NMSE increased sharply as the minimum
positive singular value decreased, whereas MOC limited this increase
across all three datasets (Figure 3b). This experiment establishes a
reconstruction benefit under controlled ill-conditioning.

Native operators revealed the accompanying bias cost (Figure 3c).
Regularization often became less favourable at higher SNR, and responses
differed among bipolar trees. As added noise vanished, Pinv approached
the exact processed field while MOC retained reconstruction bias
(Section S7). At 10 dB, the BNCI minimum-spanning bipolar tree had NMSE
0.116 with Pinv and 0.122 with MOC, illustrating dependence on the
operator's spatial structure as well as its spectrum. The separate
six-reference SNR decoding trajectories show changes in consistency and
BAC (Figure 3d). Because those references lack weak positive modes,
their classification gains cannot be attributed to correcting the
controlled experiment's inverse amplification. Spatial filtering,
regularization bias, and separately fitted models remain possible
contributors.

\subsection{3.3. Paired training clarifies the contribution of spatial
smoothing}\label{paired-training-clarifies-the-contribution-of-spatial-smoothing}

The repeated EEGNet comparison shares six reference observations across
all front ends and averages three training seeds and three noise
realizations within each held-out subject (Figure 4a). Table 2 reports
all six methods; Table 3 gives the principal paired control contrasts.
On Cho2017, MOC-minus-raw \(\Delta D\) was -0.144 (95\% CI -0.168 to
-0.112; Holm \(p=0.0156\)). On PhysioNet it was -0.107 (-0.133 to
-0.080; Holm \(p=0.0020\)). Corresponding BAC differences were 0.044
(0.031 to 0.056; Holm \(p=0.0176\)) and 0.050 (0.033 to 0.066; Holm
\(p=0.0012\)). All four meet the retained Holm threshold; their
directions also agree across the three seeds.

\begin{table*}[!t]\centering\caption{Repeated EEGNet comparison under shared 10 dB output perturbations.}
\begingroup\small\setlength{\tabcolsep}{4pt}\renewcommand{\arraystretch}{1.15}
\begin{tabular}{p{0.11\textwidth} p{0.18\textwidth} p{0.205\textwidth} p{0.205\textwidth} p{0.205\textwidth}}
\toprule
Dataset & Front end & D [95\% CI] & BAC [95\% CI] & Macro-F1 [95\% CI] \\
\midrule
Cho2017 & Raw & 0.250 [0.225, 0.275] & 0.665 [0.624, 0.707] & 0.641 [0.592, 0.688] \\
Cho2017 & Pinv & 0.125 [0.102, 0.151] & 0.702 [0.654, 0.753] & 0.694 [0.642, 0.748] \\
Cho2017 & Interpolation & 0.126 [0.103, 0.152] & 0.702 [0.654, 0.752] & 0.694 [0.641, 0.747] \\
Cho2017 & Aug. + centering & 0.115 [0.096, 0.136] & 0.699 [0.652, 0.748] & 0.690 [0.639, 0.743] \\
Cho2017 & MOC & 0.107 [0.086, 0.130] & 0.709 [0.659, 0.761] & 0.701 [0.647, 0.756] \\
Cho2017 & Fixed smoothing & 0.106 [0.086, 0.130] & 0.709 [0.659, 0.761] & 0.701 [0.647, 0.756] \\
PhysioNet & Raw & 0.277 [0.251, 0.307] & 0.627 [0.597, 0.656] & 0.586 [0.553, 0.617] \\
PhysioNet & Pinv & 0.189 [0.142, 0.242] & 0.665 [0.622, 0.709] & 0.651 [0.604, 0.697] \\
PhysioNet & Interpolation & 0.190 [0.143, 0.244] & 0.664 [0.622, 0.708] & 0.650 [0.604, 0.697] \\
PhysioNet & Aug. + centering & 0.192 [0.149, 0.241] & 0.668 [0.623, 0.712] & 0.656 [0.609, 0.702] \\
PhysioNet & MOC & 0.170 [0.124, 0.223] & 0.676 [0.632, 0.721] & 0.662 [0.615, 0.711] \\
PhysioNet & Fixed smoothing & 0.170 [0.125, 0.223] & 0.676 [0.632, 0.721] & 0.662 [0.615, 0.710] \\
\bottomrule\end{tabular}\par\endgroup\vspace{3pt}\begin{minipage}{0.98\textwidth}\footnotesize Subject means average three paired training seeds and three noise realizations before bootstrap resampling (Cho2017 n = 10; PhysioNet n = 21). Pinv and interpolation share one checkpoint per seed; MOC and fixed smoothing share another. Non-CAR test front ends remain distinct. BNCI historical descriptive results remain in Section S5. Intervals are unadjusted for multiplicity.\end{minipage}\end{table*}

\begin{table*}[!t]\centering\caption{Paired MOC-minus-control differences averaged within subject over training and noise repeats.}
\begingroup\small\setlength{\tabcolsep}{4pt}\renewcommand{\arraystretch}{1.15}
\begin{tabular}{p{0.1\textwidth} p{0.12\textwidth} p{0.2\textwidth} p{0.06\textwidth} p{0.06\textwidth} p{0.2\textwidth} p{0.06\textwidth} p{0.06\textwidth}}
\toprule
Dataset & Control & Delta D [95\% CI] & p & Holm p & Delta BAC [95\% CI] & p & Holm p \\
\midrule
Cho2017 & Raw & \shortstack[l]{-0.144\\{}[-0.168, -0.112]} & 0.0020 & 0.0156 & \shortstack[l]{0.044\\{}[0.031, 0.056]} & 0.0020 & 0.0176 \\
Cho2017 & Pinv & \shortstack[l]{-0.018\\{}[-0.021, -0.015]} & 0.0020 & 0.0156 & \shortstack[l]{0.007\\{}[0.003, 0.011]} & 0.0156 & 0.1094 \\
Cho2017 & Aug. +\newline centering & \shortstack[l]{-0.008\\{}[-0.013, -0.002]} & 0.0254 & 0.1270 & \shortstack[l]{0.010\\{}[0.002, 0.019]} & 0.0527 & 0.3164 \\
Cho2017 & Fixed\newline smoothing & \shortstack[l]{0.000259\\{}[0.000093, 0.000430]} & 0.0312 & 0.0625 & \shortstack[l]{-0.000003\\{}[-0.000315, 0.000312]} & 1.0000 & 1.0000 \\
PhysioNet & Raw & \shortstack[l]{-0.107\\{}[-0.133, -0.080]} & 0.0002 & 0.0020 & \shortstack[l]{0.050\\{}[0.033, 0.066]} & 0.0001 & 0.0012 \\
PhysioNet & Pinv & \shortstack[l]{-0.019\\{}[-0.022, -0.015]} & 0.0001 & 0.0012 & \shortstack[l]{0.011\\{}[0.006, 0.016]} & 0.0011 & 0.0110 \\
PhysioNet & Aug. +\newline centering & \shortstack[l]{-0.022\\{}[-0.036, -0.012]} & 0.0004 & 0.0036 & \shortstack[l]{0.008\\{}[0.000, 0.017]} & 0.0740 & 0.3700 \\
PhysioNet & Fixed\newline smoothing & \shortstack[l]{0.000016\\{}[-0.000423, 0.000486]} & 0.9732 & 0.9732 & \shortstack[l]{0.000094\\{}[-0.000195, 0.000363]} & 0.5395 & 1.0000 \\
\bottomrule\end{tabular}\par\endgroup\vspace{3pt}\begin{minipage}{0.98\textwidth}\footnotesize Negative Delta D favours consistency; positive Delta BAC favours accuracy. Raw, Pinv and augmentation belong to the retained twelve-test family per endpoint (three datasets, four original controls including interpolation). Fixed smoothing belongs to a separate two-test family per endpoint. Raw probabilities are two-sided paired sign-flip tests. Intervals are not multiplicity-adjusted; per-seed effects and the omitted interpolation contrasts are in Sections S5 and S13.\end{minipage}\end{table*}

Relative to Pinv, MOC reduced D by -0.018 on Cho2017 and -0.019 on
PhysioNet, with negative cohort effects at every seed (Figure 4b). BAC
differences were positive at every seed, but only the PhysioNet contrast
passed Holm adjustment (0.011; 95\% CI 0.006 to 0.016; Holm
\(p=0.0110\)). Against augmentation + centering, the PhysioNet D
contrast passed adjustment (-0.022; Holm \(p=0.0036\)); Cho2017 D and
both BAC contrasts did not. Cho2017's augmentation comparison changed
direction across seeds. The earlier single-fit ordering in which
augmentation had lower Cho2017 D, and Pinv lower PhysioNet D, did not
persist in the repeated cohort means. These earlier rankings are not
supported as stable findings; their analysis history is documented in
Section S13.

Fixed smoothing provides the most direct mechanism control. It shares
MOC's CAR-trained checkpoint, yet uses a reference-independent smoother
after Pinv at test time. MOC-minus-fixed-smoothing D was 0.000259 on
Cho2017 and 0.000016 on PhysioNet; BAC differences were -0.000003 and
0.000094. None passed the separate Holm family. The Cho2017 D interval
excluded zero before adjustment, but seed effects had mixed signs and
the adjusted probability was 0.0625. These small observed differences do
not establish equivalence. They provide no evidence here that the
operator-dependent fidelity term adds decoding benefit beyond fixed
smoothing at the specified coefficient and references.

\subsection{3.4. Fixed-checkpoint controls contextualize
disagreement}\label{fixed-checkpoint-controls-contextualize-disagreement}

Preserving nominally zero rows changed Raw D by 0.000335 (95\% CI
-0.000376 to 0.000919) on Cho2017 and -0.000807 (-0.001450 to -0.000149)
on PhysioNet. Pinv, MOC and fixed-smoothing class predictions were
unchanged. Interpolation and augmentation showed small changes because
centering alone can spread noise from a zero measurement row. The much
larger MOC-minus-raw D differences persisted under this paired control.
The complete results and unchanged-checkpoint checks are in Section S13;
the earlier one-seed sensitivity study is retained in S12.

Raw cross-reference D was 0.250 and 0.277, compared with same-operator
independent-noise D of 0.099 and 0.157 on Cho2017 and PhysioNet. For
MOC, the corresponding pairs were 0.107 versus 0.104, and 0.170 versus
0.170 (Figure 4c). The other canonical front ends showed similarly close
values. Thus a substantial disagreement level can remain when only the
noise realization changes, even after reference coordinates are aligned.
This reference helps interpret perturbation sensitivity; subtraction of
the two D values does not identify a coordinate-specific causal effect.

\begin{figure*}[!t]\centering
\includegraphics[width=\textwidth]{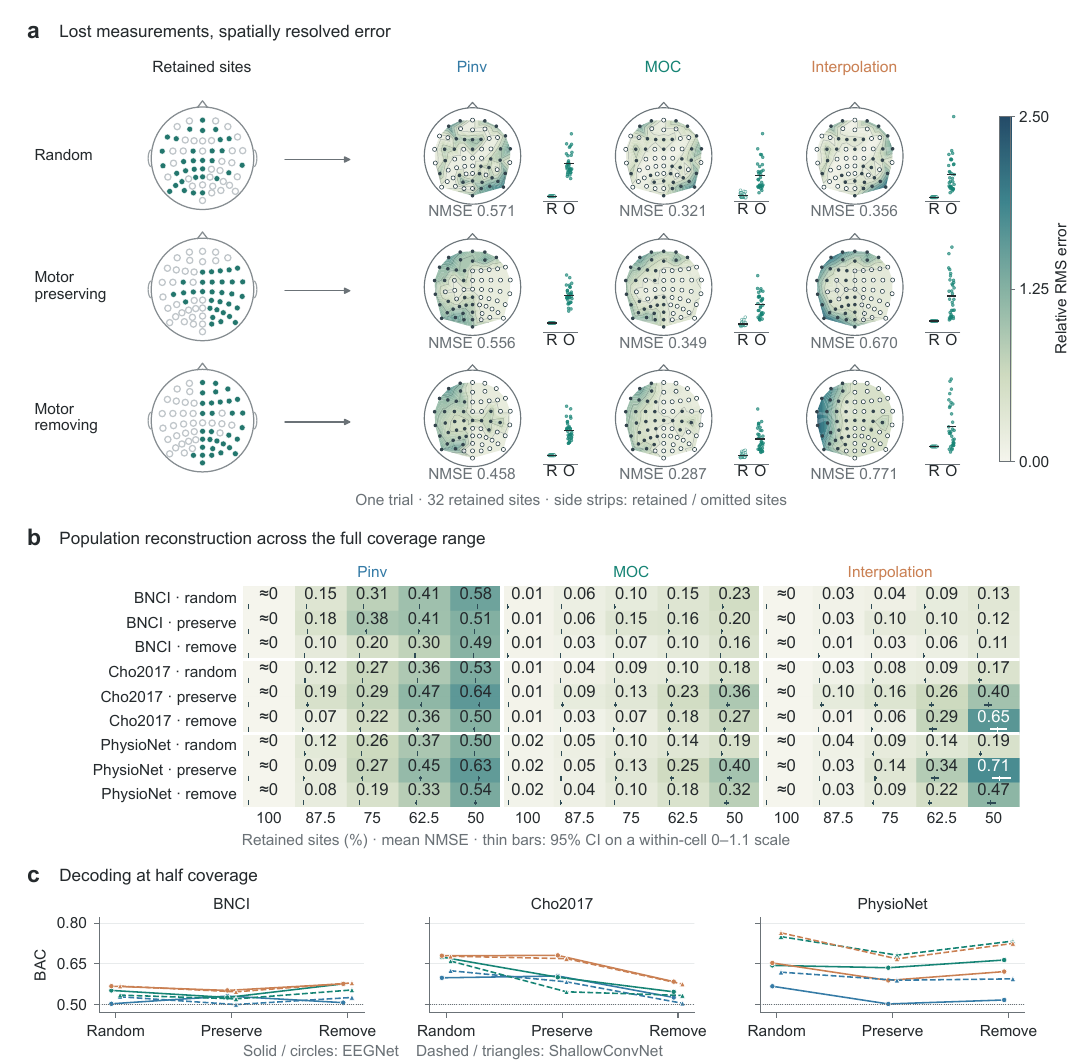}
\caption{Spatial information loss and population outcomes. (a) First predefined half-coverage Cho2017 mask in each family: 32 retained sites, rank 31, and the same trial excerpt as Figure 1. Electrode RMS error is normalized by whole recorded-field RMS. Maps interpolate within the electrode convex hull; open/dark sites indicate retained/omitted electrodes. Strips show the 32 retained (R) and 32 omitted (O) errors on a common 0–2.5 scale, with horizontal means. Under-map NMSE is the mean squared normalized error. (b) Cohort/family/method/retention NMSE means, with colour limits 0–1. Thin bars show 95\% subject bootstrap intervals on a within-cell 0–1.1 scale. Masks are averaged within subject; directed families have one unique mask per level. (c) Half-coverage BAC for full-coverage-trained EEGNet (solid/circles) and ShallowConvNet (dashed/triangles). Lines connect categorical mask families. All analyses re-center observations on the retained sites (M = P\_m S) and use no added output-channel noise. Complete intervals and actual retention counts are in Sections S2 and S8. Classification panels retain historical fits and their initialization limitation; reconstruction panels do not involve classifier training.}\label{fig:5}\end{figure*}

\subsection{3.5. Secondary analyses delimit the scope of the
evidence}\label{secondary-analyses-delimit-the-scope-of-the-evidence}

The retained architecture study uses separate historical fits and shared
observations across methods, but not the initialization and order-stream
controls of the primary repeated analysis. ShallowConvNet MOC-minus-raw
D differences were -0.057 on Cho2017 and -0.088 on PhysioNet. The
PhysioNet Conformer difference was 0.025 (95\% CI -0.018 to 0.074),
which does not establish either a decrease or a reversed response.
Chance-level canonical BNCI fits provide no useful robustness evidence.
All endpoints, including CSP and common-bipolar comparisons, remain in
Section S6.

The earlier subject-grouped analysis covered 52 Cho2017 and 104
PhysioNet subjects with three seeds. EEGNet MOC-minus-raw D was -0.126
(95\% CI -0.137 to -0.115) and -0.105 (-0.120 to -0.090); BAC
differences were 0.034 and 0.035. These results broaden within-cohort
partition coverage (Section S9). These fits used initialization before
seed resetting and method-specific output-noise draws; their estimates
describe partition sensitivity conditional on that implementation,
rather than independent paired replication.

\subsection{3.6. Missing electrodes changed the inferential
problem}\label{missing-electrodes-changed-the-inferential-problem}

At 50\% Cho2017 coverage, each illustrated mask retained 32 sites and
rank 31, yet preserved a different spatial pattern (Figure 5a). All
three masks lost relative-field components that the full 64-site domain
could observe. The reconstructed values at missing sites consequently
reflected an estimator's prior as well as the retained measurements.
Fixed-trial error maps and retained/omitted-site distributions
illustrate where the resulting errors occur, separately from the
population comparisons.

Under random removal, NMSE increased as coverage fell, and spatial
interpolation was competitive with or preferable to MOC (Figure 5b). At
half coverage, the three estimators and three mask families also
produced different decoding accuracies (Figure 5c). Filling the same
output dimensions therefore did not make their representations
equivalent. Complete retention-response curves are provided in the
supplementary digital tables.

The nonmonotonic, mask-dependent BAC responses are consistent with
differences between field recoverability and task-relevant information,
but do not identify the specific signal components responsible. Operator
rank identifies lost field dimensions; the decoding endpoint assesses
the remaining measurements for this binary task. Equal rank and equal
output size therefore provide insufficient grounds for expecting equal
classification performance.

\section{4. Discussion}\label{discussion}

Measurement definition is a distinct source of EEG prediction
variability. By holding the processed trial fixed, the experiment
without added output noise isolates changes caused by the signed rule
defining each channel. Raw decoders responded to those coordinate
changes even when the full relative field remained available. Exact
canonicalization removed this sensitivity before decoding, connecting a
standard inverse identity to an observable robustness property across
reference and bipolar representations.

The operator framework separates three questions that require different
evidence. Information-equivalent transformations of one recorded field
require coordinate alignment, including when acquisition noise is
already present. Additional output perturbations can reveal weak-mode
sensitivity in bipolar or controlled ill-conditioned maps; a spatial
prior trades that sensitivity against bias. Electrode loss removes
observable field components altogether, making reconstruction depend on
prior assumptions. Rank, absolute minimum positive singular value, and
the perturbation model jointly characterize these settings.

The complete-reference comparison locates the observed gains more
precisely than an estimator-versus-raw contrast alone. Each reference
has pseudoinverse norm one, so weak-mode amplification cannot explain
these classification results. With matched initial weights, data order
and observations, MOC improved D relative to Pinv in both cohorts, while
BAC evidence depended on the comparator and cohort. Fixed smoothing then
reproduced nearly the same observed D and BAC using the identical
trained checkpoint. This control supports spatial smoothing as a
sufficient account of the measured performance pattern in this setting;
it does not establish universal equivalence of the two estimators or
superiority of either operator-dependent estimation or ordinary
smoothing.

MOC-versus-raw differences retain their direction across all three seeds
and pass the retained correction after within-subject averaging.
Augmentation comparisons remain less consistent across endpoints, and
MOC's small fixed-smoothing contrasts do not establish an extra benefit.
Independent-noise disagreement also shows why residual prediction
variation after canonicalization should not automatically be attributed
to reference coordinates. Together, these controls provide the empirical
contribution beyond the known projector identity and beyond prior
demonstrations of reference-standardization or interpolation-based
motor-imagery decoding.

These findings motivate preserving reference weights, derivation
endpoints, and electrode-domain definitions alongside channel names.
Such metadata specify which transformations can be inverted and which
require estimation. They can support an operator-aware front end for
conventional or learned decoders and make robustness comparisons
interpretable across compatible measurement sets. The practical sequence
is to establish observability, assess inversion stability, and then
select a representation using the downstream task endpoint.

The evidence comes from retrospective transformations of motor-imagery
recordings. Independent Gaussian output perturbations remain a synthetic
stress model; the zero-row control tests one design choice without
validating an acquisition-noise mechanism. Three fixed training seeds
and three noise realizations assess repeatability within the specified
protocol, but do not characterize all training variability. Because the
held-out subjects had already been evaluated, this reanalysis corrects
procedural controls without providing an independent test of hypotheses
formed after the earlier results. Dataset-specific models also leave
cross-dataset transfer untested. Other architectures, grouped folds, SNR
decoding and missing-site classification retain historical fits and
their initialization limitations; the two-subject BNCI test set remains
descriptive. PhysioNet's fixed five-subject exclusion list lacks
subject-specific historical diagnostics, limiting the provenance of
cohort selection. Standard geometry omits individual electrode
variation. Transfer to independently acquired montages, correlated noise
and unknown measurement operators remains untested.

\section{5. Conclusion}\label{conclusion}

A controlled measurement-operator evaluation separates
reference-coordinate sensitivity, responses to additional output
perturbations, and spatial information loss in EEG decoding. The known
pseudoinverse identity supplies exact coordinate alignment. Under shared
complete-reference perturbations, paired repeated EEGNet fits support
gains relative to raw inputs, but do not establish an additional MOC
benefit over fixed graph smoothing with the same checkpoint. Weak-mode
reconstruction and missing-site estimation involve separate spectral and
observability constraints. Assessing these regimes with both consistency
and accuracy makes the limits of a decoding-robustness claim explicit.

\section{Data availability}\label{data-availability}

The original datasets are publicly available from their cited
repositories {[}15, 16, 17{]}, subject to their respective access and
usage conditions. Supplementary\_data\_and\_code.tar.xz, release FS2 (21
September 2026), supplies subject partitions, operator definitions,
derived predictions and outcomes, and statistical tables in
reports/publication\_evidence/. Its VERSION.txt identifies this
FS2 evidence release; the retained experimental phase is identified as
MR3. The ancillary archive repackages that evidence for this preprint
without additional training or changes to experimental results. Original recordings are not redistributed.

\section{Code availability}\label{code-availability}

Supplementary\_data\_and\_code.tar.xz, release FS2 (21 September 2026),
contains model and analysis code, fixed configurations, operator
matrices, a README with the reconstruction sequence, and source-hash
records. The supplementary material appended to this preprint describes
the procedures. The data and code archive is supplied as an ancillary
file with this preprint.

\section{Ethics statement}\label{ethics-statement}

This study used publicly released EEG datasets for secondary analysis
and involved no new recruitment or recording. The original acquisition
protocols are described in the cited dataset documentation {[}15, 16,
17{]}.

\section{Funding}\label{funding}

This research received no specific grant from funding agencies in the
public, commercial, or not-for-profit sectors.

\section{Declaration of competing
interest}\label{declaration-of-competing-interest}

The authors declare no competing interests.

\section{CRediT authorship contribution
statement}\label{credit-authorship-contribution-statement}

Runni Zhou: Conceptualization, Methodology, Software, Formal analysis,
Investigation, Visualization, Writing -- original draft. Rongrong Fu:
Data curation, Investigation, Validation, Writing -- review \& editing.
Xinghao Yan: Software, Validation, Visualization, Writing -- review \&
editing. Xiaoyang Li: Conceptualization, Methodology, Supervision,
Project administration, Writing -- review \& editing.

\section{Declaration of generative AI and AI-assisted technologies in
the manuscript preparation
process}\label{declaration-of-generative-ai-and-ai-assisted-technologies-in-the-manuscript-preparation-process}

Codex (OpenAI) assisted with code development, manuscript editing and
typesetting under the authors' oversight. The authors take full
responsibility for the content.

\Needspace{8\baselineskip}\section{References}\label{references}\small

{[}1{]} D. Yao, Y. Qin, S. Hu, L. Dong, M.L. Bringas Vega, P.A. Valdés
Sosa, Which reference should we use for EEG and ERP practice?, Brain
Topogr. 32 (2019) 530--549.
\url{https://doi.org/10.1007/s10548-019-00707-x}.

{[}2{]} S. Hu, D. Yao, M.L. Bringas-Vega, Y. Qin, P.A. Valdés-Sosa, The
statistics of EEG unipolar references: Derivations and properties, Brain
Topogr. 32 (2019) 696--703.
\url{https://doi.org/10.1007/s10548-019-00706-y}.

{[}3{]} D. Yao, A method to standardize a reference of scalp EEG
recordings to a point at infinity, Physiol. Meas. 22 (2001) 693--711.
\url{https://doi.org/10.1088/0967-3334/22/4/305}.

{[}4{]} S. Hu, D. Yao, P.A. Valdes-Sosa, Unified Bayesian estimator of
EEG reference at infinity: rREST (regularized reference electrode
standardization technique), Front. Neurosci. 12 (2018) 297.
\url{https://doi.org/10.3389/fnins.2018.00297}.

{[}5{]} L. Dong, L. Zhao, Y. Zhang, X. Yu, F. Li, J. Li, Y. Lai, T. Liu,
D. Yao, Reference electrode standardization interpolation technique
(RESIT): A novel interpolation method for scalp EEG, Brain Topogr. 34
(2021) 403--414. \url{https://doi.org/10.1007/s10548-021-00844-2}.

{[}6{]} F. Perrin, J. Pernier, O. Bertrand, J.F. Echallier, Spherical
splines for scalp potential and current density mapping,
Electroencephalogr. Clin. Neurophysiol. 72 (1989) 184--187.
\url{https://pubmed.ncbi.nlm.nih.gov/2464490/}.

{[}7{]} Y. Zhao, J. Han, Y. Chen, H. Sun, J. Chen, A. Ke, Y. Han, P.
Zhang, Y. Zhang, J. Zhou, C. Wang, Improving generalization based on
l1-norm regularization for EEG-based motor imagery classification,
Front. Neurosci. 12 (2018) 272.
\url{https://doi.org/10.3389/fnins.2018.00272}.

{[}8{]} M. Xu, J. Jiao, D. Chen, Y. Ding, Q. Chen, J. Wu, P. Gu, Y. Pan,
X. Peng, N. Xiao, B. Yang, Q. Li, J. Guo, REI-Net: A reference electrode
standardization interpolation technique based 3D CNN for motor imagery
classification, IEEE J. Biomed. Health Inform. 29 (2025) 2136--2147.
\url{https://doi.org/10.1109/JBHI.2024.3498916}.

{[}9{]} C. Yang, M.B. Westover, J. Sun, BIOT: Cross-data biosignal
learning in the wild, arXiv:2305.10351 (2023).
\url{https://arxiv.org/abs/2305.10351}.

{[}10{]} J. Wang, S. Zhao, Z. Luo, Y. Zhou, H. Jiang, S. Li, T. Li, G.
Pan, CBraMod: A criss-cross brain foundation model for EEG decoding,
International Conference on Learning Representations (2025).
\url{https://arxiv.org/abs/2412.07236}.

{[}11{]} B. Döner, T.M. Ingolfsson, L. Benini, Y. Li, LUNA: Efficient
and topology-agnostic foundation model for EEG signal analysis,
arXiv:2510.22257, version 1 (2025).
\url{https://arxiv.org/abs/2510.22257v1}.

{[}12{]} X.-H. Liu, B.-L. Lu, W.-L. Zheng, EEGMirror: Leveraging EEG
data in the wild via montage-agnostic self-supervision for EEG to video
decoding, Proceedings of the IEEE/CVF International Conference on
Computer Vision (2025) 18273--18283.
\url{https://openaccess.thecvf.com/content/ICCV2025/html/Liu_EEGMirror_Leveraging_EEG_Data_in_the_Wild_via_Montage-Agnostic_Self-Supervision_ICCV_2025_paper.html}.

{[}13{]} M. Jin, L. Yu, X. Wei, K. Li, E2EG2G: Latent node generation
and Graph-to-Grid conversion for unified EEG decoding, Inf. Fusion 136
(2026) 104580. \url{https://doi.org/10.1016/j.inffus.2026.104580}.

{[}14{]} H. Yin, X. Yin, Montage-guided fusion network for long-term
scalp electroencephalography seizure detection, Front. Neuroinform. 20
(2026) 1942480. \url{https://doi.org/10.3389/fninf.2026.1942480}.

{[}15{]} BCI Competition IV, Data set 2a description (2008).
\url{https://bbci.de/competition/iv/desc_2a.pdf}.

{[}16{]} H. Cho, M. Ahn, S. Ahn, M. Kwon, S.C. Jun, EEG datasets for
motor imagery brain--computer interface, GigaScience 6 (2017) gix034.
\url{https://doi.org/10.1093/gigascience/gix034}.

{[}17{]} PhysioNet, EEG Motor Movement/Imagery Dataset, version 1.0.0
(2009). \url{https://physionet.org/content/eegmmidb/1.0.0/}.

{[}18{]} V. Jayaram, A. Barachant, MOABB: Trustworthy algorithm
benchmarking for BCIs, J. Neural Eng. 15 (2018) 066011.
\url{https://arxiv.org/abs/1805.06427}.

{[}19{]} V.J. Lawhern, A.J. Solon, N.R. Waytowich, S.M. Gordon, C.P.
Hung, B.J. Lance, EEGNet: A compact convolutional neural network for
EEG-based brain--computer interfaces, J. Neural Eng. 15 (2018) 056013.
\url{https://doi.org/10.1088/1741-2552/aace8c}.

{[}20{]} R.T. Schirrmeister, J.T. Springenberg, L.D.J. Fiederer, M.
Glasstetter, K. Eggensperger, M. Tangermann, F. Hutter, W. Burgard, T.
Ball, Deep learning with convolutional neural networks for EEG decoding
and visualization, Hum. Brain Mapp. 38 (2017) 5391--5420.
\url{https://doi.org/10.1002/hbm.23730}.

{[}21{]} Y. Song, Q. Zheng, B. Liu, X. Gao, EEG Conformer: Convolutional
transformer for EEG decoding and visualization, IEEE Trans. Neural Syst.
Rehabil. Eng. 31 (2023) 710--719.
\url{https://ieeexplore.ieee.org/document/9991178}.

\end{document}


\maketitle

\section{S1. Cohorts, preprocessing, and
partitions}\label{s1.-cohorts-preprocessing-and-partitions}

This supplement reports the processed left/right motor-imagery cohorts.
The three primary test sets contain 2, 10, and 21 subjects. All views of
one trial retain its subject partition. The accompanying subject-trial
table gives class counts for every subject. No subject is selected based
on decoder performance. In experimental tables, the compact label
``Noiseless'' means without added output-channel noise; existing
recording noise remains in the processed target.

MOABB LeftRightImagery applies a 4--38 Hz bandpass and resampling to 128
Hz. The configured intervals are 2--6 s for BNCI2014-001 and 0--3 s for
Cho2017 and PhysioNet. The analyzed arrays contain 513, 384, and 385
time samples, respectively. EEG amplitudes are those returned by the
MOABB pipeline; arrays are converted to float32 and mean-centered across
electrode sites. Models do not apply a separate per-trial amplitude
normalization. Dataset-native acquisition references are not
reconstructed; the common-average projected array defines the empirical
field used to create every measurement view. Recording noise already
present in this array remains in the reconstruction target.

The frozen PhysioNet protocol excludes subjects 88, 89, 92, 100, and
104. The loader and partition records establish these identifiers, but
individual diagnostic records were not found in the retained materials.
Recording incompatibility or corruption therefore cannot be verified as
the reason for these exclusions. All other cached subjects were
included. This release preserves the original list without rescreening
or assigning retrospective reasons; the incomplete selection provenance
remains a cohort limitation.

\subsection{BNCI2014-001 subject
identifiers}\label{bnci2014-001-subject-identifiers}

\textbf{Training:} 1, 2, 3, 8, 9.

\textbf{Validation:} 5, 6.

\textbf{Test:} 4, 7.

\subsection{Cho2017 subject
identifiers}\label{cho2017-subject-identifiers}

\textbf{Training:} 2, 3, 5, 8, 9, 10, 12, 13, 14, 16, 17, 19, 20, 21,
23, 25, 28, 29, 32, 33, 34, 35, 38, 39, 40, 42, 43, 44, 48, 50, 51, 52.

\textbf{Validation:} 6, 7, 11, 18, 27, 30, 31, 37, 41, 49.

\textbf{Test:} 1, 4, 15, 22, 24, 26, 36, 45, 46, 47.

\subsection{PhysioNet subject
identifiers}\label{physionet-subject-identifiers}

\textbf{Training:} 2, 3, 4, 5, 6, 8, 9, 11, 13, 14, 17, 21, 22, 23, 25,
28, 33, 34, 35, 37, 38, 39, 41, 42, 43, 44, 47, 49, 50, 52, 53, 54, 55,
61, 62, 64, 66, 69, 71, 72, 74, 76, 77, 78, 79, 80, 81, 82, 83, 84, 85,
90, 91, 93, 95, 96, 97, 99, 103, 106, 108, 109.

\textbf{Validation:} 10, 12, 16, 19, 20, 29, 32, 48, 56, 57, 58, 60, 67,
73, 75, 87, 94, 98, 102, 105, 107.

\textbf{Test:} 1, 7, 15, 18, 24, 26, 27, 30, 31, 36, 40, 45, 46, 51, 59,
63, 65, 68, 70, 86, 101.

\section{S2. Operator construction, spectra, and
masks}\label{s2.-operator-construction-spectra-and-masks}

The electrode order is given in the machine-readable electrode-domain
file. CAR is \(P_K=I-11^T/K\). A single reference is \(I-1e_r^T\).
Weighted references use \(I-1w^T\), with independent uniform positive
draws normalized to sum to one. The two weighted-reference random seeds
are 20260950 and 20260981. All measurement rows sum to zero.

Connected bipolar matrices have one +1 and one -1 per row. Four spanning
trees are used: a Euclidean minimum-spanning tree; an anterior-posterior
preferred tree completed to connectivity; and two distance-biased random
spanning trees generated with seeds 20260936 and 20260953. The trees are
computational constructions on standard electrode coordinates. Their
incidence matrices have K-1 rows and rank K-1.

Controlled operators are \(M=U\Sigma B^T\), where \(B\) is an
orthonormal mean-zero basis, \(U\) has orthonormal columns from a
Gaussian QR decomposition, and the K-1 positive singular values are
geometrically spaced from 1 to \(1/\kappa\). Conditions are 1, 3, 10,
30, 100, 300, and 1000; operator seeds are 20260919--20260923. The
smallest positive singular value is recorded separately from the gauge
zero. All 135 full-quotient-rank matrices, their positive spectra, and
the graph Laplacians are provided as numerical files. This includes 30
reference/bipolar operators and 105 controlled matrices.

The spatial penalty uses the unnormalized Laplacian \(L=D-W\). Four
nearest nonself neighbors per electrode define a symmetric adjacency
union. For the Gaussian affinity bandwidth, the implementation takes the
median of index 4 after partitioning distances with self-distance
replaced by the maximum distance; this corresponds to the fifth ordered
nonself distance when the self replacement is largest. The supplied
graph matrices retain this bandwidth definition. Weights on adjacent
sites are \(\exp[-d_{ij}^2/(2\sigma^2)]\). The neighborhood size and
bandwidth rule are fixed geometry hyperparameters chosen independently
of test decoding performance; they are not selected by a decoding
sensitivity search. No individualized digitized coordinates are used.

Partial-site measurements use \(M=P_qS\), where S selects q retained
electrodes. Thus rank(M)=q-1 and the null space has dimension K-q+1.
Interpolation and MOC return K outputs but do not increase this
measurement rank. Random masks have five unique realizations below full
coverage. Motor-preserving and motor-removing masks are deterministic
orderings relative to the defined motor-electrode set, with one unique
mask per level. The five stored copies of each directed mask are
collapsed before estimating subject-level uncertainty.

\begingroup\small\setlength{\tabcolsep}{4pt}\renewcommand{\arraystretch}{1.15}
\begin{longtable}{l l l l l l}
\caption{Reference and bipolar operator diagnostics. The gauge zero is excluded from the condition number.}\\
\toprule
Dataset & Operator & Rows & Rank & Min. positive & Condition \\
\midrule\endfirsthead
\toprule
Dataset & Operator & Rows & Rank & Min. positive & Condition \\
\midrule\endhead
BNCI2014-001 & car & 22 & 21 & 1.000 & 1.000 \\
BNCI2014-001 & bipolar mst & 21 & 21 & 0.221 & 10.211 \\
BNCI2014-001 & bipolar longitudinal & 21 & 21 & 0.143 & 13.982 \\
BNCI2014-001 & ref Cz & 22 & 21 & 1.000 & 4.690 \\
BNCI2014-001 & ref Pz & 22 & 21 & 1.000 & 4.690 \\
BNCI2014-001 & ref C3 & 22 & 21 & 1.000 & 4.690 \\
BNCI2014-001 & bipolar tree 0 & 21 & 21 & 0.203 & 11.377 \\
BNCI2014-001 & bipolar tree 1 & 21 & 21 & 0.254 & 9.565 \\
BNCI2014-001 & weighted ref 0 & 22 & 21 & 1.000 & 1.173 \\
BNCI2014-001 & weighted ref 1 & 22 & 21 & 1.000 & 1.172 \\
Cho2017 & car & 64 & 63 & 1.000 & 1.000 \\
Cho2017 & bipolar mst & 63 & 63 & 0.089 & 26.512 \\
Cho2017 & bipolar longitudinal & 63 & 63 & 0.049 & 40.735 \\
Cho2017 & ref Cz & 64 & 63 & 1.000 & 8.000 \\
Cho2017 & ref Pz & 64 & 63 & 1.000 & 8.000 \\
Cho2017 & ref Oz & 64 & 63 & 1.000 & 8.000 \\
Cho2017 & bipolar tree 0 & 63 & 63 & 0.113 & 22.440 \\
Cho2017 & bipolar tree 1 & 63 & 63 & 0.133 & 19.253 \\
Cho2017 & weighted ref 0 & 64 & 63 & 1.000 & 1.136 \\
Cho2017 & weighted ref 1 & 64 & 63 & 1.000 & 1.175 \\
PhysioNet & car & 64 & 63 & 1.000 & 1.000 \\
PhysioNet & bipolar mst & 63 & 63 & 0.089 & 26.304 \\
PhysioNet & bipolar longitudinal & 63 & 63 & 0.049 & 40.735 \\
PhysioNet & ref Cz & 64 & 63 & 1.000 & 8.000 \\
PhysioNet & ref Pz & 64 & 63 & 1.000 & 8.000 \\
PhysioNet & ref Oz & 64 & 63 & 1.000 & 8.000 \\
PhysioNet & bipolar tree 0 & 63 & 63 & 0.088 & 28.212 \\
PhysioNet & bipolar tree 1 & 63 & 63 & 0.088 & 27.828 \\
PhysioNet & weighted ref 0 & 64 & 63 & 1.000 & 1.136 \\
PhysioNet & weighted ref 1 & 64 & 63 & 1.000 & 1.175 \\
\bottomrule\end{longtable}\endgroup

\begingroup\small\setlength{\tabcolsep}{4pt}\renewcommand{\arraystretch}{1.15}
\begin{longtable}{l l l l l}
\caption{Actual site counts and ranks for all mask families. The last column counts unique random masks; directed families have one unique mask.}\\
\toprule
Dataset & Nominal retention & Sites & Rank & Random masks \\
\midrule\endfirsthead
\toprule
Dataset & Nominal retention & Sites & Rank & Random masks \\
\midrule\endhead
BNCI2014-001 & 100\% & 22 & 21 & 1 \\
BNCI2014-001 & 87.5\% & 19 & 18 & 5 \\
BNCI2014-001 & 75\% & 16 & 15 & 5 \\
BNCI2014-001 & 62.5\% & 14 & 13 & 5 \\
BNCI2014-001 & 50\% & 11 & 10 & 5 \\
Cho2017 & 100\% & 64 & 63 & 1 \\
Cho2017 & 87.5\% & 56 & 55 & 5 \\
Cho2017 & 75\% & 48 & 47 & 5 \\
Cho2017 & 62.5\% & 40 & 39 & 5 \\
Cho2017 & 50\% & 32 & 31 & 5 \\
PhysioNet & 100\% & 64 & 63 & 1 \\
PhysioNet & 87.5\% & 56 & 55 & 5 \\
PhysioNet & 75\% & 48 & 47 & 5 \\
PhysioNet & 62.5\% & 40 & 39 & 5 \\
PhysioNet & 50\% & 32 & 31 & 5 \\
\bottomrule\end{longtable}\endgroup

\section{S3. Estimation, training, and
software}\label{s3.-estimation-training-and-software}

MOC solves the constrained quadratic objective in the main text with
identity noise precision and a graph penalty. Its coefficient is 0.1 for
the main decoding and response analyses. The exploratory validation grid
was 0.0001, 0.001, 0.01, 0.1, and 1.0, with selection emphasizing
condition numbers at least 100 and SNR at most 10 dB. Selection used
validation reconstruction error. The exploratory PhysioNet
reconstruction analysis used 0.01 and is reported separately. Spherical
interpolation uses MNE's spherical-spline matrix with regularization
\(10^{-5}\), followed by mean centering. Observed sites are copied,
missing sites are interpolated, and site interpolation is undefined for
arbitrary bipolar channels.

The neural optimizer is Adam with learning rate 0.001 and weight decay
0.0001, cross-entropy loss, CUDA batch size 128 (CPU fallback 32), and
minimum validation-loss checkpoint selection. Ideal sensitivity runs use
20 epochs; later noisy comparisons, response studies, and grouped
analyses use 15. Training on CAR is used for reference comparisons and
full-coverage training for missing-site decoding. In the
controlled-mixture decoding experiment, each condition uses the operator
with seed 20260919 for training and validation, with all five operator
seeds used for test views; training and test measurement distributions
therefore vary with condition.

EEGNet uses eight temporal filters, depth multiplier two, sixteen output
feature maps, temporal kernel length 32, separable temporal length 16,
pooling factors 4 and 8, and dropout 0.25. ShallowConvNet uses forty
temporal/spatial feature maps, temporal kernel length 13, square
nonlinearity, average pooling length 35 with stride 7, log nonlinearity,
and dropout 0.25. Both terminate in a linear classifier. EEG Conformer
uses Braindecode 1.8.1 defaults: forty temporal filters, kernel length
25, pooling length 75 and stride 15, dropout 0.5, six attention layers,
ten heads, attention dropout 0.5, ELU convolution activation, GELU
transformer activation, and automatic final feature length. It has two
outputs and dataset-specific channel/time dimensions. The architecture
used these defaults without additional tuning.

CSP receives the measurement/front-end arrays in the numerical amplitude
scale inherited from MOABB. MOABB extracts MNE epoch arrays in volts and
multiplies them by the dataset unit factor \(10^6\) for BNCI2014-001,
Cho2017, and PhysioNet, giving microvolt-scaled electrode inputs. The
project loader casts these arrays to float32 and projects to the
mean-zero domain; the measurement operators and linear front ends
introduce no additional global amplitude rescaling. There is no
per-trial amplitude normalization or input standardization before CSP.
For each float64 trial \(X\) with \(T\) time samples, the implementation
subtracts each channel temporal mean and computes
\(C_X=(X-\bar X)(X-\bar X)^T/T\). It averages these matrices within each
class without trace normalization, then adds \(10^{-6}I\) to each class
covariance. This is an absolute loading in squared numerical input
units, not a fraction of the covariance trace; for the site-potential
inputs the underlying scale is microvolt squared. The combined
covariance in the generalized eigenproblem contains twice that loading.
CSP uses diagonal loading rather than explicit quotient-rank reduction.
The generalized eigenproblem is solved by scipy.linalg.eigh with finite
checking. NumPy pseudoinverses elsewhere use relative SVD cutoff 1e-15;
reduced MOC Hessians use np.linalg.inv and fall back to np.linalg.pinv
on LinAlgError. There is no test-dependent numerical threshold or
noise-precision fitting. CSP uses four filters, comprising the two
lowest and two highest eigenvectors of \(C_0w=\mu(C_0+C_1)w\). For each
selected filter, projection is applied to the trial array and NumPy
temporal variance uses divisor \(T\) (ddof = 0). The \(10^{-10}\) floor
is applied to these CSP-projected variances immediately before the
natural logarithm; it is not a floor on raw electrode variance.
Generalized-eigenvector normalization determines the projected numerical
scale. StandardScaler is fitted to the resulting training log-variance
features, followed by sklearn LinearDiscriminantAnalysis with its
default SVD solver; those fitted transforms are applied to test
features. All reported CSP results use this symmetric generalized
eigensystem, and interpolation matrices map observed-site potentials to
target sites. Implementation checks and superseded outputs are
documented in the reproducibility archive.

The base random seed is 20260919. Ideal-study optimization seeds are
drawn sequentially from that base generator; grouped validation uses
20260919, 20260920, and 20260921. Later response training uses 20260919.
Historical model construction precedes seed resetting within the fitting
function, and CUDA benchmarking is enabled. Saved predictions reproduce
the reported statistics. Neural retraining may yield different weights
because initialization precedes seed resetting. Subject intervals
condition on the fitted models.

The output-perturbation model is \(y_j=M_jv+\epsilon_j\), with nominal
variance
\(\sigma_\epsilon^2=\operatorname{mean}[(M_jv)^2]10^{-\mathrm{SNR}/10}\).
Gaussian entries are independent across output channels and time; the
realized power is not normalized afterward. Noise also enters nominally
zero reference rows. All means include every channel and time sample in
the processing unit below; blocks follow stored trial order within a
split and can cross subject boundaries. This construction differs from
re-referencing a shared noisy field, \(M_j(v+\eta)\), whose covariance
transforms with the operator and whose pseudoinverse remains exactly
aligned.

In the primary five-pipeline comparison, each operator and each
train/validation/test split is processed in consecutive blocks of at
most 256 trials. The last shorter block uses its own mean power. Native
reconstruction and controlled-conditioning reconstruction also use
blocks of at most 256 test trials per operator (and per
controlled-operator seed). Within these blocks, all front ends share the
same noisy observations. The exploratory reconstruction selection
likewise used blocks of at most 256 trials.

The separate architecture comparison applies the noise function once to
each complete split/operator array, sharing that array across Raw, Pinv,
and MOC. SNR-response decoding and controlled-conditioning decoding
compute power over each complete split/operator array separately for
each representation, so noise draws differ across methods.
Subject-grouped decoding computes power over the complete fold
split/operator array per representation. The main 256-trial processing
blocks are distinct from the optimizer's 128-trial CUDA minibatches.

Augmentation + centering constructs one fixed array before fitting: each
training or validation trial independently selects CAR or an available
Cz/Pz/Oz/C3 reference, receives output noise scaled by that trial's
complete channel-by-time power, and is interpolated and centered. There
is no epoch-wise resampling. Full-coverage centering cancels the
reference offset because \(PM=P\). Only the added noise scale and
realization, floating-point differences, and any downstream training
effects remain; the deterministic reference signal variation itself is
absent.

In the historical MR2 primary experiment, Pinv and full-coverage
interpolation + centering shared the same CAR observations, including
the same noise, and hence have algebraically identical CAR inputs. Both
use 15 epochs, optimization seed 20260919, Adam, and
minimum-validation-loss selection. The network constructor is called
before the seed reset inside the fitting function; its initial weights
are therefore not controlled by that reset. CUDA benchmarking is also
enabled. Saved configuration and code establish these differences in
control, but retained outputs do not identify the exact contribution of
initialization versus numerical training variation. Those historical
pipeline score differences cannot isolate a CAR front-end effect. The
MR3 primary comparison replaces the Cho2017 and PhysioNet cells with
three paired seeds, explicit random streams and checkpoint sharing,
documented in Section S13. The separate response analyses additionally
use independent noise draws. Missing-site analyses use no added
output-channel noise.

\Needspace{12\baselineskip}

\begingroup\small\setlength{\tabcolsep}{4pt}\renewcommand{\arraystretch}{1.15}
\begin{longtable}{p{0.27\linewidth} p{0.65\linewidth}}
\caption{Environment used for the final analysis and correction reruns. Historical scores use the archived local research environment.}\\
\toprule
Component & Version \\
\midrule\endfirsthead
\toprule
Component & Version \\
\midrule\endhead
numpy & 2.3.5 \\
scipy & 1.16.3 \\
pandas & 2.3.3 \\
scikit-learn & 1.7.2 \\
torch & 2.11.0+cu128 \\
mne & 1.12.1 \\
moabb & 1.7.1 \\
braindecode & 1.8.1 \\
matplotlib & 3.10.6 \\
PyMuPDF & 1.28.2 \\
Python & 3.13.9 \\
platform & Windows-11-10.0.26200-SP0 \\
GPU & NVIDIA GeForce RTX 5060 Ti, 16 GB \\
\bottomrule\end{longtable}\endgroup

\section{S4. Ideal sensitivity and operator
correctness}\label{s4.-ideal-sensitivity-and-operator-correctness}

Disagreement is the mean fraction of unequal trial predictions over
unordered operator pairs. Raw fixed-channel models use six site-like
operators and fifteen pairs; inverse-canonical models use all ten
information-equivalent operators and forty-five pairs. The noiseless
suite evaluates exact canonical consistency across these compatible
inputs. Canonical disagreement is zero for every included subject and
decoder; classification endpoints describe task performance.

\begingroup\small\setlength{\tabcolsep}{4pt}\renewcommand{\arraystretch}{1.15}
\begin{longtable}{l l l l l l}
\caption{Complete ideal-sensitivity means. Subject-level values and intervals for every endpoint are provided in the digital tables.}\\
\toprule
Dataset & Decoder & Method & D & BAC & Macro-F1 \\
\midrule\endfirsthead
\toprule
Dataset & Decoder & Method & D & BAC & Macro-F1 \\
\midrule\endhead
BNCI2014-001 & EEG Conformer & Pinv & 0.000 & 0.609 & 0.608 \\
BNCI2014-001 & EEG Conformer & Raw & 0.004 & 0.502 & 0.340 \\
BNCI2014-001 & CSP + LDA & Pinv & 0.000 & 0.542 & 0.486 \\
BNCI2014-001 & CSP + LDA & Raw & 0.001 & 0.541 & 0.486 \\
BNCI2014-001 & EEGNet & Pinv & 0.000 & 0.595 & 0.580 \\
BNCI2014-001 & EEGNet & Raw & 0.252 & 0.595 & 0.562 \\
BNCI2014-001 & ShallowConvNet & Pinv & 0.000 & 0.585 & 0.521 \\
BNCI2014-001 & ShallowConvNet & Raw & 0.091 & 0.543 & 0.475 \\
Cho2017 & EEG Conformer & Pinv & 0.000 & 0.709 & 0.695 \\
Cho2017 & EEG Conformer & Raw & 0.084 & 0.717 & 0.699 \\
Cho2017 & CSP + LDA & Pinv & 0.000 & 0.577 & 0.518 \\
Cho2017 & CSP + LDA & Raw & 0.096 & 0.513 & 0.364 \\
Cho2017 & EEGNet & Pinv & 0.000 & 0.728 & 0.716 \\
Cho2017 & EEGNet & Raw & 0.274 & 0.670 & 0.649 \\
Cho2017 & ShallowConvNet & Pinv & 0.000 & 0.670 & 0.655 \\
Cho2017 & ShallowConvNet & Raw & 0.119 & 0.660 & 0.647 \\
PhysioNet & EEG Conformer & Pinv & 0.000 & 0.641 & 0.600 \\
PhysioNet & EEG Conformer & Raw & 0.092 & 0.577 & 0.520 \\
PhysioNet & CSP + LDA & Pinv & 0.000 & 0.596 & 0.547 \\
PhysioNet & CSP + LDA & Raw & 0.299 & 0.555 & 0.460 \\
PhysioNet & EEGNet & Pinv & 0.000 & 0.713 & 0.697 \\
PhysioNet & EEGNet & Raw & 0.172 & 0.638 & 0.571 \\
PhysioNet & ShallowConvNet & Pinv & 0.000 & 0.777 & 0.773 \\
PhysioNet & ShallowConvNet & Raw & 0.089 & 0.758 & 0.750 \\
\bottomrule\end{longtable}\endgroup

\section{S5. Matched five-method results and
inference}\label{s5.-matched-five-method-results-and-inference}

These tables retain the original five-method family. Cho2017 and
PhysioNet now use the MR3 repeated paired EEGNet comparison at 10 dB;
BNCI retains its historical descriptive fits. The comparison uses the
same six reference operators and noisy test arrays for every method. The
augmentation + centering decoder is evaluated on the
interpolation-and-centering view. Every subject contributes fifteen
pairs. The MR3 cells average three training seeds and three noise
realizations within subject first. Independent sample sizes remain 10
and 21. Fixed smoothing is reported separately in Section S13. Main
Figure 4 presents aligned disagreement and accuracy endpoints and paired
differences; the following tables report complete means and macro-F1.

Percentile bootstrap intervals use 10,000 subject resamples with seed
20260919. Paired intervals resample subject differences, not unpaired
method means. Two-sided random-sign tests enumerate all sign
combinations for n at most 16, and use 10,000 draws with a plus-one
correction otherwise (seed 20260920). Numerical equality uses tolerance
\(10^{-14}\). Holm correction treats every dataset/decoder/baseline
contrast for one endpoint within an analysis family as one family:
twelve tests per endpoint in the five-method comparison, twenty-four in
the architecture comparison, and twelve in grouped validation. These
retrospective analyses report effect estimates and intervals alongside
adjusted probabilities.

\begingroup\small\setlength{\tabcolsep}{4pt}\renewcommand{\arraystretch}{1.15}
\begin{longtable}{l l l l l l}
\caption{Five-method EEGNet results; common operators, 10 dB.}\\
\toprule
Dataset & Decoder & Method & D & BAC & Macro-F1 \\
\midrule\endfirsthead
\toprule
Dataset & Decoder & Method & D & BAC & Macro-F1 \\
\midrule\endhead
BNCI2014-001 & EEGNet & Aug. + centering & 0.136 & 0.579 & 0.566 \\
BNCI2014-001 & EEGNet & Interpolation & 0.104 & 0.554 & 0.529 \\
BNCI2014-001 & EEGNet & MOC & 0.088 & 0.549 & 0.521 \\
BNCI2014-001 & EEGNet & Pinv & 0.097 & 0.552 & 0.526 \\
BNCI2014-001 & EEGNet & Raw & 0.265 & 0.557 & 0.499 \\
Cho2017 & EEGNet & Aug. + centering & 0.115 & 0.699 & 0.690 \\
Cho2017 & EEGNet & Interpolation & 0.126 & 0.702 & 0.694 \\
Cho2017 & EEGNet & MOC & 0.107 & 0.709 & 0.701 \\
Cho2017 & EEGNet & Pinv & 0.125 & 0.702 & 0.694 \\
Cho2017 & EEGNet & Raw & 0.250 & 0.665 & 0.641 \\
PhysioNet & EEGNet & Aug. + centering & 0.192 & 0.668 & 0.656 \\
PhysioNet & EEGNet & Interpolation & 0.190 & 0.664 & 0.650 \\
PhysioNet & EEGNet & MOC & 0.170 & 0.676 & 0.662 \\
PhysioNet & EEGNet & Pinv & 0.189 & 0.665 & 0.651 \\
PhysioNet & EEGNet & Raw & 0.277 & 0.627 & 0.586 \\
\bottomrule\end{longtable}\endgroup

\begingroup\small\setlength{\tabcolsep}{4pt}\renewcommand{\arraystretch}{1.15}
\begin{longtable}{l l l l l l}
\caption{Paired five-method contrasts for D; 95\% subject intervals.}\\
\toprule
Dataset & Decoder & Baseline & MOC minus baseline & p & Holm p \\
\midrule\endfirsthead
\toprule
Dataset & Decoder & Baseline & MOC minus baseline & p & Holm p \\
\midrule\endhead
BNCI2014-001 & EEGNet & Raw & -0.177 [-0.223, -0.131] & 0.5000 & 1.0000 \\
BNCI2014-001 & EEGNet & Pinv & -0.010 [-0.015, -0.005] & 0.5000 & 1.0000 \\
BNCI2014-001 & EEGNet & Interpolation & -0.017 [-0.023, -0.010] & 0.5000 & 1.0000 \\
BNCI2014-001 & EEGNet & Aug. + centering & -0.048 [-0.053, -0.044] & 0.5000 & 1.0000 \\
Cho2017 & EEGNet & Raw & -0.144 [-0.168, -0.112] & 0.0020 & 0.0156 \\
Cho2017 & EEGNet & Pinv & -0.018 [-0.021, -0.015] & 0.0020 & 0.0156 \\
Cho2017 & EEGNet & Interpolation & -0.019 [-0.022, -0.016] & 0.0020 & 0.0156 \\
Cho2017 & EEGNet & Aug. + centering & -0.008 [-0.013, -0.002] & 0.0254 & 0.1270 \\
PhysioNet & EEGNet & Raw & -0.107 [-0.133, -0.080] & 0.0002 & 0.0020 \\
PhysioNet & EEGNet & Pinv & -0.019 [-0.022, -0.015] & 0.0001 & 0.0012 \\
PhysioNet & EEGNet & Interpolation & -0.020 [-0.024, -0.016] & 0.0001 & 0.0012 \\
PhysioNet & EEGNet & Aug. + centering & -0.022 [-0.036, -0.012] & 0.0004 & 0.0036 \\
\bottomrule\end{longtable}\endgroup

\begingroup\small\setlength{\tabcolsep}{4pt}\renewcommand{\arraystretch}{1.15}
\begin{longtable}{l l l l l l}
\caption{Paired five-method contrasts for BAC; 95\% subject intervals.}\\
\toprule
Dataset & Decoder & Baseline & MOC minus baseline & p & Holm p \\
\midrule\endfirsthead
\toprule
Dataset & Decoder & Baseline & MOC minus baseline & p & Holm p \\
\midrule\endhead
BNCI2014-001 & EEGNet & Raw & -0.008 [-0.012, -0.003] & 0.5000 & 1.0000 \\
BNCI2014-001 & EEGNet & Pinv & -0.003 [-0.007, 0.002] & 1.0000 & 1.0000 \\
BNCI2014-001 & EEGNet & Interpolation & -0.004 [-0.010, 0.002] & 1.0000 & 1.0000 \\
BNCI2014-001 & EEGNet & Aug. + centering & -0.030 [-0.057, -0.002] & 0.5000 & 1.0000 \\
Cho2017 & EEGNet & Raw & 0.044 [0.031, 0.056] & 0.0020 & 0.0176 \\
Cho2017 & EEGNet & Pinv & 0.007 [0.003, 0.011] & 0.0156 & 0.1094 \\
Cho2017 & EEGNet & Interpolation & 0.007 [0.003, 0.012] & 0.0117 & 0.0938 \\
Cho2017 & EEGNet & Aug. + centering & 0.010 [0.002, 0.019] & 0.0527 & 0.3164 \\
PhysioNet & EEGNet & Raw & 0.050 [0.033, 0.066] & 0.0001 & 0.0012 \\
PhysioNet & EEGNet & Pinv & 0.011 [0.006, 0.016] & 0.0011 & 0.0110 \\
PhysioNet & EEGNet & Interpolation & 0.012 [0.006, 0.017] & 0.0007 & 0.0077 \\
PhysioNet & EEGNet & Aug. + centering & 0.008 [0.000, 0.017] & 0.0740 & 0.3700 \\
\bottomrule\end{longtable}\endgroup

\begingroup\small\setlength{\tabcolsep}{4pt}\renewcommand{\arraystretch}{1.15}
\begin{longtable}{l l l l l l}
\caption{Paired five-method contrasts for macro-F1; 95\% subject intervals.}\\
\toprule
Dataset & Decoder & Baseline & MOC minus baseline & p & Holm p \\
\midrule\endfirsthead
\toprule
Dataset & Decoder & Baseline & MOC minus baseline & p & Holm p \\
\midrule\endhead
BNCI2014-001 & EEGNet & Raw & 0.022 [-0.034, 0.079] & 1.0000 & 1.0000 \\
BNCI2014-001 & EEGNet & Pinv & -0.004 [-0.010, 0.002] & 1.0000 & 1.0000 \\
BNCI2014-001 & EEGNet & Interpolation & -0.008 [-0.017, 0.001] & 1.0000 & 1.0000 \\
BNCI2014-001 & EEGNet & Aug. + centering & -0.045 [-0.089, 0.000] & 1.0000 & 1.0000 \\
Cho2017 & EEGNet & Raw & 0.060 [0.043, 0.076] & 0.0020 & 0.0176 \\
Cho2017 & EEGNet & Pinv & 0.007 [0.003, 0.011] & 0.0117 & 0.0820 \\
Cho2017 & EEGNet & Interpolation & 0.007 [0.003, 0.012] & 0.0078 & 0.0625 \\
Cho2017 & EEGNet & Aug. + centering & 0.011 [-0.000, 0.022] & 0.1113 & 0.6680 \\
PhysioNet & EEGNet & Raw & 0.077 [0.057, 0.097] & 0.0001 & 0.0012 \\
PhysioNet & EEGNet & Pinv & 0.012 [0.006, 0.018] & 0.0014 & 0.0140 \\
PhysioNet & EEGNet & Interpolation & 0.012 [0.006, 0.018] & 0.0011 & 0.0121 \\
PhysioNet & EEGNet & Aug. + centering & 0.006 [-0.003, 0.016] & 0.1871 & 0.9354 \\
\bottomrule\end{longtable}\endgroup

\section{S6. Architecture dependence and bipolar
coverage}\label{s6.-architecture-dependence-and-bipolar-coverage}

The architecture comparison retains historical models, fitted separately
from the MR3 primary EEGNet experiment. It shares observations across
methods but uses initialization before seed resetting and CUDA
benchmarking. It is conditional descriptive evidence, not a repeat under
the MR3 paired protocol. All method contrasts below use the six shared
reference operators. Bipolar-only results use the same four trees for
Pinv and MOC; Raw is undefined. In the full operator suite, CSP on
PhysioNet has mean D 0.082 for Raw and 0.094 for MOC across six and ten
operators, respectively. The common-reference tables below align the
comparison on six operators and include all decoder outcomes.

\begingroup\small\setlength{\tabcolsep}{4pt}\renewcommand{\arraystretch}{1.15}
\begin{longtable}{l l l l l l}
\caption{Architecture comparison on six shared references at 10 dB.}\\
\toprule
Dataset & Decoder & Method & D & BAC & Macro-F1 \\
\midrule\endfirsthead
\toprule
Dataset & Decoder & Method & D & BAC & Macro-F1 \\
\midrule\endhead
BNCI2014-001 & EEG Conformer & MOC & 0.000 & 0.500 & 0.333 \\
BNCI2014-001 & EEG Conformer & Pinv & 0.000 & 0.500 & 0.333 \\
BNCI2014-001 & EEG Conformer & Raw & 0.095 & 0.531 & 0.439 \\
BNCI2014-001 & CSP + LDA & MOC & 0.103 & 0.504 & 0.432 \\
BNCI2014-001 & CSP + LDA & Pinv & 0.101 & 0.504 & 0.431 \\
BNCI2014-001 & CSP + LDA & Raw & 0.450 & 0.524 & 0.443 \\
BNCI2014-001 & EEGNet & MOC & 0.097 & 0.601 & 0.598 \\
BNCI2014-001 & EEGNet & Pinv & 0.116 & 0.593 & 0.592 \\
BNCI2014-001 & EEGNet & Raw & 0.362 & 0.576 & 0.520 \\
BNCI2014-001 & ShallowConvNet & MOC & 0.054 & 0.568 & 0.504 \\
BNCI2014-001 & ShallowConvNet & Pinv & 0.053 & 0.558 & 0.488 \\
BNCI2014-001 & ShallowConvNet & Raw & 0.098 & 0.546 & 0.462 \\
Cho2017 & EEG Conformer & MOC & 0.113 & 0.686 & 0.672 \\
Cho2017 & EEG Conformer & Pinv & 0.124 & 0.688 & 0.670 \\
Cho2017 & EEG Conformer & Raw & 0.173 & 0.673 & 0.644 \\
Cho2017 & CSP + LDA & MOC & 0.063 & 0.555 & 0.491 \\
Cho2017 & CSP + LDA & Pinv & 0.064 & 0.555 & 0.490 \\
Cho2017 & CSP + LDA & Raw & 0.078 & 0.554 & 0.487 \\
Cho2017 & EEGNet & MOC & 0.120 & 0.710 & 0.701 \\
Cho2017 & EEGNet & Pinv & 0.142 & 0.709 & 0.701 \\
Cho2017 & EEGNet & Raw & 0.198 & 0.687 & 0.671 \\
Cho2017 & ShallowConvNet & MOC & 0.071 & 0.676 & 0.666 \\
Cho2017 & ShallowConvNet & Pinv & 0.078 & 0.676 & 0.666 \\
Cho2017 & ShallowConvNet & Raw & 0.128 & 0.685 & 0.678 \\
PhysioNet & EEG Conformer & MOC & 0.158 & 0.640 & 0.614 \\
PhysioNet & EEG Conformer & Pinv & 0.176 & 0.644 & 0.623 \\
PhysioNet & EEG Conformer & Raw & 0.132 & 0.564 & 0.514 \\
PhysioNet & CSP + LDA & MOC & 0.048 & 0.559 & 0.480 \\
PhysioNet & CSP + LDA & Pinv & 0.048 & 0.558 & 0.479 \\
PhysioNet & CSP + LDA & Raw & 0.082 & 0.558 & 0.481 \\
PhysioNet & EEGNet & MOC & 0.093 & 0.657 & 0.621 \\
PhysioNet & EEGNet & Pinv & 0.106 & 0.656 & 0.623 \\
PhysioNet & EEGNet & Raw & 0.339 & 0.608 & 0.543 \\
PhysioNet & ShallowConvNet & MOC & 0.096 & 0.792 & 0.784 \\
PhysioNet & ShallowConvNet & Pinv & 0.114 & 0.775 & 0.771 \\
PhysioNet & ShallowConvNet & Raw & 0.184 & 0.750 & 0.745 \\
\bottomrule\end{longtable}\endgroup

\begingroup\small\setlength{\tabcolsep}{4pt}\renewcommand{\arraystretch}{1.15}
\begin{longtable}{l l l l l l}
\caption{Bipolar-only comparison on four connected trees at 10 dB. No Raw or site-interpolation baseline is defined.}\\
\toprule
Dataset & Decoder & Method & D & BAC & Macro-F1 \\
\midrule\endfirsthead
\toprule
Dataset & Decoder & Method & D & BAC & Macro-F1 \\
\midrule\endhead
BNCI2014-001 & EEG Conformer & MOC & 0.000 & 0.500 & 0.333 \\
BNCI2014-001 & EEG Conformer & Pinv & 0.000 & 0.500 & 0.333 \\
BNCI2014-001 & CSP + LDA & MOC & 0.083 & 0.506 & 0.411 \\
BNCI2014-001 & CSP + LDA & Pinv & 0.523 & 0.537 & 0.437 \\
BNCI2014-001 & EEGNet & MOC & 0.227 & 0.589 & 0.581 \\
BNCI2014-001 & EEGNet & Pinv & 0.228 & 0.579 & 0.575 \\
BNCI2014-001 & ShallowConvNet & MOC & 0.139 & 0.558 & 0.486 \\
BNCI2014-001 & ShallowConvNet & Pinv & 0.164 & 0.559 & 0.488 \\
Cho2017 & EEG Conformer & MOC & 0.219 & 0.661 & 0.642 \\
Cho2017 & EEG Conformer & Pinv & 0.293 & 0.650 & 0.638 \\
Cho2017 & CSP + LDA & MOC & 0.315 & 0.542 & 0.467 \\
Cho2017 & CSP + LDA & Pinv & 0.046 & 0.503 & 0.347 \\
Cho2017 & EEGNet & MOC & 0.198 & 0.695 & 0.684 \\
Cho2017 & EEGNet & Pinv & 0.333 & 0.640 & 0.612 \\
Cho2017 & ShallowConvNet & MOC & 0.182 & 0.662 & 0.650 \\
Cho2017 & ShallowConvNet & Pinv & 0.287 & 0.632 & 0.597 \\
PhysioNet & EEG Conformer & MOC & 0.257 & 0.624 & 0.598 \\
PhysioNet & EEG Conformer & Pinv & 0.359 & 0.613 & 0.592 \\
PhysioNet & CSP + LDA & MOC & 0.144 & 0.559 & 0.481 \\
PhysioNet & CSP + LDA & Pinv & 0.150 & 0.534 & 0.423 \\
PhysioNet & EEGNet & MOC & 0.172 & 0.643 & 0.612 \\
PhysioNet & EEGNet & Pinv & 0.323 & 0.622 & 0.581 \\
PhysioNet & ShallowConvNet & MOC & 0.203 & 0.760 & 0.747 \\
PhysioNet & ShallowConvNet & Pinv & 0.291 & 0.710 & 0.692 \\
\bottomrule\end{longtable}\endgroup

\begingroup\small\setlength{\tabcolsep}{4pt}\renewcommand{\arraystretch}{1.15}
\begin{longtable}{l l l l l l}
\caption{Architecture-specific paired contrasts for D.}\\
\toprule
Dataset & Decoder & Baseline & MOC minus baseline & p & Holm p \\
\midrule\endfirsthead
\toprule
Dataset & Decoder & Baseline & MOC minus baseline & p & Holm p \\
\midrule\endhead
BNCI2014-001 & EEG Conformer & Pinv & 0.000 [0.000, 0.000] & 1.0000 & 1.0000 \\
BNCI2014-001 & EEG Conformer & Raw & -0.095 [-0.150, -0.041] & 0.5000 & 1.0000 \\
BNCI2014-001 & CSP + LDA & Pinv & 0.002 [0.002, 0.002] & 0.5000 & 1.0000 \\
BNCI2014-001 & CSP + LDA & Raw & -0.347 [-0.420, -0.273] & 0.5000 & 1.0000 \\
BNCI2014-001 & EEGNet & Pinv & -0.019 [-0.029, -0.008] & 0.5000 & 1.0000 \\
BNCI2014-001 & EEGNet & Raw & -0.265 [-0.287, -0.244] & 0.5000 & 1.0000 \\
BNCI2014-001 & ShallowConvNet & Pinv & 0.000 [-0.002, 0.003] & 1.0000 & 1.0000 \\
BNCI2014-001 & ShallowConvNet & Raw & -0.045 [-0.077, -0.013] & 0.5000 & 1.0000 \\
Cho2017 & EEG Conformer & Pinv & -0.010 [-0.026, 0.006] & 0.2402 & 1.0000 \\
Cho2017 & EEG Conformer & Raw & -0.059 [-0.084, -0.030] & 0.0039 & 0.0781 \\
Cho2017 & CSP + LDA & Pinv & -0.001 [-0.002, 0.000] & 0.2109 & 1.0000 \\
Cho2017 & CSP + LDA & Raw & -0.014 [-0.023, -0.007] & 0.0039 & 0.0781 \\
Cho2017 & EEGNet & Pinv & -0.022 [-0.033, -0.013] & 0.0039 & 0.0781 \\
Cho2017 & EEGNet & Raw & -0.078 [-0.097, -0.059] & 0.0020 & 0.0410 \\
Cho2017 & ShallowConvNet & Pinv & -0.007 [-0.013, -0.001] & 0.0684 & 0.8887 \\
Cho2017 & ShallowConvNet & Raw & -0.057 [-0.082, -0.036] & 0.0039 & 0.0781 \\
PhysioNet & EEG Conformer & Pinv & -0.019 [-0.037, -0.001] & 0.0541 & 0.7573 \\
PhysioNet & EEG Conformer & Raw & 0.025 [-0.018, 0.074] & 0.3130 & 1.0000 \\
PhysioNet & CSP + LDA & Pinv & 0.000 [-0.002, 0.002] & 0.9785 & 1.0000 \\
PhysioNet & CSP + LDA & Raw & -0.034 [-0.057, -0.012] & 0.0065 & 0.0975 \\
PhysioNet & EEGNet & Pinv & -0.013 [-0.020, -0.007] & 0.0011 & 0.0242 \\
PhysioNet & EEGNet & Raw & -0.246 [-0.281, -0.209] & 0.0001 & 0.0024 \\
PhysioNet & ShallowConvNet & Pinv & -0.018 [-0.030, -0.007] & 0.0051 & 0.0816 \\
PhysioNet & ShallowConvNet & Raw & -0.088 [-0.101, -0.075] & 0.0001 & 0.0024 \\
\bottomrule\end{longtable}\endgroup

\begingroup\small\setlength{\tabcolsep}{4pt}\renewcommand{\arraystretch}{1.15}
\begin{longtable}{l l l l l l}
\caption{Architecture-specific paired contrasts for BAC.}\\
\toprule
Dataset & Decoder & Baseline & MOC minus baseline & p & Holm p \\
\midrule\endfirsthead
\toprule
Dataset & Decoder & Baseline & MOC minus baseline & p & Holm p \\
\midrule\endhead
BNCI2014-001 & EEG Conformer & Pinv & 0.000 [0.000, 0.000] & 1.0000 & 1.0000 \\
BNCI2014-001 & EEG Conformer & Raw & -0.031 [-0.038, -0.024] & 0.5000 & 1.0000 \\
BNCI2014-001 & CSP + LDA & Pinv & 0.000 [0.000, 0.001] & 1.0000 & 1.0000 \\
BNCI2014-001 & CSP + LDA & Raw & -0.020 [-0.034, -0.006] & 0.5000 & 1.0000 \\
BNCI2014-001 & EEGNet & Pinv & 0.007 [-0.005, 0.019] & 1.0000 & 1.0000 \\
BNCI2014-001 & EEGNet & Raw & 0.025 [0.024, 0.025] & 0.5000 & 1.0000 \\
BNCI2014-001 & ShallowConvNet & Pinv & 0.010 [0.009, 0.010] & 0.5000 & 1.0000 \\
BNCI2014-001 & ShallowConvNet & Raw & 0.022 [0.014, 0.030] & 0.5000 & 1.0000 \\
Cho2017 & EEG Conformer & Pinv & -0.001 [-0.032, 0.028] & 0.9473 & 1.0000 \\
Cho2017 & EEG Conformer & Raw & 0.013 [-0.009, 0.038] & 0.3184 & 1.0000 \\
Cho2017 & CSP + LDA & Pinv & 0.001 [-0.000, 0.001] & 0.2656 & 1.0000 \\
Cho2017 & CSP + LDA & Raw & 0.001 [-0.002, 0.004] & 0.8125 & 1.0000 \\
Cho2017 & EEGNet & Pinv & 0.002 [-0.008, 0.013] & 0.8066 & 1.0000 \\
Cho2017 & EEGNet & Raw & 0.024 [0.013, 0.036] & 0.0059 & 0.1289 \\
Cho2017 & ShallowConvNet & Pinv & -0.000 [-0.008, 0.009] & 0.9766 & 1.0000 \\
Cho2017 & ShallowConvNet & Raw & -0.009 [-0.033, 0.011] & 0.5332 & 1.0000 \\
PhysioNet & EEG Conformer & Pinv & -0.005 [-0.014, 0.005] & 0.3774 & 1.0000 \\
PhysioNet & EEG Conformer & Raw & 0.075 [0.039, 0.111] & 0.0004 & 0.0096 \\
PhysioNet & CSP + LDA & Pinv & 0.001 [-0.000, 0.002] & 0.2558 & 1.0000 \\
PhysioNet & CSP + LDA & Raw & 0.001 [-0.010, 0.013] & 0.8484 & 1.0000 \\
PhysioNet & EEGNet & Pinv & 0.001 [-0.009, 0.010] & 0.8676 & 1.0000 \\
PhysioNet & EEGNet & Raw & 0.049 [0.014, 0.083] & 0.0178 & 0.3738 \\
PhysioNet & ShallowConvNet & Pinv & 0.018 [-0.001, 0.036] & 0.0960 & 1.0000 \\
PhysioNet & ShallowConvNet & Raw & 0.042 [0.024, 0.059] & 0.0004 & 0.0096 \\
\bottomrule\end{longtable}\endgroup

\begingroup\small\setlength{\tabcolsep}{4pt}\renewcommand{\arraystretch}{1.15}
\begin{longtable}{l l l l l l}
\caption{Architecture-specific paired contrasts for macro-F1.}\\
\toprule
Dataset & Decoder & Baseline & MOC minus baseline & p & Holm p \\
\midrule\endfirsthead
\toprule
Dataset & Decoder & Baseline & MOC minus baseline & p & Holm p \\
\midrule\endhead
BNCI2014-001 & EEG Conformer & Pinv & 0.000 [0.000, 0.000] & 1.0000 & 1.0000 \\
BNCI2014-001 & EEG Conformer & Raw & -0.105 [-0.123, -0.088] & 0.5000 & 1.0000 \\
BNCI2014-001 & CSP + LDA & Pinv & 0.001 [-0.000, 0.002] & 1.0000 & 1.0000 \\
BNCI2014-001 & CSP + LDA & Raw & -0.011 [-0.063, 0.042] & 1.0000 & 1.0000 \\
BNCI2014-001 & EEGNet & Pinv & 0.006 [-0.007, 0.019] & 1.0000 & 1.0000 \\
BNCI2014-001 & EEGNet & Raw & 0.078 [0.053, 0.104] & 0.5000 & 1.0000 \\
BNCI2014-001 & ShallowConvNet & Pinv & 0.016 [0.015, 0.017] & 0.5000 & 1.0000 \\
BNCI2014-001 & ShallowConvNet & Raw & 0.042 [0.031, 0.052] & 0.5000 & 1.0000 \\
Cho2017 & EEG Conformer & Pinv & 0.002 [-0.038, 0.044] & 0.9199 & 1.0000 \\
Cho2017 & EEG Conformer & Raw & 0.028 [-0.004, 0.063] & 0.1836 & 1.0000 \\
Cho2017 & CSP + LDA & Pinv & 0.001 [0.000, 0.002] & 0.0859 & 1.0000 \\
Cho2017 & CSP + LDA & Raw & 0.003 [0.000, 0.007] & 0.1035 & 1.0000 \\
Cho2017 & EEGNet & Pinv & -0.000 [-0.014, 0.014] & 0.9727 & 1.0000 \\
Cho2017 & EEGNet & Raw & 0.030 [0.018, 0.042] & 0.0020 & 0.0449 \\
Cho2017 & ShallowConvNet & Pinv & 0.000 [-0.009, 0.013] & 0.9844 & 1.0000 \\
Cho2017 & ShallowConvNet & Raw & -0.012 [-0.039, 0.012] & 0.4590 & 1.0000 \\
PhysioNet & EEG Conformer & Pinv & -0.009 [-0.021, 0.004] & 0.2025 & 1.0000 \\
PhysioNet & EEG Conformer & Raw & 0.100 [0.057, 0.141] & 0.0003 & 0.0072 \\
PhysioNet & CSP + LDA & Pinv & 0.000 [-0.001, 0.002] & 0.5693 & 1.0000 \\
PhysioNet & CSP + LDA & Raw & -0.001 [-0.018, 0.016] & 0.8702 & 1.0000 \\
PhysioNet & EEGNet & Pinv & -0.001 [-0.012, 0.009] & 0.8241 & 1.0000 \\
PhysioNet & EEGNet & Raw & 0.079 [0.030, 0.125] & 0.0066 & 0.1452 \\
PhysioNet & ShallowConvNet & Pinv & 0.013 [-0.014, 0.037] & 0.3372 & 1.0000 \\
PhysioNet & ShallowConvNet & Raw & 0.039 [0.011, 0.063] & 0.0079 & 0.1659 \\
\bottomrule\end{longtable}\endgroup

\section{S7. Noise and spectral
response}\label{s7.-noise-and-spectral-response}

The complete digital tables contain every SNR/condition/method/decoder
endpoint and its subject interval. SNR levels are no added noise, 30,
20, 10, and 5 dB. Spectral reconstruction uses all seven condition
numbers and five operator seeds; decoding uses the conditions retained
by its fixed design, as listed below. Operator seeds are averaged within
subject before calculating intervals. Figure 3 maps relative NMSE and
visualizes the EEGNet SNR response from these complete data. The main
SNR figure includes all five EEGNet methods; ShallowConvNet endpoints
are retained in the annex.

\begingroup\small\setlength{\tabcolsep}{4pt}\renewcommand{\arraystretch}{1.15}
\begin{longtable}{l l l l l l}
\caption{BNCI2014-001 decoding response to SNR.}\\
\toprule
SNR & Decoder & Method & D & BAC & Macro-F1 \\
\midrule\endfirsthead
\toprule
SNR & Decoder & Method & D & BAC & Macro-F1 \\
\midrule\endhead
5 & EEGNet & Aug. + centering & 0.177 & 0.586 & 0.574 \\
10 & EEGNet & Aug. + centering & 0.121 & 0.582 & 0.565 \\
20 & EEGNet & Aug. + centering & 0.041 & 0.582 & 0.566 \\
30 & EEGNet & Aug. + centering & 0.009 & 0.584 & 0.569 \\
Noiseless & EEGNet & Aug. + centering & 0.000 & 0.550 & 0.524 \\
5 & EEGNet & Interpolation & 0.206 & 0.569 & 0.553 \\
10 & EEGNet & Interpolation & 0.112 & 0.564 & 0.541 \\
20 & EEGNet & Interpolation & 0.046 & 0.582 & 0.567 \\
30 & EEGNet & Interpolation & 0.011 & 0.587 & 0.572 \\
Noiseless & EEGNet & Interpolation & 0.000 & 0.580 & 0.563 \\
5 & EEGNet & MOC & 0.189 & 0.551 & 0.520 \\
10 & EEGNet & MOC & 0.104 & 0.565 & 0.548 \\
20 & EEGNet & MOC & 0.037 & 0.568 & 0.549 \\
30 & EEGNet & MOC & 0.013 & 0.565 & 0.547 \\
Noiseless & EEGNet & MOC & 0.012 & 0.564 & 0.543 \\
5 & EEGNet & Pinv & 0.141 & 0.554 & 0.522 \\
10 & EEGNet & Pinv & 0.090 & 0.556 & 0.527 \\
20 & EEGNet & Pinv & 0.043 & 0.578 & 0.561 \\
30 & EEGNet & Pinv & 0.011 & 0.583 & 0.567 \\
Noiseless & EEGNet & Pinv & 0.000 & 0.580 & 0.565 \\
5 & EEGNet & Raw & 0.278 & 0.537 & 0.463 \\
10 & EEGNet & Raw & 0.310 & 0.534 & 0.459 \\
20 & EEGNet & Raw & 0.218 & 0.538 & 0.459 \\
30 & EEGNet & Raw & 0.248 & 0.541 & 0.468 \\
Noiseless & EEGNet & Raw & 0.378 & 0.539 & 0.450 \\
5 & ShallowConvNet & Aug. + centering & 0.131 & 0.573 & 0.571 \\
10 & ShallowConvNet & Aug. + centering & 0.055 & 0.564 & 0.534 \\
20 & ShallowConvNet & Aug. + centering & 0.017 & 0.567 & 0.540 \\
30 & ShallowConvNet & Aug. + centering & 0.006 & 0.567 & 0.539 \\
Noiseless & ShallowConvNet & Aug. + centering & 0.000 & 0.559 & 0.530 \\
5 & ShallowConvNet & Interpolation & 0.141 & 0.561 & 0.555 \\
10 & ShallowConvNet & Interpolation & 0.077 & 0.561 & 0.555 \\
20 & ShallowConvNet & Interpolation & 0.017 & 0.570 & 0.545 \\
30 & ShallowConvNet & Interpolation & 0.006 & 0.570 & 0.542 \\
Noiseless & ShallowConvNet & Interpolation & 0.000 & 0.568 & 0.541 \\
5 & ShallowConvNet & MOC & 0.120 & 0.549 & 0.542 \\
10 & ShallowConvNet & MOC & 0.075 & 0.564 & 0.559 \\
20 & ShallowConvNet & MOC & 0.017 & 0.560 & 0.541 \\
30 & ShallowConvNet & MOC & 0.006 & 0.568 & 0.551 \\
Noiseless & ShallowConvNet & MOC & 0.004 & 0.567 & 0.551 \\
5 & ShallowConvNet & Pinv & 0.148 & 0.556 & 0.550 \\
10 & ShallowConvNet & Pinv & 0.061 & 0.590 & 0.581 \\
20 & ShallowConvNet & Pinv & 0.019 & 0.565 & 0.538 \\
30 & ShallowConvNet & Pinv & 0.009 & 0.566 & 0.539 \\
Noiseless & ShallowConvNet & Pinv & 0.000 & 0.568 & 0.541 \\
5 & ShallowConvNet & Raw & 0.252 & 0.571 & 0.552 \\
10 & ShallowConvNet & Raw & 0.192 & 0.596 & 0.579 \\
20 & ShallowConvNet & Raw & 0.188 & 0.576 & 0.554 \\
30 & ShallowConvNet & Raw & 0.177 & 0.576 & 0.556 \\
Noiseless & ShallowConvNet & Raw & 0.180 & 0.577 & 0.556 \\
\bottomrule\end{longtable}\endgroup

\begingroup\small\setlength{\tabcolsep}{4pt}\renewcommand{\arraystretch}{1.15}
\begin{longtable}{l l l l l l}
\caption{Cho2017 decoding response to SNR.}\\
\toprule
SNR & Decoder & Method & D & BAC & Macro-F1 \\
\midrule\endfirsthead
\toprule
SNR & Decoder & Method & D & BAC & Macro-F1 \\
\midrule\endhead
5 & EEGNet & Aug. + centering & 0.179 & 0.697 & 0.692 \\
10 & EEGNet & Aug. + centering & 0.110 & 0.712 & 0.706 \\
20 & EEGNet & Aug. + centering & 0.043 & 0.718 & 0.712 \\
30 & EEGNet & Aug. + centering & 0.015 & 0.718 & 0.711 \\
Noiseless & EEGNet & Aug. + centering & 0.000 & 0.723 & 0.717 \\
5 & EEGNet & Interpolation & 0.161 & 0.682 & 0.672 \\
10 & EEGNet & Interpolation & 0.115 & 0.705 & 0.699 \\
20 & EEGNet & Interpolation & 0.042 & 0.716 & 0.709 \\
30 & EEGNet & Interpolation & 0.016 & 0.716 & 0.710 \\
Noiseless & EEGNet & Interpolation & 0.000 & 0.718 & 0.712 \\
5 & EEGNet & MOC & 0.182 & 0.684 & 0.678 \\
10 & EEGNet & MOC & 0.121 & 0.698 & 0.692 \\
20 & EEGNet & MOC & 0.044 & 0.714 & 0.708 \\
30 & EEGNet & MOC & 0.019 & 0.716 & 0.709 \\
Noiseless & EEGNet & MOC & 0.006 & 0.717 & 0.710 \\
5 & EEGNet & Pinv & 0.189 & 0.678 & 0.671 \\
10 & EEGNet & Pinv & 0.146 & 0.702 & 0.696 \\
20 & EEGNet & Pinv & 0.046 & 0.716 & 0.709 \\
30 & EEGNet & Pinv & 0.017 & 0.717 & 0.711 \\
Noiseless & EEGNet & Pinv & 0.000 & 0.719 & 0.713 \\
5 & EEGNet & Raw & 0.316 & 0.639 & 0.611 \\
10 & EEGNet & Raw & 0.321 & 0.629 & 0.578 \\
20 & EEGNet & Raw & 0.289 & 0.630 & 0.577 \\
30 & EEGNet & Raw & 0.291 & 0.629 & 0.571 \\
Noiseless & EEGNet & Raw & 0.301 & 0.638 & 0.580 \\
5 & ShallowConvNet & Aug. + centering & 0.130 & 0.670 & 0.662 \\
10 & ShallowConvNet & Aug. + centering & 0.077 & 0.676 & 0.667 \\
20 & ShallowConvNet & Aug. + centering & 0.022 & 0.689 & 0.679 \\
30 & ShallowConvNet & Aug. + centering & 0.005 & 0.690 & 0.680 \\
Noiseless & ShallowConvNet & Aug. + centering & 0.000 & 0.682 & 0.673 \\
5 & ShallowConvNet & Interpolation & 0.134 & 0.676 & 0.668 \\
10 & ShallowConvNet & Interpolation & 0.074 & 0.676 & 0.665 \\
20 & ShallowConvNet & Interpolation & 0.030 & 0.708 & 0.703 \\
30 & ShallowConvNet & Interpolation & 0.007 & 0.691 & 0.680 \\
Noiseless & ShallowConvNet & Interpolation & 0.000 & 0.691 & 0.680 \\
5 & ShallowConvNet & MOC & 0.107 & 0.661 & 0.651 \\
10 & ShallowConvNet & MOC & 0.066 & 0.665 & 0.654 \\
20 & ShallowConvNet & MOC & 0.026 & 0.696 & 0.686 \\
30 & ShallowConvNet & MOC & 0.006 & 0.675 & 0.665 \\
Noiseless & ShallowConvNet & MOC & 0.002 & 0.678 & 0.668 \\
5 & ShallowConvNet & Pinv & 0.112 & 0.660 & 0.647 \\
10 & ShallowConvNet & Pinv & 0.056 & 0.665 & 0.646 \\
20 & ShallowConvNet & Pinv & 0.022 & 0.683 & 0.671 \\
30 & ShallowConvNet & Pinv & 0.005 & 0.691 & 0.680 \\
Noiseless & ShallowConvNet & Pinv & 0.000 & 0.691 & 0.680 \\
5 & ShallowConvNet & Raw & 0.203 & 0.668 & 0.654 \\
10 & ShallowConvNet & Raw & 0.185 & 0.689 & 0.682 \\
20 & ShallowConvNet & Raw & 0.162 & 0.671 & 0.654 \\
30 & ShallowConvNet & Raw & 0.145 & 0.677 & 0.660 \\
Noiseless & ShallowConvNet & Raw & 0.144 & 0.678 & 0.661 \\
\bottomrule\end{longtable}\endgroup

\begingroup\small\setlength{\tabcolsep}{4pt}\renewcommand{\arraystretch}{1.15}
\begin{longtable}{l l l l l l}
\caption{PhysioNet decoding response to SNR.}\\
\toprule
SNR & Decoder & Method & D & BAC & Macro-F1 \\
\midrule\endfirsthead
\toprule
SNR & Decoder & Method & D & BAC & Macro-F1 \\
\midrule\endhead
5 & EEGNet & Aug. + centering & 0.243 & 0.632 & 0.609 \\
10 & EEGNet & Aug. + centering & 0.165 & 0.657 & 0.632 \\
20 & EEGNet & Aug. + centering & 0.045 & 0.672 & 0.643 \\
30 & EEGNet & Aug. + centering & 0.013 & 0.679 & 0.651 \\
Noiseless & EEGNet & Aug. + centering & 0.000 & 0.677 & 0.650 \\
5 & EEGNet & Interpolation & 0.164 & 0.630 & 0.596 \\
10 & EEGNet & Interpolation & 0.118 & 0.643 & 0.612 \\
20 & EEGNet & Interpolation & 0.047 & 0.670 & 0.641 \\
30 & EEGNet & Interpolation & 0.012 & 0.673 & 0.642 \\
Noiseless & EEGNet & Interpolation & 0.000 & 0.681 & 0.654 \\
5 & EEGNet & MOC & 0.205 & 0.638 & 0.614 \\
10 & EEGNet & MOC & 0.116 & 0.661 & 0.631 \\
20 & EEGNet & MOC & 0.032 & 0.678 & 0.647 \\
30 & EEGNet & MOC & 0.009 & 0.682 & 0.654 \\
Noiseless & EEGNet & MOC & 0.001 & 0.675 & 0.644 \\
5 & EEGNet & Pinv & 0.186 & 0.631 & 0.605 \\
10 & EEGNet & Pinv & 0.117 & 0.663 & 0.635 \\
20 & EEGNet & Pinv & 0.034 & 0.672 & 0.646 \\
30 & EEGNet & Pinv & 0.012 & 0.682 & 0.656 \\
Noiseless & EEGNet & Pinv & 0.000 & 0.681 & 0.654 \\
5 & EEGNet & Raw & 0.292 & 0.574 & 0.499 \\
10 & EEGNet & Raw & 0.240 & 0.594 & 0.519 \\
20 & EEGNet & Raw & 0.190 & 0.605 & 0.526 \\
30 & EEGNet & Raw & 0.195 & 0.602 & 0.520 \\
Noiseless & EEGNet & Raw & 0.120 & 0.651 & 0.616 \\
5 & ShallowConvNet & Aug. + centering & 0.154 & 0.718 & 0.705 \\
10 & ShallowConvNet & Aug. + centering & 0.092 & 0.758 & 0.751 \\
20 & ShallowConvNet & Aug. + centering & 0.038 & 0.779 & 0.775 \\
30 & ShallowConvNet & Aug. + centering & 0.015 & 0.797 & 0.788 \\
Noiseless & ShallowConvNet & Aug. + centering & 0.000 & 0.766 & 0.760 \\
5 & ShallowConvNet & Interpolation & 0.193 & 0.731 & 0.725 \\
10 & ShallowConvNet & Interpolation & 0.120 & 0.767 & 0.764 \\
20 & ShallowConvNet & Interpolation & 0.040 & 0.792 & 0.788 \\
30 & ShallowConvNet & Interpolation & 0.011 & 0.792 & 0.788 \\
Noiseless & ShallowConvNet & Interpolation & 0.000 & 0.794 & 0.784 \\
5 & ShallowConvNet & MOC & 0.150 & 0.749 & 0.741 \\
10 & ShallowConvNet & MOC & 0.108 & 0.790 & 0.786 \\
20 & ShallowConvNet & MOC & 0.035 & 0.765 & 0.759 \\
30 & ShallowConvNet & MOC & 0.014 & 0.781 & 0.777 \\
Noiseless & ShallowConvNet & MOC & 0.002 & 0.790 & 0.786 \\
5 & ShallowConvNet & Pinv & 0.193 & 0.743 & 0.736 \\
10 & ShallowConvNet & Pinv & 0.112 & 0.760 & 0.754 \\
20 & ShallowConvNet & Pinv & 0.037 & 0.791 & 0.787 \\
30 & ShallowConvNet & Pinv & 0.011 & 0.801 & 0.793 \\
Noiseless & ShallowConvNet & Pinv & 0.000 & 0.794 & 0.784 \\
5 & ShallowConvNet & Raw & 0.244 & 0.718 & 0.709 \\
10 & ShallowConvNet & Raw & 0.201 & 0.747 & 0.736 \\
20 & ShallowConvNet & Raw & 0.122 & 0.775 & 0.764 \\
30 & ShallowConvNet & Raw & 0.070 & 0.740 & 0.722 \\
Noiseless & ShallowConvNet & Raw & 0.081 & 0.775 & 0.761 \\
\bottomrule\end{longtable}\endgroup

\begingroup\small\setlength{\tabcolsep}{4pt}\renewcommand{\arraystretch}{1.15}
\begin{longtable}{l l l l l l l}
\caption{BNCI2014-001 decoding response to synthetic conditioning.}\\
\toprule
Condition & SNR & Decoder & Method & D & BAC & Macro-F1 \\
\midrule\endfirsthead
\toprule
Condition & SNR & Decoder & Method & D & BAC & Macro-F1 \\
\midrule\endhead
1 & 10 & EEGNet & MOC & 0.072 & 0.547 & 0.515 \\
1 & 20 & EEGNet & MOC & 0.039 & 0.564 & 0.545 \\
1 & Noiseless & EEGNet & MOC & 0.000 & 0.562 & 0.541 \\
1 & 10 & EEGNet & Pinv & 0.080 & 0.549 & 0.522 \\
1 & 20 & EEGNet & Pinv & 0.037 & 0.575 & 0.558 \\
1 & Noiseless & EEGNet & Pinv & 0.000 & 0.517 & 0.460 \\
1 & 10 & ShallowConvNet & MOC & 0.053 & 0.606 & 0.600 \\
1 & 20 & ShallowConvNet & MOC & 0.011 & 0.564 & 0.549 \\
1 & Noiseless & ShallowConvNet & MOC & 0.000 & 0.568 & 0.551 \\
1 & 10 & ShallowConvNet & Pinv & 0.043 & 0.589 & 0.579 \\
1 & 20 & ShallowConvNet & Pinv & 0.016 & 0.565 & 0.545 \\
1 & Noiseless & ShallowConvNet & Pinv & 0.000 & 0.568 & 0.541 \\
3 & 10 & EEGNet & MOC & 0.083 & 0.559 & 0.536 \\
3 & 20 & EEGNet & MOC & 0.035 & 0.578 & 0.567 \\
3 & Noiseless & EEGNet & MOC & 0.000 & 0.583 & 0.574 \\
3 & 10 & EEGNet & Pinv & 0.096 & 0.543 & 0.508 \\
3 & 20 & EEGNet & Pinv & 0.047 & 0.573 & 0.557 \\
3 & Noiseless & EEGNet & Pinv & 0.000 & 0.576 & 0.559 \\
3 & 10 & ShallowConvNet & MOC & 0.059 & 0.586 & 0.579 \\
3 & 20 & ShallowConvNet & MOC & 0.019 & 0.572 & 0.565 \\
3 & Noiseless & ShallowConvNet & MOC & 0.000 & 0.580 & 0.572 \\
3 & 10 & ShallowConvNet & Pinv & 0.094 & 0.575 & 0.572 \\
3 & 20 & ShallowConvNet & Pinv & 0.015 & 0.566 & 0.547 \\
3 & Noiseless & ShallowConvNet & Pinv & 0.000 & 0.568 & 0.541 \\
10 & 10 & EEGNet & MOC & 0.110 & 0.579 & 0.565 \\
10 & 20 & EEGNet & MOC & 0.055 & 0.596 & 0.588 \\
10 & Noiseless & EEGNet & MOC & 0.000 & 0.606 & 0.598 \\
10 & 10 & EEGNet & Pinv & 0.092 & 0.531 & 0.458 \\
10 & 20 & EEGNet & Pinv & 0.076 & 0.555 & 0.529 \\
10 & Noiseless & EEGNet & Pinv & 0.000 & 0.575 & 0.557 \\
10 & 10 & ShallowConvNet & MOC & 0.031 & 0.525 & 0.436 \\
10 & 20 & ShallowConvNet & MOC & 0.013 & 0.532 & 0.441 \\
10 & Noiseless & ShallowConvNet & MOC & 0.000 & 0.530 & 0.423 \\
10 & 10 & ShallowConvNet & Pinv & 0.159 & 0.593 & 0.584 \\
10 & 20 & ShallowConvNet & Pinv & 0.035 & 0.583 & 0.573 \\
10 & Noiseless & ShallowConvNet & Pinv & 0.000 & 0.568 & 0.541 \\
30 & 10 & EEGNet & MOC & 0.086 & 0.537 & 0.483 \\
30 & 20 & EEGNet & MOC & 0.050 & 0.571 & 0.556 \\
30 & Noiseless & EEGNet & MOC & 0.000 & 0.562 & 0.546 \\
30 & 10 & EEGNet & Pinv & 0.205 & 0.556 & 0.532 \\
30 & 20 & EEGNet & Pinv & 0.142 & 0.546 & 0.515 \\
30 & Noiseless & EEGNet & Pinv & 0.000 & 0.576 & 0.559 \\
30 & 10 & ShallowConvNet & MOC & 0.057 & 0.529 & 0.527 \\
30 & 20 & ShallowConvNet & MOC & 0.014 & 0.524 & 0.445 \\
30 & Noiseless & ShallowConvNet & MOC & 0.000 & 0.523 & 0.441 \\
30 & 10 & ShallowConvNet & Pinv & 0.207 & 0.564 & 0.547 \\
30 & 20 & ShallowConvNet & Pinv & 0.108 & 0.575 & 0.569 \\
30 & Noiseless & ShallowConvNet & Pinv & 0.000 & 0.568 & 0.541 \\
100 & 10 & EEGNet & MOC & 0.141 & 0.535 & 0.501 \\
100 & 20 & EEGNet & MOC & 0.055 & 0.581 & 0.557 \\
100 & Noiseless & EEGNet & MOC & 0.000 & 0.573 & 0.551 \\
100 & 10 & EEGNet & Pinv & 0.433 & 0.511 & 0.511 \\
100 & 20 & EEGNet & Pinv & 0.140 & 0.538 & 0.489 \\
100 & Noiseless & EEGNet & Pinv & 0.000 & 0.576 & 0.559 \\
100 & 10 & ShallowConvNet & MOC & 0.062 & 0.539 & 0.536 \\
100 & 20 & ShallowConvNet & MOC & 0.005 & 0.538 & 0.522 \\
100 & Noiseless & ShallowConvNet & MOC & 0.000 & 0.535 & 0.519 \\
100 & 10 & ShallowConvNet & Pinv & 0.146 & 0.505 & 0.413 \\
100 & 20 & ShallowConvNet & Pinv & 0.206 & 0.572 & 0.561 \\
100 & Noiseless & ShallowConvNet & Pinv & 0.000 & 0.568 & 0.541 \\
300 & 10 & EEGNet & MOC & 0.140 & 0.569 & 0.558 \\
300 & 20 & EEGNet & MOC & 0.042 & 0.570 & 0.552 \\
300 & Noiseless & EEGNet & MOC & 0.000 & 0.571 & 0.557 \\
300 & 10 & EEGNet & Pinv & 0.282 & 0.514 & 0.456 \\
300 & 20 & EEGNet & Pinv & 0.207 & 0.510 & 0.446 \\
300 & Noiseless & EEGNet & Pinv & 0.000 & 0.587 & 0.572 \\
300 & 10 & ShallowConvNet & MOC & 0.054 & 0.535 & 0.504 \\
300 & 20 & ShallowConvNet & MOC & 0.015 & 0.540 & 0.503 \\
300 & Noiseless & ShallowConvNet & MOC & 0.000 & 0.543 & 0.499 \\
300 & 10 & ShallowConvNet & Pinv & 0.000 & 0.500 & 0.333 \\
300 & 20 & ShallowConvNet & Pinv & 0.249 & 0.549 & 0.514 \\
300 & Noiseless & ShallowConvNet & Pinv & 0.000 & 0.568 & 0.541 \\
1000 & 10 & EEGNet & MOC & 0.033 & 0.529 & 0.488 \\
1000 & 20 & EEGNet & MOC & 0.010 & 0.532 & 0.491 \\
1000 & Noiseless & EEGNet & MOC & 0.000 & 0.531 & 0.490 \\
1000 & 10 & EEGNet & Pinv & 0.495 & 0.510 & 0.507 \\
1000 & 20 & EEGNet & Pinv & 0.383 & 0.506 & 0.478 \\
1000 & Noiseless & EEGNet & Pinv & 0.000 & 0.589 & 0.575 \\
1000 & 10 & ShallowConvNet & MOC & 0.025 & 0.505 & 0.422 \\
1000 & 20 & ShallowConvNet & MOC & 0.012 & 0.512 & 0.464 \\
1000 & Noiseless & ShallowConvNet & MOC & 0.000 & 0.510 & 0.465 \\
1000 & 10 & ShallowConvNet & Pinv & 0.442 & 0.500 & 0.485 \\
1000 & 20 & ShallowConvNet & Pinv & 0.040 & 0.504 & 0.356 \\
1000 & Noiseless & ShallowConvNet & Pinv & 0.000 & 0.568 & 0.541 \\
\bottomrule\end{longtable}\endgroup

\begingroup\small\setlength{\tabcolsep}{4pt}\renewcommand{\arraystretch}{1.15}
\begin{longtable}{l l l l l l l}
\caption{Cho2017 decoding response to synthetic conditioning.}\\
\toprule
Condition & SNR & Decoder & Method & D & BAC & Macro-F1 \\
\midrule\endfirsthead
\toprule
Condition & SNR & Decoder & Method & D & BAC & Macro-F1 \\
\midrule\endhead
1 & 10 & EEGNet & MOC & 0.104 & 0.713 & 0.708 \\
1 & 20 & EEGNet & MOC & 0.046 & 0.714 & 0.707 \\
1 & Noiseless & EEGNet & MOC & 0.000 & 0.719 & 0.713 \\
1 & 10 & EEGNet & Pinv & 0.114 & 0.704 & 0.697 \\
1 & 20 & EEGNet & Pinv & 0.040 & 0.713 & 0.705 \\
1 & Noiseless & EEGNet & Pinv & 0.000 & 0.711 & 0.702 \\
1 & 10 & ShallowConvNet & MOC & 0.061 & 0.677 & 0.668 \\
1 & 20 & ShallowConvNet & MOC & 0.018 & 0.675 & 0.664 \\
1 & Noiseless & ShallowConvNet & MOC & 0.000 & 0.679 & 0.668 \\
1 & 10 & ShallowConvNet & Pinv & 0.044 & 0.662 & 0.647 \\
1 & 20 & ShallowConvNet & Pinv & 0.016 & 0.686 & 0.675 \\
1 & Noiseless & ShallowConvNet & Pinv & 0.000 & 0.691 & 0.680 \\
3 & 10 & EEGNet & MOC & 0.105 & 0.707 & 0.702 \\
3 & 20 & EEGNet & MOC & 0.042 & 0.718 & 0.714 \\
3 & Noiseless & EEGNet & MOC & 0.000 & 0.716 & 0.711 \\
3 & 10 & EEGNet & Pinv & 0.131 & 0.684 & 0.676 \\
3 & 20 & EEGNet & Pinv & 0.048 & 0.714 & 0.706 \\
3 & Noiseless & EEGNet & Pinv & 0.000 & 0.718 & 0.711 \\
3 & 10 & ShallowConvNet & MOC & 0.082 & 0.687 & 0.682 \\
3 & 20 & ShallowConvNet & MOC & 0.025 & 0.686 & 0.679 \\
3 & Noiseless & ShallowConvNet & MOC & 0.000 & 0.686 & 0.677 \\
3 & 10 & ShallowConvNet & Pinv & 0.083 & 0.678 & 0.670 \\
3 & 20 & ShallowConvNet & Pinv & 0.021 & 0.688 & 0.676 \\
3 & Noiseless & ShallowConvNet & Pinv & 0.000 & 0.691 & 0.680 \\
10 & 10 & EEGNet & MOC & 0.124 & 0.681 & 0.677 \\
10 & 20 & EEGNet & MOC & 0.032 & 0.683 & 0.673 \\
10 & Noiseless & EEGNet & MOC & 0.000 & 0.710 & 0.706 \\
10 & 10 & EEGNet & Pinv & 0.176 & 0.697 & 0.692 \\
10 & 20 & EEGNet & Pinv & 0.076 & 0.710 & 0.704 \\
10 & Noiseless & EEGNet & Pinv & 0.000 & 0.719 & 0.712 \\
10 & 10 & ShallowConvNet & MOC & 0.067 & 0.654 & 0.645 \\
10 & 20 & ShallowConvNet & MOC & 0.020 & 0.659 & 0.646 \\
10 & Noiseless & ShallowConvNet & MOC & 0.000 & 0.662 & 0.647 \\
10 & 10 & ShallowConvNet & Pinv & 0.122 & 0.658 & 0.644 \\
10 & 20 & ShallowConvNet & Pinv & 0.056 & 0.698 & 0.691 \\
10 & Noiseless & ShallowConvNet & Pinv & 0.000 & 0.691 & 0.680 \\
30 & 10 & EEGNet & MOC & 0.090 & 0.665 & 0.657 \\
30 & 20 & EEGNet & MOC & 0.036 & 0.670 & 0.662 \\
30 & Noiseless & EEGNet & MOC & 0.000 & 0.670 & 0.663 \\
30 & 10 & EEGNet & Pinv & 0.226 & 0.674 & 0.670 \\
30 & 20 & EEGNet & Pinv & 0.141 & 0.705 & 0.699 \\
30 & Noiseless & EEGNet & Pinv & 0.000 & 0.717 & 0.711 \\
30 & 10 & ShallowConvNet & MOC & 0.075 & 0.630 & 0.614 \\
30 & 20 & ShallowConvNet & MOC & 0.024 & 0.649 & 0.634 \\
30 & Noiseless & ShallowConvNet & MOC & 0.000 & 0.645 & 0.633 \\
30 & 10 & ShallowConvNet & Pinv & 0.204 & 0.642 & 0.633 \\
30 & 20 & ShallowConvNet & Pinv & 0.130 & 0.695 & 0.688 \\
30 & Noiseless & ShallowConvNet & Pinv & 0.000 & 0.691 & 0.680 \\
100 & 10 & EEGNet & MOC & 0.088 & 0.652 & 0.645 \\
100 & 20 & EEGNet & MOC & 0.038 & 0.658 & 0.651 \\
100 & Noiseless & EEGNet & MOC & 0.000 & 0.655 & 0.648 \\
100 & 10 & EEGNet & Pinv & 0.402 & 0.523 & 0.523 \\
100 & 20 & EEGNet & Pinv & 0.202 & 0.651 & 0.641 \\
100 & Noiseless & EEGNet & Pinv & 0.000 & 0.718 & 0.711 \\
100 & 10 & ShallowConvNet & MOC & 0.069 & 0.612 & 0.587 \\
100 & 20 & ShallowConvNet & MOC & 0.021 & 0.616 & 0.591 \\
100 & Noiseless & ShallowConvNet & MOC & 0.000 & 0.616 & 0.591 \\
100 & 10 & ShallowConvNet & Pinv & 0.274 & 0.569 & 0.553 \\
100 & 20 & ShallowConvNet & Pinv & 0.216 & 0.648 & 0.639 \\
100 & Noiseless & ShallowConvNet & Pinv & 0.000 & 0.691 & 0.680 \\
300 & 10 & EEGNet & MOC & 0.092 & 0.653 & 0.646 \\
300 & 20 & EEGNet & MOC & 0.032 & 0.656 & 0.648 \\
300 & Noiseless & EEGNet & MOC & 0.000 & 0.653 & 0.646 \\
300 & 10 & EEGNet & Pinv & 0.328 & 0.498 & 0.452 \\
300 & 20 & EEGNet & Pinv & 0.367 & 0.541 & 0.540 \\
300 & Noiseless & EEGNet & Pinv & 0.000 & 0.717 & 0.709 \\
300 & 10 & ShallowConvNet & MOC & 0.066 & 0.617 & 0.606 \\
300 & 20 & ShallowConvNet & MOC & 0.023 & 0.613 & 0.599 \\
300 & Noiseless & ShallowConvNet & MOC & 0.000 & 0.621 & 0.610 \\
300 & 10 & ShallowConvNet & Pinv & 0.417 & 0.525 & 0.513 \\
300 & 20 & ShallowConvNet & Pinv & 0.296 & 0.616 & 0.609 \\
300 & Noiseless & ShallowConvNet & Pinv & 0.000 & 0.691 & 0.680 \\
1000 & 10 & EEGNet & MOC & 0.102 & 0.635 & 0.628 \\
1000 & 20 & EEGNet & MOC & 0.038 & 0.640 & 0.631 \\
1000 & Noiseless & EEGNet & MOC & 0.000 & 0.638 & 0.630 \\
1000 & 10 & EEGNet & Pinv & 0.499 & 0.506 & 0.504 \\
1000 & 20 & EEGNet & Pinv & 0.488 & 0.504 & 0.502 \\
1000 & Noiseless & EEGNet & Pinv & 0.000 & 0.718 & 0.711 \\
1000 & 10 & ShallowConvNet & MOC & 0.075 & 0.596 & 0.583 \\
1000 & 20 & ShallowConvNet & MOC & 0.019 & 0.595 & 0.577 \\
1000 & Noiseless & ShallowConvNet & MOC & 0.000 & 0.596 & 0.574 \\
1000 & 10 & ShallowConvNet & Pinv & 0.489 & 0.503 & 0.500 \\
1000 & 20 & ShallowConvNet & Pinv & 0.438 & 0.515 & 0.506 \\
1000 & Noiseless & ShallowConvNet & Pinv & 0.000 & 0.691 & 0.680 \\
\bottomrule\end{longtable}\endgroup

\begingroup\small\setlength{\tabcolsep}{4pt}\renewcommand{\arraystretch}{1.15}
\begin{longtable}{l l l l l l l}
\caption{PhysioNet decoding response to synthetic conditioning.}\\
\toprule
Condition & SNR & Decoder & Method & D & BAC & Macro-F1 \\
\midrule\endfirsthead
\toprule
Condition & SNR & Decoder & Method & D & BAC & Macro-F1 \\
\midrule\endhead
1 & 10 & EEGNet & MOC & 0.083 & 0.676 & 0.645 \\
1 & 20 & EEGNet & MOC & 0.032 & 0.676 & 0.647 \\
1 & Noiseless & EEGNet & MOC & 0.000 & 0.674 & 0.643 \\
1 & 10 & EEGNet & Pinv & 0.101 & 0.646 & 0.609 \\
1 & 20 & EEGNet & Pinv & 0.037 & 0.679 & 0.654 \\
1 & Noiseless & EEGNet & Pinv & 0.000 & 0.672 & 0.642 \\
1 & 10 & ShallowConvNet & MOC & 0.095 & 0.766 & 0.757 \\
1 & 20 & ShallowConvNet & MOC & 0.029 & 0.780 & 0.775 \\
1 & Noiseless & ShallowConvNet & MOC & 0.000 & 0.791 & 0.787 \\
1 & 10 & ShallowConvNet & Pinv & 0.098 & 0.773 & 0.765 \\
1 & 20 & ShallowConvNet & Pinv & 0.030 & 0.773 & 0.767 \\
1 & Noiseless & ShallowConvNet & Pinv & 0.000 & 0.794 & 0.784 \\
3 & 10 & EEGNet & MOC & 0.105 & 0.663 & 0.635 \\
3 & 20 & EEGNet & MOC & 0.029 & 0.678 & 0.646 \\
3 & Noiseless & EEGNet & MOC & 0.000 & 0.678 & 0.646 \\
3 & 10 & EEGNet & Pinv & 0.132 & 0.641 & 0.611 \\
3 & 20 & EEGNet & Pinv & 0.036 & 0.680 & 0.655 \\
3 & Noiseless & EEGNet & Pinv & 0.000 & 0.681 & 0.654 \\
3 & 10 & ShallowConvNet & MOC & 0.079 & 0.748 & 0.736 \\
3 & 20 & ShallowConvNet & MOC & 0.021 & 0.788 & 0.781 \\
3 & Noiseless & ShallowConvNet & MOC & 0.000 & 0.793 & 0.789 \\
3 & 10 & ShallowConvNet & Pinv & 0.097 & 0.761 & 0.750 \\
3 & 20 & ShallowConvNet & Pinv & 0.035 & 0.787 & 0.784 \\
3 & Noiseless & ShallowConvNet & Pinv & 0.000 & 0.794 & 0.784 \\
10 & 10 & EEGNet & MOC & 0.089 & 0.670 & 0.637 \\
10 & 20 & EEGNet & MOC & 0.026 & 0.677 & 0.642 \\
10 & Noiseless & EEGNet & MOC & 0.000 & 0.679 & 0.647 \\
10 & 10 & EEGNet & Pinv & 0.137 & 0.614 & 0.575 \\
10 & 20 & EEGNet & Pinv & 0.096 & 0.662 & 0.635 \\
10 & Noiseless & EEGNet & Pinv & 0.000 & 0.681 & 0.654 \\
10 & 10 & ShallowConvNet & MOC & 0.065 & 0.734 & 0.719 \\
10 & 20 & ShallowConvNet & MOC & 0.031 & 0.783 & 0.773 \\
10 & Noiseless & ShallowConvNet & MOC & 0.000 & 0.778 & 0.766 \\
10 & 10 & ShallowConvNet & Pinv & 0.189 & 0.735 & 0.732 \\
10 & 20 & ShallowConvNet & Pinv & 0.070 & 0.773 & 0.769 \\
10 & Noiseless & ShallowConvNet & Pinv & 0.000 & 0.794 & 0.784 \\
30 & 10 & EEGNet & MOC & 0.062 & 0.654 & 0.616 \\
30 & 20 & EEGNet & MOC & 0.024 & 0.682 & 0.661 \\
30 & Noiseless & EEGNet & MOC & 0.000 & 0.673 & 0.643 \\
30 & 10 & EEGNet & Pinv & 0.259 & 0.596 & 0.570 \\
30 & 20 & EEGNet & Pinv & 0.128 & 0.645 & 0.610 \\
30 & Noiseless & EEGNet & Pinv & 0.000 & 0.680 & 0.652 \\
30 & 10 & ShallowConvNet & MOC & 0.080 & 0.765 & 0.755 \\
30 & 20 & ShallowConvNet & MOC & 0.025 & 0.767 & 0.756 \\
30 & Noiseless & ShallowConvNet & MOC & 0.000 & 0.759 & 0.746 \\
30 & 10 & ShallowConvNet & Pinv & 0.261 & 0.647 & 0.635 \\
30 & 20 & ShallowConvNet & Pinv & 0.109 & 0.770 & 0.765 \\
30 & Noiseless & ShallowConvNet & Pinv & 0.000 & 0.794 & 0.784 \\
100 & 10 & EEGNet & MOC & 0.081 & 0.623 & 0.572 \\
100 & 20 & EEGNet & MOC & 0.020 & 0.623 & 0.563 \\
100 & Noiseless & EEGNet & MOC & 0.000 & 0.643 & 0.596 \\
100 & 10 & EEGNet & Pinv & 0.192 & 0.500 & 0.408 \\
100 & 20 & EEGNet & Pinv & 0.265 & 0.623 & 0.603 \\
100 & Noiseless & EEGNet & Pinv & 0.000 & 0.677 & 0.646 \\
100 & 10 & ShallowConvNet & MOC & 0.078 & 0.729 & 0.717 \\
100 & 20 & ShallowConvNet & MOC & 0.027 & 0.720 & 0.705 \\
100 & Noiseless & ShallowConvNet & MOC & 0.000 & 0.725 & 0.711 \\
100 & 10 & ShallowConvNet & Pinv & 0.391 & 0.561 & 0.551 \\
100 & 20 & ShallowConvNet & Pinv & 0.230 & 0.714 & 0.710 \\
100 & Noiseless & ShallowConvNet & Pinv & 0.000 & 0.794 & 0.784 \\
300 & 10 & EEGNet & MOC & 0.058 & 0.607 & 0.554 \\
300 & 20 & EEGNet & MOC & 0.014 & 0.598 & 0.539 \\
300 & Noiseless & EEGNet & MOC & 0.000 & 0.602 & 0.544 \\
300 & 10 & EEGNet & Pinv & 0.162 & 0.507 & 0.402 \\
300 & 20 & EEGNet & Pinv & 0.440 & 0.521 & 0.515 \\
300 & Noiseless & EEGNet & Pinv & 0.000 & 0.680 & 0.652 \\
300 & 10 & ShallowConvNet & MOC & 0.080 & 0.700 & 0.687 \\
300 & 20 & ShallowConvNet & MOC & 0.028 & 0.710 & 0.701 \\
300 & Noiseless & ShallowConvNet & MOC & 0.000 & 0.706 & 0.698 \\
300 & 10 & ShallowConvNet & Pinv & 0.456 & 0.507 & 0.495 \\
300 & 20 & ShallowConvNet & Pinv & 0.322 & 0.596 & 0.590 \\
300 & Noiseless & ShallowConvNet & Pinv & 0.000 & 0.794 & 0.784 \\
1000 & 10 & EEGNet & MOC & 0.066 & 0.588 & 0.540 \\
1000 & 20 & EEGNet & MOC & 0.023 & 0.579 & 0.526 \\
1000 & Noiseless & EEGNet & MOC & 0.000 & 0.575 & 0.516 \\
1000 & 10 & EEGNet & Pinv & 0.187 & 0.499 & 0.400 \\
1000 & 20 & EEGNet & Pinv & 0.484 & 0.499 & 0.491 \\
1000 & Noiseless & EEGNet & Pinv & 0.000 & 0.678 & 0.648 \\
1000 & 10 & ShallowConvNet & MOC & 0.082 & 0.670 & 0.660 \\
1000 & 20 & ShallowConvNet & MOC & 0.030 & 0.661 & 0.649 \\
1000 & Noiseless & ShallowConvNet & MOC & 0.000 & 0.678 & 0.664 \\
1000 & 10 & ShallowConvNet & Pinv & 0.481 & 0.495 & 0.488 \\
1000 & 20 & ShallowConvNet & Pinv & 0.472 & 0.518 & 0.514 \\
1000 & Noiseless & ShallowConvNet & Pinv & 0.000 & 0.794 & 0.784 \\
\bottomrule\end{longtable}\endgroup

\subsection{Reconstruction tables}\label{reconstruction-tables}

The digital noise-reconstruction table is operator-specific and includes
all reference and bipolar cells. Its NMSE denominator is each trial's
processed field energy, with a numerical offset of \(10^{-12}\).
Reconstruction intervals resample subjects after within-subject
aggregation. For compact print reporting, the following table gives each
operator at 10 dB; all SNR values and intervals remain in the annex.

\begingroup\small\setlength{\tabcolsep}{4pt}\renewcommand{\arraystretch}{1.15}
\begin{longtable}{l l l l}
\caption{BNCI2014-001 operator-specific reconstruction at 10 dB: NMSE [95\% interval].}\\
\toprule
Operator & Pinv & Interpolation & MOC \\
\midrule\endfirsthead
\toprule
Operator & Pinv & Interpolation & MOC \\
\midrule\endhead
bipolar longitudinal & 0.550 [0.518, 0.583] & Undefined & 0.332 [0.328, 0.337] \\
bipolar mst & 0.116 [0.112, 0.119] & Undefined & 0.122 [0.120, 0.124] \\
bipolar tree 0 & 0.545 [0.544, 0.545] & Undefined & 0.126 [0.126, 0.126] \\
bipolar tree 1 & 0.423 [0.418, 0.429] & Undefined & 0.123 [0.119, 0.127] \\
car & 0.099 [0.098, 0.100] & 0.099 [0.098, 0.100] & 0.071 [0.069, 0.072] \\
ref C3 & 0.168 [0.161, 0.175] & 0.176 [0.169, 0.183] & 0.111 [0.106, 0.116] \\
ref Cz & 0.142 [0.136, 0.148] & 0.149 [0.143, 0.155] & 0.095 [0.091, 0.100] \\
ref Pz & 0.204 [0.199, 0.210] & 0.214 [0.208, 0.220] & 0.132 [0.130, 0.135] \\
weighted ref 0 & 0.099 [0.097, 0.100] & 0.100 [0.099, 0.101] & 0.070 [0.069, 0.072] \\
weighted ref 1 & 0.098 [0.097, 0.100] & 0.100 [0.098, 0.101] & 0.070 [0.069, 0.072] \\
\bottomrule\end{longtable}\endgroup

\begingroup\small\setlength{\tabcolsep}{4pt}\renewcommand{\arraystretch}{1.15}
\begin{longtable}{l l l l}
\caption{Cho2017 operator-specific reconstruction at 10 dB: NMSE [95\% interval].}\\
\toprule
Operator & Pinv & Interpolation & MOC \\
\midrule\endfirsthead
\toprule
Operator & Pinv & Interpolation & MOC \\
\midrule\endhead
bipolar longitudinal & 1.561 [1.359, 1.807] & Undefined & 0.425 [0.392, 0.459] \\
bipolar mst & 0.285 [0.237, 0.339] & Undefined & 0.143 [0.129, 0.157] \\
bipolar tree 0 & 0.763 [0.687, 0.870] & Undefined & 0.133 [0.121, 0.146] \\
bipolar tree 1 & 0.734 [0.655, 0.851] & Undefined & 0.132 [0.120, 0.146] \\
car & 0.114 [0.102, 0.129] & 0.114 [0.102, 0.129] & 0.082 [0.076, 0.090] \\
ref Cz & 0.159 [0.143, 0.179] & 0.162 [0.146, 0.182] & 0.109 [0.101, 0.119] \\
ref Oz & 0.245 [0.218, 0.274] & 0.249 [0.221, 0.278] & 0.160 [0.146, 0.174] \\
ref Pz & 0.184 [0.164, 0.210] & 0.187 [0.167, 0.213] & 0.124 [0.114, 0.137] \\
weighted ref 0 & 0.114 [0.102, 0.129] & 0.114 [0.103, 0.130] & 0.082 [0.076, 0.090] \\
weighted ref 1 & 0.113 [0.102, 0.129] & 0.114 [0.103, 0.129] & 0.082 [0.076, 0.090] \\
\bottomrule\end{longtable}\endgroup

\begingroup\small\setlength{\tabcolsep}{4pt}\renewcommand{\arraystretch}{1.15}
\begin{longtable}{l l l l}
\caption{PhysioNet operator-specific reconstruction at 10 dB: NMSE [95\% interval].}\\
\toprule
Operator & Pinv & Interpolation & MOC \\
\midrule\endfirsthead
\toprule
Operator & Pinv & Interpolation & MOC \\
\midrule\endhead
bipolar longitudinal & 6.793 [2.126, 13.009] & Undefined & 0.701 [0.447, 1.039] \\
bipolar mst & 1.388 [0.443, 2.611] & Undefined & 0.339 [0.181, 0.543] \\
bipolar tree 0 & 4.262 [1.384, 7.912] & Undefined & 0.473 [0.185, 0.838] \\
bipolar tree 1 & 3.835 [1.244, 7.142] & Undefined & 0.431 [0.170, 0.764] \\
car & 0.483 [0.157, 0.897] & 0.483 [0.157, 0.897] & 0.303 [0.110, 0.547] \\
ref Cz & 0.722 [0.232, 1.350] & 0.733 [0.236, 1.371] & 0.444 [0.155, 0.813] \\
ref Oz & 1.012 [0.327, 1.887] & 1.028 [0.333, 1.918] & 0.615 [0.211, 1.133] \\
ref Pz & 0.757 [0.250, 1.394] & 0.769 [0.254, 1.416] & 0.465 [0.165, 0.841] \\
weighted ref 0 & 0.484 [0.157, 0.898] & 0.486 [0.158, 0.901] & 0.304 [0.110, 0.548] \\
weighted ref 1 & 0.484 [0.157, 0.898] & 0.486 [0.158, 0.902] & 0.304 [0.110, 0.548] \\
\bottomrule\end{longtable}\endgroup

\section{S8. Information-loss
results}\label{s8.-information-loss-results}

All missing-site results use fixed subject partitions and no added
noise. Models are trained on complete CAR coverage. Reconstruction and
decoding endpoints are averaged within subject across masks. The main
figure shows both EEGNet and ShallowConvNet at half coverage; the
complete tables below include every retention level. Directed-mask
families contain one unique mask per retention level, so the tables
report accuracy and macro-F1; their repeated-copy disagreement values
are omitted.

\begingroup\small\setlength{\tabcolsep}{4pt}\renewcommand{\arraystretch}{1.15}
\begin{longtable}{l l l l l l}
\caption{BNCI2014-001 complete information-loss decoding means.}\\
\toprule
Mask & Retained & Decoder & Method & BAC & Macro-F1 \\
\midrule\endfirsthead
\toprule
Mask & Retained & Decoder & Method & BAC & Macro-F1 \\
\midrule\endhead
Motor preserving & 50\% & EEGNet & Interpolation & 0.552 & 0.531 \\
Motor preserving & 50\% & EEGNet & MOC & 0.526 & 0.459 \\
Motor preserving & 50\% & EEGNet & Pinv & 0.531 & 0.494 \\
Motor preserving & 50\% & ShallowConvNet & Interpolation & 0.547 & 0.498 \\
Motor preserving & 50\% & ShallowConvNet & MOC & 0.521 & 0.407 \\
Motor preserving & 50\% & ShallowConvNet & Pinv & 0.500 & 0.333 \\
Motor preserving & 62.5\% & EEGNet & Interpolation & 0.576 & 0.564 \\
Motor preserving & 62.5\% & EEGNet & MOC & 0.578 & 0.566 \\
Motor preserving & 62.5\% & EEGNet & Pinv & 0.533 & 0.474 \\
Motor preserving & 62.5\% & ShallowConvNet & Interpolation & 0.561 & 0.519 \\
Motor preserving & 62.5\% & ShallowConvNet & MOC & 0.536 & 0.467 \\
Motor preserving & 62.5\% & ShallowConvNet & Pinv & 0.517 & 0.388 \\
Motor preserving & 75\% & EEGNet & Interpolation & 0.559 & 0.541 \\
Motor preserving & 75\% & EEGNet & MOC & 0.556 & 0.523 \\
Motor preserving & 75\% & EEGNet & Pinv & 0.540 & 0.479 \\
Motor preserving & 75\% & ShallowConvNet & Interpolation & 0.562 & 0.519 \\
Motor preserving & 75\% & ShallowConvNet & MOC & 0.533 & 0.460 \\
Motor preserving & 75\% & ShallowConvNet & Pinv & 0.528 & 0.420 \\
Motor preserving & 87.5\% & EEGNet & Interpolation & 0.561 & 0.541 \\
Motor preserving & 87.5\% & EEGNet & MOC & 0.540 & 0.485 \\
Motor preserving & 87.5\% & EEGNet & Pinv & 0.566 & 0.539 \\
Motor preserving & 87.5\% & ShallowConvNet & Interpolation & 0.566 & 0.527 \\
Motor preserving & 87.5\% & ShallowConvNet & MOC & 0.547 & 0.503 \\
Motor preserving & 87.5\% & ShallowConvNet & Pinv & 0.536 & 0.455 \\
Motor preserving & 100\% & EEGNet & Interpolation & 0.564 & 0.544 \\
Motor preserving & 100\% & EEGNet & MOC & 0.569 & 0.548 \\
Motor preserving & 100\% & EEGNet & Pinv & 0.590 & 0.579 \\
Motor preserving & 100\% & ShallowConvNet & Interpolation & 0.568 & 0.541 \\
Motor preserving & 100\% & ShallowConvNet & MOC & 0.568 & 0.551 \\
Motor preserving & 100\% & ShallowConvNet & Pinv & 0.568 & 0.541 \\
Motor removing & 50\% & EEGNet & Interpolation & 0.576 & 0.547 \\
Motor removing & 50\% & EEGNet & MOC & 0.576 & 0.571 \\
Motor removing & 50\% & EEGNet & Pinv & 0.507 & 0.424 \\
Motor removing & 50\% & ShallowConvNet & Interpolation & 0.580 & 0.572 \\
Motor removing & 50\% & ShallowConvNet & MOC & 0.554 & 0.506 \\
Motor removing & 50\% & ShallowConvNet & Pinv & 0.526 & 0.452 \\
Motor removing & 62.5\% & EEGNet & Interpolation & 0.564 & 0.549 \\
Motor removing & 62.5\% & EEGNet & MOC & 0.540 & 0.486 \\
Motor removing & 62.5\% & EEGNet & Pinv & 0.493 & 0.333 \\
Motor removing & 62.5\% & ShallowConvNet & Interpolation & 0.566 & 0.547 \\
Motor removing & 62.5\% & ShallowConvNet & MOC & 0.521 & 0.417 \\
Motor removing & 62.5\% & ShallowConvNet & Pinv & 0.502 & 0.340 \\
Motor removing & 75\% & EEGNet & Interpolation & 0.556 & 0.539 \\
Motor removing & 75\% & EEGNet & MOC & 0.535 & 0.462 \\
Motor removing & 75\% & EEGNet & Pinv & 0.502 & 0.363 \\
Motor removing & 75\% & ShallowConvNet & Interpolation & 0.556 & 0.529 \\
Motor removing & 75\% & ShallowConvNet & MOC & 0.554 & 0.524 \\
Motor removing & 75\% & ShallowConvNet & Pinv & 0.524 & 0.444 \\
Motor removing & 87.5\% & EEGNet & Interpolation & 0.562 & 0.543 \\
Motor removing & 87.5\% & EEGNet & MOC & 0.549 & 0.521 \\
Motor removing & 87.5\% & EEGNet & Pinv & 0.531 & 0.427 \\
Motor removing & 87.5\% & ShallowConvNet & Interpolation & 0.559 & 0.531 \\
Motor removing & 87.5\% & ShallowConvNet & MOC & 0.561 & 0.545 \\
Motor removing & 87.5\% & ShallowConvNet & Pinv & 0.569 & 0.536 \\
Motor removing & 100\% & EEGNet & Interpolation & 0.564 & 0.544 \\
Motor removing & 100\% & EEGNet & MOC & 0.569 & 0.548 \\
Motor removing & 100\% & EEGNet & Pinv & 0.590 & 0.579 \\
Motor removing & 100\% & ShallowConvNet & Interpolation & 0.568 & 0.541 \\
Motor removing & 100\% & ShallowConvNet & MOC & 0.568 & 0.551 \\
Motor removing & 100\% & ShallowConvNet & Pinv & 0.568 & 0.541 \\
Random & 50\% & EEGNet & Interpolation & 0.568 & 0.545 \\
Random & 50\% & EEGNet & MOC & 0.552 & 0.479 \\
Random & 50\% & EEGNet & Pinv & 0.503 & 0.375 \\
Random & 50\% & ShallowConvNet & Interpolation & 0.568 & 0.539 \\
Random & 50\% & ShallowConvNet & MOC & 0.535 & 0.439 \\
Random & 50\% & ShallowConvNet & Pinv & 0.528 & 0.435 \\
Random & 62.5\% & EEGNet & Interpolation & 0.570 & 0.549 \\
Random & 62.5\% & EEGNet & MOC & 0.520 & 0.395 \\
Random & 62.5\% & EEGNet & Pinv & 0.531 & 0.451 \\
Random & 62.5\% & ShallowConvNet & Interpolation & 0.569 & 0.543 \\
Random & 62.5\% & ShallowConvNet & MOC & 0.539 & 0.444 \\
Random & 62.5\% & ShallowConvNet & Pinv & 0.532 & 0.421 \\
Random & 75\% & EEGNet & Interpolation & 0.567 & 0.550 \\
Random & 75\% & EEGNet & MOC & 0.561 & 0.520 \\
Random & 75\% & EEGNet & Pinv & 0.530 & 0.450 \\
Random & 75\% & ShallowConvNet & Interpolation & 0.573 & 0.549 \\
Random & 75\% & ShallowConvNet & MOC & 0.560 & 0.525 \\
Random & 75\% & ShallowConvNet & Pinv & 0.549 & 0.465 \\
Random & 87.5\% & EEGNet & Interpolation & 0.558 & 0.535 \\
Random & 87.5\% & EEGNet & MOC & 0.583 & 0.549 \\
Random & 87.5\% & EEGNet & Pinv & 0.541 & 0.467 \\
Random & 87.5\% & ShallowConvNet & Interpolation & 0.564 & 0.535 \\
Random & 87.5\% & ShallowConvNet & MOC & 0.567 & 0.535 \\
Random & 87.5\% & ShallowConvNet & Pinv & 0.541 & 0.457 \\
Random & 100\% & EEGNet & Interpolation & 0.564 & 0.544 \\
Random & 100\% & EEGNet & MOC & 0.569 & 0.548 \\
Random & 100\% & EEGNet & Pinv & 0.590 & 0.579 \\
Random & 100\% & ShallowConvNet & Interpolation & 0.568 & 0.541 \\
Random & 100\% & ShallowConvNet & MOC & 0.568 & 0.551 \\
Random & 100\% & ShallowConvNet & Pinv & 0.568 & 0.541 \\
\bottomrule\end{longtable}\endgroup

\begingroup\small\setlength{\tabcolsep}{4pt}\renewcommand{\arraystretch}{1.15}
\begin{longtable}{l l l l l}
\caption{BNCI2014-001 information-loss NMSE means. Full intervals are in the digital annex.}\\
\toprule
Mask & Retained & Pinv & Interpolation & MOC \\
\midrule\endfirsthead
\toprule
Mask & Retained & Pinv & Interpolation & MOC \\
\midrule\endhead
Motor preserving & 50\% & 0.512 & 0.120 & 0.197 \\
Motor preserving & 62.5\% & 0.407 & 0.102 & 0.162 \\
Motor preserving & 75\% & 0.377 & 0.100 & 0.148 \\
Motor preserving & 87.5\% & 0.180 & 0.027 & 0.063 \\
Motor preserving & 100\% & 7.8e-15 & 2.2e-16 & 0.012 \\
Motor removing & 50\% & 0.491 & 0.108 & 0.164 \\
Motor removing & 62.5\% & 0.299 & 0.056 & 0.103 \\
Motor removing & 75\% & 0.202 & 0.032 & 0.068 \\
Motor removing & 87.5\% & 0.101 & 0.011 & 0.031 \\
Motor removing & 100\% & 7.8e-15 & 2.2e-16 & 0.012 \\
Random & 50\% & 0.579 & 0.133 & 0.227 \\
Random & 62.5\% & 0.411 & 0.089 & 0.148 \\
Random & 75\% & 0.310 & 0.041 & 0.097 \\
Random & 87.5\% & 0.152 & 0.027 & 0.057 \\
Random & 100\% & 7.8e-15 & 2.2e-16 & 0.012 \\
\bottomrule\end{longtable}\endgroup

\begingroup\small\setlength{\tabcolsep}{4pt}\renewcommand{\arraystretch}{1.15}
\begin{longtable}{l l l l l l}
\caption{Cho2017 complete information-loss decoding means.}\\
\toprule
Mask & Retained & Decoder & Method & BAC & Macro-F1 \\
\midrule\endfirsthead
\toprule
Mask & Retained & Decoder & Method & BAC & Macro-F1 \\
\midrule\endhead
Motor preserving & 50\% & EEGNet & Interpolation & 0.681 & 0.663 \\
Motor preserving & 50\% & EEGNet & MOC & 0.601 & 0.547 \\
Motor preserving & 50\% & EEGNet & Pinv & 0.605 & 0.567 \\
Motor preserving & 50\% & ShallowConvNet & Interpolation & 0.669 & 0.655 \\
Motor preserving & 50\% & ShallowConvNet & MOC & 0.547 & 0.446 \\
Motor preserving & 50\% & ShallowConvNet & Pinv & 0.584 & 0.524 \\
Motor preserving & 62.5\% & EEGNet & Interpolation & 0.702 & 0.689 \\
Motor preserving & 62.5\% & EEGNet & MOC & 0.641 & 0.607 \\
Motor preserving & 62.5\% & EEGNet & Pinv & 0.603 & 0.565 \\
Motor preserving & 62.5\% & ShallowConvNet & Interpolation & 0.663 & 0.647 \\
Motor preserving & 62.5\% & ShallowConvNet & MOC & 0.594 & 0.536 \\
Motor preserving & 62.5\% & ShallowConvNet & Pinv & 0.556 & 0.463 \\
Motor preserving & 75\% & EEGNet & Interpolation & 0.705 & 0.692 \\
Motor preserving & 75\% & EEGNet & MOC & 0.675 & 0.656 \\
Motor preserving & 75\% & EEGNet & Pinv & 0.607 & 0.572 \\
Motor preserving & 75\% & ShallowConvNet & Interpolation & 0.675 & 0.661 \\
Motor preserving & 75\% & ShallowConvNet & MOC & 0.657 & 0.643 \\
Motor preserving & 75\% & ShallowConvNet & Pinv & 0.570 & 0.497 \\
Motor preserving & 87.5\% & EEGNet & Interpolation & 0.712 & 0.702 \\
Motor preserving & 87.5\% & EEGNet & MOC & 0.704 & 0.693 \\
Motor preserving & 87.5\% & EEGNet & Pinv & 0.649 & 0.621 \\
Motor preserving & 87.5\% & ShallowConvNet & Interpolation & 0.679 & 0.669 \\
Motor preserving & 87.5\% & ShallowConvNet & MOC & 0.668 & 0.657 \\
Motor preserving & 87.5\% & ShallowConvNet & Pinv & 0.662 & 0.647 \\
Motor preserving & 100\% & EEGNet & Interpolation & 0.715 & 0.704 \\
Motor preserving & 100\% & EEGNet & MOC & 0.717 & 0.711 \\
Motor preserving & 100\% & EEGNet & Pinv & 0.724 & 0.714 \\
Motor preserving & 100\% & ShallowConvNet & Interpolation & 0.691 & 0.680 \\
Motor preserving & 100\% & ShallowConvNet & MOC & 0.679 & 0.668 \\
Motor preserving & 100\% & ShallowConvNet & Pinv & 0.691 & 0.680 \\
Motor removing & 50\% & EEGNet & Interpolation & 0.585 & 0.506 \\
Motor removing & 50\% & EEGNet & MOC & 0.547 & 0.446 \\
Motor removing & 50\% & EEGNet & Pinv & 0.526 & 0.444 \\
Motor removing & 50\% & ShallowConvNet & Interpolation & 0.576 & 0.491 \\
Motor removing & 50\% & ShallowConvNet & MOC & 0.534 & 0.427 \\
Motor removing & 50\% & ShallowConvNet & Pinv & 0.505 & 0.367 \\
Motor removing & 62.5\% & EEGNet & Interpolation & 0.673 & 0.653 \\
Motor removing & 62.5\% & EEGNet & MOC & 0.621 & 0.582 \\
Motor removing & 62.5\% & EEGNet & Pinv & 0.504 & 0.353 \\
Motor removing & 62.5\% & ShallowConvNet & Interpolation & 0.646 & 0.626 \\
Motor removing & 62.5\% & ShallowConvNet & MOC & 0.618 & 0.595 \\
Motor removing & 62.5\% & ShallowConvNet & Pinv & 0.553 & 0.499 \\
Motor removing & 75\% & EEGNet & Interpolation & 0.684 & 0.670 \\
Motor removing & 75\% & EEGNet & MOC & 0.677 & 0.661 \\
Motor removing & 75\% & EEGNet & Pinv & 0.541 & 0.424 \\
Motor removing & 75\% & ShallowConvNet & Interpolation & 0.673 & 0.660 \\
Motor removing & 75\% & ShallowConvNet & MOC & 0.654 & 0.640 \\
Motor removing & 75\% & ShallowConvNet & Pinv & 0.601 & 0.583 \\
Motor removing & 87.5\% & EEGNet & Interpolation & 0.698 & 0.686 \\
Motor removing & 87.5\% & EEGNet & MOC & 0.700 & 0.692 \\
Motor removing & 87.5\% & EEGNet & Pinv & 0.661 & 0.646 \\
Motor removing & 87.5\% & ShallowConvNet & Interpolation & 0.688 & 0.678 \\
Motor removing & 87.5\% & ShallowConvNet & MOC & 0.671 & 0.659 \\
Motor removing & 87.5\% & ShallowConvNet & Pinv & 0.644 & 0.628 \\
Motor removing & 100\% & EEGNet & Interpolation & 0.715 & 0.704 \\
Motor removing & 100\% & EEGNet & MOC & 0.717 & 0.711 \\
Motor removing & 100\% & EEGNet & Pinv & 0.724 & 0.714 \\
Motor removing & 100\% & ShallowConvNet & Interpolation & 0.691 & 0.680 \\
Motor removing & 100\% & ShallowConvNet & MOC & 0.679 & 0.668 \\
Motor removing & 100\% & ShallowConvNet & Pinv & 0.691 & 0.680 \\
Random & 50\% & EEGNet & Interpolation & 0.680 & 0.664 \\
Random & 50\% & EEGNet & MOC & 0.675 & 0.654 \\
Random & 50\% & EEGNet & Pinv & 0.599 & 0.542 \\
Random & 50\% & ShallowConvNet & Interpolation & 0.678 & 0.667 \\
Random & 50\% & ShallowConvNet & MOC & 0.661 & 0.649 \\
Random & 50\% & ShallowConvNet & Pinv & 0.625 & 0.583 \\
Random & 62.5\% & EEGNet & Interpolation & 0.686 & 0.673 \\
Random & 62.5\% & EEGNet & MOC & 0.693 & 0.682 \\
Random & 62.5\% & EEGNet & Pinv & 0.587 & 0.512 \\
Random & 62.5\% & ShallowConvNet & Interpolation & 0.679 & 0.668 \\
Random & 62.5\% & ShallowConvNet & MOC & 0.665 & 0.654 \\
Random & 62.5\% & ShallowConvNet & Pinv & 0.626 & 0.585 \\
Random & 75\% & EEGNet & Interpolation & 0.698 & 0.688 \\
Random & 75\% & EEGNet & MOC & 0.704 & 0.692 \\
Random & 75\% & EEGNet & Pinv & 0.636 & 0.591 \\
Random & 75\% & ShallowConvNet & Interpolation & 0.686 & 0.675 \\
Random & 75\% & ShallowConvNet & MOC & 0.670 & 0.658 \\
Random & 75\% & ShallowConvNet & Pinv & 0.661 & 0.644 \\
Random & 87.5\% & EEGNet & Interpolation & 0.707 & 0.696 \\
Random & 87.5\% & EEGNet & MOC & 0.713 & 0.707 \\
Random & 87.5\% & EEGNet & Pinv & 0.703 & 0.691 \\
Random & 87.5\% & ShallowConvNet & Interpolation & 0.688 & 0.677 \\
Random & 87.5\% & ShallowConvNet & MOC & 0.672 & 0.662 \\
Random & 87.5\% & ShallowConvNet & Pinv & 0.677 & 0.665 \\
Random & 100\% & EEGNet & Interpolation & 0.715 & 0.704 \\
Random & 100\% & EEGNet & MOC & 0.717 & 0.711 \\
Random & 100\% & EEGNet & Pinv & 0.724 & 0.714 \\
Random & 100\% & ShallowConvNet & Interpolation & 0.691 & 0.680 \\
Random & 100\% & ShallowConvNet & MOC & 0.679 & 0.668 \\
Random & 100\% & ShallowConvNet & Pinv & 0.691 & 0.680 \\
\bottomrule\end{longtable}\endgroup

\begingroup\small\setlength{\tabcolsep}{4pt}\renewcommand{\arraystretch}{1.15}
\begin{longtable}{l l l l l}
\caption{Cho2017 information-loss NMSE means. Full intervals are in the digital annex.}\\
\toprule
Mask & Retained & Pinv & Interpolation & MOC \\
\midrule\endfirsthead
\toprule
Mask & Retained & Pinv & Interpolation & MOC \\
\midrule\endhead
Motor preserving & 50\% & 0.638 & 0.399 & 0.364 \\
Motor preserving & 62.5\% & 0.467 & 0.256 & 0.228 \\
Motor preserving & 75\% & 0.287 & 0.156 & 0.127 \\
Motor preserving & 87.5\% & 0.185 & 0.097 & 0.087 \\
Motor preserving & 100\% & 2.1e-14 & 1.5e-16 & 0.014 \\
Motor removing & 50\% & 0.500 & 0.654 & 0.275 \\
Motor removing & 62.5\% & 0.362 & 0.287 & 0.177 \\
Motor removing & 75\% & 0.217 & 0.055 & 0.073 \\
Motor removing & 87.5\% & 0.067 & 0.014 & 0.030 \\
Motor removing & 100\% & 2.1e-14 & 1.5e-16 & 0.014 \\
Random & 50\% & 0.528 & 0.170 & 0.180 \\
Random & 62.5\% & 0.360 & 0.093 & 0.105 \\
Random & 75\% & 0.267 & 0.076 & 0.091 \\
Random & 87.5\% & 0.124 & 0.028 & 0.042 \\
Random & 100\% & 2.1e-14 & 1.5e-16 & 0.014 \\
\bottomrule\end{longtable}\endgroup

\begingroup\small\setlength{\tabcolsep}{4pt}\renewcommand{\arraystretch}{1.15}
\begin{longtable}{l l l l l l}
\caption{PhysioNet complete information-loss decoding means.}\\
\toprule
Mask & Retained & Decoder & Method & BAC & Macro-F1 \\
\midrule\endfirsthead
\toprule
Mask & Retained & Decoder & Method & BAC & Macro-F1 \\
\midrule\endhead
Motor preserving & 50\% & EEGNet & Interpolation & 0.589 & 0.505 \\
Motor preserving & 50\% & EEGNet & MOC & 0.636 & 0.596 \\
Motor preserving & 50\% & EEGNet & Pinv & 0.502 & 0.340 \\
Motor preserving & 50\% & ShallowConvNet & Interpolation & 0.669 & 0.634 \\
Motor preserving & 50\% & ShallowConvNet & MOC & 0.682 & 0.665 \\
Motor preserving & 50\% & ShallowConvNet & Pinv & 0.590 & 0.570 \\
Motor preserving & 62.5\% & EEGNet & Interpolation & 0.661 & 0.616 \\
Motor preserving & 62.5\% & EEGNet & MOC & 0.684 & 0.662 \\
Motor preserving & 62.5\% & EEGNet & Pinv & 0.559 & 0.490 \\
Motor preserving & 62.5\% & ShallowConvNet & Interpolation & 0.750 & 0.729 \\
Motor preserving & 62.5\% & ShallowConvNet & MOC & 0.759 & 0.749 \\
Motor preserving & 62.5\% & ShallowConvNet & Pinv & 0.667 & 0.640 \\
Motor preserving & 75\% & EEGNet & Interpolation & 0.690 & 0.661 \\
Motor preserving & 75\% & EEGNet & MOC & 0.663 & 0.620 \\
Motor preserving & 75\% & EEGNet & Pinv & 0.541 & 0.435 \\
Motor preserving & 75\% & ShallowConvNet & Interpolation & 0.778 & 0.771 \\
Motor preserving & 75\% & ShallowConvNet & MOC & 0.775 & 0.766 \\
Motor preserving & 75\% & ShallowConvNet & Pinv & 0.696 & 0.683 \\
Motor preserving & 87.5\% & EEGNet & Interpolation & 0.680 & 0.647 \\
Motor preserving & 87.5\% & EEGNet & MOC & 0.686 & 0.652 \\
Motor preserving & 87.5\% & EEGNet & Pinv & 0.669 & 0.643 \\
Motor preserving & 87.5\% & ShallowConvNet & Interpolation & 0.795 & 0.788 \\
Motor preserving & 87.5\% & ShallowConvNet & MOC & 0.794 & 0.790 \\
Motor preserving & 87.5\% & ShallowConvNet & Pinv & 0.757 & 0.747 \\
Motor preserving & 100\% & EEGNet & Interpolation & 0.681 & 0.654 \\
Motor preserving & 100\% & EEGNet & MOC & 0.674 & 0.642 \\
Motor preserving & 100\% & EEGNet & Pinv & 0.669 & 0.642 \\
Motor preserving & 100\% & ShallowConvNet & Interpolation & 0.794 & 0.784 \\
Motor preserving & 100\% & ShallowConvNet & MOC & 0.791 & 0.787 \\
Motor preserving & 100\% & ShallowConvNet & Pinv & 0.794 & 0.784 \\
Motor removing & 50\% & EEGNet & Interpolation & 0.622 & 0.572 \\
Motor removing & 50\% & EEGNet & MOC & 0.664 & 0.634 \\
Motor removing & 50\% & EEGNet & Pinv & 0.517 & 0.415 \\
Motor removing & 50\% & ShallowConvNet & Interpolation & 0.726 & 0.703 \\
Motor removing & 50\% & ShallowConvNet & MOC & 0.734 & 0.719 \\
Motor removing & 50\% & ShallowConvNet & Pinv & 0.595 & 0.531 \\
Motor removing & 62.5\% & EEGNet & Interpolation & 0.671 & 0.628 \\
Motor removing & 62.5\% & EEGNet & MOC & 0.649 & 0.603 \\
Motor removing & 62.5\% & EEGNet & Pinv & 0.540 & 0.456 \\
Motor removing & 62.5\% & ShallowConvNet & Interpolation & 0.764 & 0.752 \\
Motor removing & 62.5\% & ShallowConvNet & MOC & 0.749 & 0.743 \\
Motor removing & 62.5\% & ShallowConvNet & Pinv & 0.644 & 0.627 \\
Motor removing & 75\% & EEGNet & Interpolation & 0.666 & 0.621 \\
Motor removing & 75\% & EEGNet & MOC & 0.667 & 0.627 \\
Motor removing & 75\% & EEGNet & Pinv & 0.631 & 0.582 \\
Motor removing & 75\% & ShallowConvNet & Interpolation & 0.796 & 0.784 \\
Motor removing & 75\% & ShallowConvNet & MOC & 0.786 & 0.784 \\
Motor removing & 75\% & ShallowConvNet & Pinv & 0.762 & 0.756 \\
Motor removing & 87.5\% & EEGNet & Interpolation & 0.666 & 0.635 \\
Motor removing & 87.5\% & EEGNet & MOC & 0.672 & 0.640 \\
Motor removing & 87.5\% & EEGNet & Pinv & 0.656 & 0.627 \\
Motor removing & 87.5\% & ShallowConvNet & Interpolation & 0.788 & 0.776 \\
Motor removing & 87.5\% & ShallowConvNet & MOC & 0.787 & 0.784 \\
Motor removing & 87.5\% & ShallowConvNet & Pinv & 0.772 & 0.756 \\
Motor removing & 100\% & EEGNet & Interpolation & 0.681 & 0.654 \\
Motor removing & 100\% & EEGNet & MOC & 0.674 & 0.642 \\
Motor removing & 100\% & EEGNet & Pinv & 0.669 & 0.642 \\
Motor removing & 100\% & ShallowConvNet & Interpolation & 0.794 & 0.784 \\
Motor removing & 100\% & ShallowConvNet & MOC & 0.791 & 0.787 \\
Motor removing & 100\% & ShallowConvNet & Pinv & 0.794 & 0.784 \\
Random & 50\% & EEGNet & Interpolation & 0.653 & 0.616 \\
Random & 50\% & EEGNet & MOC & 0.644 & 0.598 \\
Random & 50\% & EEGNet & Pinv & 0.567 & 0.495 \\
Random & 50\% & ShallowConvNet & Interpolation & 0.765 & 0.754 \\
Random & 50\% & ShallowConvNet & MOC & 0.751 & 0.737 \\
Random & 50\% & ShallowConvNet & Pinv & 0.619 & 0.552 \\
Random & 62.5\% & EEGNet & Interpolation & 0.673 & 0.647 \\
Random & 62.5\% & EEGNet & MOC & 0.641 & 0.590 \\
Random & 62.5\% & EEGNet & Pinv & 0.540 & 0.428 \\
Random & 62.5\% & ShallowConvNet & Interpolation & 0.778 & 0.770 \\
Random & 62.5\% & ShallowConvNet & MOC & 0.748 & 0.735 \\
Random & 62.5\% & ShallowConvNet & Pinv & 0.625 & 0.560 \\
Random & 75\% & EEGNet & Interpolation & 0.672 & 0.643 \\
Random & 75\% & EEGNet & MOC & 0.668 & 0.639 \\
Random & 75\% & EEGNet & Pinv & 0.602 & 0.540 \\
Random & 75\% & ShallowConvNet & Interpolation & 0.782 & 0.770 \\
Random & 75\% & ShallowConvNet & MOC & 0.774 & 0.766 \\
Random & 75\% & ShallowConvNet & Pinv & 0.706 & 0.678 \\
Random & 87.5\% & EEGNet & Interpolation & 0.672 & 0.644 \\
Random & 87.5\% & EEGNet & MOC & 0.674 & 0.644 \\
Random & 87.5\% & EEGNet & Pinv & 0.649 & 0.615 \\
Random & 87.5\% & ShallowConvNet & Interpolation & 0.783 & 0.775 \\
Random & 87.5\% & ShallowConvNet & MOC & 0.782 & 0.779 \\
Random & 87.5\% & ShallowConvNet & Pinv & 0.746 & 0.727 \\
Random & 100\% & EEGNet & Interpolation & 0.681 & 0.654 \\
Random & 100\% & EEGNet & MOC & 0.674 & 0.642 \\
Random & 100\% & EEGNet & Pinv & 0.669 & 0.642 \\
Random & 100\% & ShallowConvNet & Interpolation & 0.794 & 0.784 \\
Random & 100\% & ShallowConvNet & MOC & 0.791 & 0.787 \\
Random & 100\% & ShallowConvNet & Pinv & 0.794 & 0.784 \\
\bottomrule\end{longtable}\endgroup

\begingroup\small\setlength{\tabcolsep}{4pt}\renewcommand{\arraystretch}{1.15}
\begin{longtable}{l l l l l}
\caption{PhysioNet information-loss NMSE means. Full intervals are in the digital annex.}\\
\toprule
Mask & Retained & Pinv & Interpolation & MOC \\
\midrule\endfirsthead
\toprule
Mask & Retained & Pinv & Interpolation & MOC \\
\midrule\endhead
Motor preserving & 50\% & 0.628 & 0.709 & 0.397 \\
Motor preserving & 62.5\% & 0.449 & 0.342 & 0.246 \\
Motor preserving & 75\% & 0.275 & 0.138 & 0.127 \\
Motor preserving & 87.5\% & 0.093 & 0.030 & 0.046 \\
Motor preserving & 100\% & 1.9e-14 & 1.7e-16 & 0.017 \\
Motor removing & 50\% & 0.540 & 0.474 & 0.321 \\
Motor removing & 62.5\% & 0.329 & 0.219 & 0.180 \\
Motor removing & 75\% & 0.187 & 0.090 & 0.100 \\
Motor removing & 87.5\% & 0.085 & 0.028 & 0.043 \\
Motor removing & 100\% & 1.9e-14 & 1.7e-16 & 0.017 \\
Random & 50\% & 0.501 & 0.191 & 0.191 \\
Random & 62.5\% & 0.374 & 0.135 & 0.140 \\
Random & 75\% & 0.255 & 0.092 & 0.100 \\
Random & 87.5\% & 0.116 & 0.039 & 0.051 \\
Random & 100\% & 1.9e-14 & 1.7e-16 & 0.017 \\
\bottomrule\end{longtable}\endgroup

\section{S9. Subject-grouped
validation}\label{s9.-subject-grouped-validation}

Secondary analysis uses leave-one-subject-out folds for BNCI and five
subject-grouped folds for Cho2017 and PhysioNet. EEGNet and
ShallowConvNet use three training seeds (20260919--20260921). Seed
results are first averaged within subject. Fold membership is fully
specified in the accompanying partition file. Test subjects are excluded
from each fold's training and validation sets, but training sets overlap
between folds. The analyses assess within-cohort partition sensitivity.
Audit of the retained code confirms that model construction preceded
seed resetting and the method views used distinct output-noise draws.
Three seeds and grouped folds therefore do not make this a controlled
replication of MR3; all values are retained as historical sensitivity
evidence. Confidence intervals summarize subjects conditional on the
fitted cross-validation procedure.

\begingroup\small\setlength{\tabcolsep}{4pt}\renewcommand{\arraystretch}{1.15}
\begin{longtable}{l l l l l l}
\caption{Subject-grouped paired contrasts for D.}\\
\toprule
Dataset & Decoder & Baseline & MOC minus baseline & p & Holm p \\
\midrule\endfirsthead
\toprule
Dataset & Decoder & Baseline & MOC minus baseline & p & Holm p \\
\midrule\endhead
BNCI2014-001 & EEGNet & Pinv & -0.001 [-0.011, 0.009] & 0.8203 & 0.8203 \\
BNCI2014-001 & EEGNet & Raw & -0.160 [-0.189, -0.130] & 0.0039 & 0.0195 \\
BNCI2014-001 & ShallowConvNet & Pinv & -0.005 [-0.013, 0.004] & 0.3555 & 0.7109 \\
BNCI2014-001 & ShallowConvNet & Raw & -0.117 [-0.148, -0.088] & 0.0039 & 0.0195 \\
Cho2017 & EEGNet & Pinv & -0.006 [-0.010, -0.002] & 0.0090 & 0.0270 \\
Cho2017 & EEGNet & Raw & -0.126 [-0.137, -0.115] & 0.0001 & 0.0012 \\
Cho2017 & ShallowConvNet & Pinv & -0.007 [-0.010, -0.004] & 0.0001 & 0.0012 \\
Cho2017 & ShallowConvNet & Raw & -0.059 [-0.067, -0.052] & 0.0001 & 0.0012 \\
PhysioNet & EEGNet & Pinv & -0.020 [-0.026, -0.014] & 0.0001 & 0.0012 \\
PhysioNet & EEGNet & Raw & -0.105 [-0.120, -0.090] & 0.0001 & 0.0012 \\
PhysioNet & ShallowConvNet & Pinv & -0.020 [-0.025, -0.015] & 0.0001 & 0.0012 \\
PhysioNet & ShallowConvNet & Raw & -0.087 [-0.095, -0.080] & 0.0001 & 0.0012 \\
\bottomrule\end{longtable}\endgroup

\begingroup\small\setlength{\tabcolsep}{4pt}\renewcommand{\arraystretch}{1.15}
\begin{longtable}{l l l l l l}
\caption{Subject-grouped paired contrasts for BAC.}\\
\toprule
Dataset & Decoder & Baseline & MOC minus baseline & p & Holm p \\
\midrule\endfirsthead
\toprule
Dataset & Decoder & Baseline & MOC minus baseline & p & Holm p \\
\midrule\endhead
BNCI2014-001 & EEGNet & Pinv & 0.002 [-0.007, 0.011] & 0.6719 & 1.0000 \\
BNCI2014-001 & EEGNet & Raw & 0.065 [0.040, 0.093] & 0.0039 & 0.0312 \\
BNCI2014-001 & ShallowConvNet & Pinv & -0.000 [-0.019, 0.019] & 0.9570 & 1.0000 \\
BNCI2014-001 & ShallowConvNet & Raw & 0.023 [0.012, 0.034] & 0.0156 & 0.0938 \\
Cho2017 & EEGNet & Pinv & -0.000 [-0.004, 0.004] & 0.9622 & 1.0000 \\
Cho2017 & EEGNet & Raw & 0.034 [0.025, 0.042] & 0.0001 & 0.0012 \\
Cho2017 & ShallowConvNet & Pinv & 0.000 [-0.004, 0.004] & 0.9554 & 1.0000 \\
Cho2017 & ShallowConvNet & Raw & 0.008 [0.003, 0.014] & 0.0080 & 0.0560 \\
PhysioNet & EEGNet & Pinv & 0.004 [-0.000, 0.008] & 0.0859 & 0.4295 \\
PhysioNet & EEGNet & Raw & 0.035 [0.029, 0.042] & 0.0001 & 0.0012 \\
PhysioNet & ShallowConvNet & Pinv & 0.012 [0.006, 0.018] & 0.0004 & 0.0036 \\
PhysioNet & ShallowConvNet & Raw & 0.031 [0.025, 0.038] & 0.0001 & 0.0012 \\
\bottomrule\end{longtable}\endgroup

\begingroup\small\setlength{\tabcolsep}{4pt}\renewcommand{\arraystretch}{1.15}
\begin{longtable}{l l l l l l}
\caption{Subject-grouped paired contrasts for macro-F1.}\\
\toprule
Dataset & Decoder & Baseline & MOC minus baseline & p & Holm p \\
\midrule\endfirsthead
\toprule
Dataset & Decoder & Baseline & MOC minus baseline & p & Holm p \\
\midrule\endhead
BNCI2014-001 & EEGNet & Pinv & 0.003 [-0.005, 0.012] & 0.5078 & 1.0000 \\
BNCI2014-001 & EEGNet & Raw & 0.103 [0.082, 0.128] & 0.0039 & 0.0273 \\
BNCI2014-001 & ShallowConvNet & Pinv & -0.002 [-0.025, 0.022] & 0.8672 & 1.0000 \\
BNCI2014-001 & ShallowConvNet & Raw & 0.030 [0.015, 0.044] & 0.0156 & 0.0938 \\
Cho2017 & EEGNet & Pinv & -0.001 [-0.006, 0.004] & 0.7237 & 1.0000 \\
Cho2017 & EEGNet & Raw & 0.047 [0.036, 0.059] & 0.0001 & 0.0012 \\
Cho2017 & ShallowConvNet & Pinv & 0.000 [-0.004, 0.005] & 0.8743 & 1.0000 \\
Cho2017 & ShallowConvNet & Raw & 0.013 [0.005, 0.020] & 0.0029 & 0.0232 \\
PhysioNet & EEGNet & Pinv & 0.002 [-0.003, 0.007] & 0.3705 & 1.0000 \\
PhysioNet & EEGNet & Raw & 0.049 [0.041, 0.058] & 0.0001 & 0.0012 \\
PhysioNet & ShallowConvNet & Pinv & 0.013 [0.006, 0.019] & 0.0005 & 0.0045 \\
PhysioNet & ShallowConvNet & Raw & 0.036 [0.029, 0.044] & 0.0001 & 0.0012 \\
\bottomrule\end{longtable}\endgroup

\section{S10. Algebraic and computational
checks}\label{s10.-algebraic-and-computational-checks}

For every rank-(K-1) zero-sum matrix, the null space is span(1), so the
row-space projector is \(M^\dagger M=P\). Both referential and
connected-tree operators satisfy this identity. The supplied diagnostics
report the Frobenius projector residual, zero-sum residual, rank, and
complete singular spectrum. If rank is below K-1, the projector has a
smaller range and exact recovery of every mean-zero field is impossible.
If two fields differ by a nonconstant null vector, their observations
agree; no deterministic data-only estimator can return both correctly.

With output noise, \(\widehat v-Pv=M^\dagger\epsilon\). The operator
norm equals the reciprocal smallest positive singular value, yielding
the main-text bound. For isotropic noise, taking the trace of
\(\sigma_\epsilon^2M^\dagger(M^\dagger)^T\) gives the summed
inverse-square singular-value formula. The regularized estimator follows
by substituting \(v=Bz\), differentiating the quadratic objective, and
solving the reduced normal equations. Its numerical implementation is
checked against this constrained system. Rank deficiency requires a
prior or a generalized inverse and is not removed by the penalty.

For \(M=I-\mathbf{1}w^T\) and \(B^T\mathbf{1}=0\),
\(MB=B-\mathbf{1}w^TB\). Expanding \((MB)^TMB\) gives
\(I+K(B^Tw)(B^Tw)^T\), a rank-one update. Since
\(\|B^Tw\|^2=\|w\|^2-1/K\), its eigenvalues are 1 (multiplicity \(K-2\))
and \(K\|w\|^2\). The unscaled complete references thus have minimum
positive singular value and pseudoinverse norm both equal to 1. CAR
graph regularization is \((I+\lambda L)^{-1}P\). The additional
mechanism audit checks these identities against all eighteen stored
complete-reference matrices, verifies shared-field-noise alignment,
reproduces the saved graph matrices, and verifies \(P_mS\) against every
unique retained-site mask. These are algebraic and implementation
checks, not new classifier training runs.

The final software check suite contains 35 tests covering operator
algebra, information equivalence, rank loss, partition isolation,
matched-operator aggregation, inverse behavior, and the corrected
CSP/interpolation implementations. In addition, 264
subject/decoder/representation disagreement values were independently
reconstructed from saved trial predictions and matched the pair-table
aggregation to absolute tolerance \(10^{-12}\). Primary statistics and
figure inputs are rebuilt from immutable scored outputs. The digital
audit records source hashes, selectors, comparison counts, spectra, and
the exact publication-number mapping.

\section{S11. Digital annex and
reproduction}\label{s11.-digital-annex-and-reproduction}

The digital annex, Supplementary\_data\_and\_code.tar.xz, release FS2 (21
September 2026), is supplied as an ancillary file with this preprint.
VERSION.txt identifies the evidence release and README.md gives the
reconstruction sequence. The arXiv packaging release (17 November 2026)
removes journal-only correspondence and document-building assets; all
experimental results and analysis inputs are retained unchanged.
reports/publication\_evidence/ contains subject-level values and dataset
summaries for ideal sensitivity, matched methods, architecture
comparisons, bipolar coverage, noise, conditioning, information loss,
and grouped validation. Summary files include counts, bootstrap
intervals, paired differences, and raw and Holm-adjusted probabilities.
Undefined comparisons are omitted. The archive also supplies masks,
illustrative-trial selection, geometry, signed operators, spectra,
Laplacians, source hashes, the additional mechanism audit, and the
protocol, checkpoints and paired predictions for the separate zero-row
sensitivity study in Section S12. Publication calculations use the
estimator implementations specified in Section S3. Raw recordings are
not redistributed.

\clearpage

\section{S12. Earlier one-seed zero-row sensitivity
(MR2)}\label{s12.-earlier-one-seed-zero-row-sensitivity-mr2}

Before MR3, a supplementary paired check used the existing Cho2017 and
PhysioNet partitions and six complete references. Three EEGNet pipelines
(Raw, Pinv, MOC) were fitted once per dataset on CAR at 10 dB, using the
same 15-epoch budget, Adam settings, minimum-validation-loss selection,
and MOC coefficient 0.1. The seed was set before network construction
and again before each fit; initial parameter and buffer hashes agreed
across the three methods within each dataset. Deterministic CUDA
algorithms were enabled and benchmarking was disabled. The protocol was
fixed before the run, and checkpoints and loss histories were retained.

The observation generator followed the original split, operator and
256-trial-block sequence. Each fitted decoder was then evaluated on two
paired versions of the same six test views: noise on all rows (All), and
the same arrays with nominally zero measurement rows set exactly to zero
(Zero). Noise on every other row was unchanged, without rescaling. CAR
training and validation arrays have no zero rows, so this comparison
required no condition-specific refitting. D averages fifteen reference
pairs; BAC and macro-F1 average six views within subject, followed by
equal-subject means.

\begingroup\small\setlength{\tabcolsep}{4pt}\renewcommand{\arraystretch}{1.15}
\begin{longtable}{p{0.135\linewidth} p{0.08\linewidth} p{0.085\linewidth} p{0.085\linewidth} p{0.1\linewidth} p{0.1\linewidth} p{0.1\linewidth} p{0.1\linewidth}}
\caption{Historical MR2 zero-row sensitivity in six controlled EEGNet fits. All and Zero refer to test-row perturbations. These are distinct from the repeated MR3 fits in main Table 2.}\\
\toprule
Dataset & Method & D All & D Zero & BAC All & BAC Zero & F1 All & F1 Zero \\
\midrule\endfirsthead
\toprule
Dataset & Method & D All & D Zero & BAC All & BAC Zero & F1 All & F1 Zero \\
\midrule\endhead
Cho2017 & Raw & 0.313 & 0.311 & 0.654 & 0.653 & 0.628 & 0.628 \\
Cho2017 & Pinv & 0.153 & 0.153 & 0.700 & 0.700 & 0.696 & 0.696 \\
Cho2017 & MOC & 0.133 & 0.133 & 0.700 & 0.700 & 0.695 & 0.695 \\
PhysioNet & Raw & 0.250 & 0.251 & 0.653 & 0.652 & 0.641 & 0.640 \\
PhysioNet & Pinv & 0.173 & 0.173 & 0.673 & 0.673 & 0.663 & 0.663 \\
PhysioNet & MOC & 0.143 & 0.143 & 0.689 & 0.689 & 0.678 & 0.678 \\
\bottomrule\end{longtable}\endgroup

MOC-minus-raw D and paired 95\% subject-bootstrap intervals were
Cho2017: -0.180 {[}-0.213, -0.141{]} (All) and -0.179 {[}-0.212,
-0.140{]} (Zero); PhysioNet: -0.107 {[}-0.124, -0.090{]} (All) and
-0.108 {[}-0.124, -0.093{]} (Zero). Intervals use 10,000 resamples and
describe these supplementary fits; they are exploratory and unadjusted.
Complete subject values, all three endpoint contrasts and within-model
changes are supplied in the digital annex.

Pinv and MOC predictions were identical across the two row conditions;
Raw predictions changed at some single-reference views. The row
perturbation therefore affects Raw, while its removal leaves the inverse
inputs and predictions effectively unchanged. The new fits retained
lower MOC D than Raw under both conditions. Their PhysioNet Pinv/MOC D
ranking differed from the original MR2 main table, illustrating fit
dependence. These estimates describe within-fit sensitivity in one new
controlled fit per method and dataset. The historical weights underlying
the original MR2 main table were not retained, so this check cannot
assign the original pipeline differences to initialization or zero-row
noise. It did not compare deterministic Pinv and interpolation fits.
This earlier implementation correctly set initialization and CUDA
determinism, but did not isolate a dedicated DataLoader order stream
from validation iteration. The MR3 protocol addresses that remaining
requirement and retains this check as prior evidence rather than
silently pooling the fits.

\section{S13. Analysis history, paired training, and mechanism
controls}\label{s13.-analysis-history-paired-training-and-mechanism-controls}

FS2 (21 September 2026) is a document and evidence-packaging release
with zero additional training. MR2 and MR3 are internal analysis-phase
identifiers, not journal revision rounds; no journal peer review is
claimed. Earlier results on the same held-out subjects were available
before the MR3 paired repeated analysis. That phase corrected
initialization and random-stream control and added checkpoint-sharing
and perturbation controls while retaining the partitions. It was a
reanalysis, not an independent new test-cohort validation or a
study-wide preregistered evaluation. Models were fitted separately
within each dataset.

Earlier single-fit results placed augmentation below MOC in Cho2017
disagreement and Pinv below MOC in PhysioNet disagreement. These
rankings did not persist in the paired repeated analysis and are not
retained as stable findings. Historical initialization and noise-sharing
limitations remain specified in Sections S3, S6, and S9; the earlier
controlled one-seed check remains distinct in S12. The original
numerical tables are preserved in reports/legacy\_mr2.

The MR3 protocol in experiments/configs/mr3\_protocol.json and its
recorded numerical preflight amendment were fixed before training in
that repeated-analysis phase. The training ledger records 26 completed
fits and the preflight/evaluation records retain the corresponding
hashes. These records do not establish that test outcomes were unseen
when the study was initially designed. It preserves the subject
partitions, preprocessing, operators, lambda 0.1, 15 epochs, batch size
128, Adam learning rate 0.001 and weight decay 0.0001, cross-entropy,
and first strictly lowest validation-loss checkpoint. It fixes training
seeds 20260919, 20260920, 20260921; three test-noise realizations; all
six front ends; D and BAC as primary endpoints; and the original
twelve-test and separate two-test Holm families. Macro-F1 remains
secondary. The experimental-phase ledger counted every attempted fit,
including reproducibility repeats, against its then-authorized ceiling
of 36; that allowance is not a training budget for FS2. No result-based
seed selection or hyperparameter search was performed.

Initialization is performed on CPU after setting the training seed and
within an isolated Torch random-state context. The identical initial
parameter/buffer hash is required across the four training groups at
each cohort/seed. NumPy SeedSequence derives independent streams from
{[}training seed, stream identifier{]}: 101 for the dedicated Torch CPU
DataLoader order generator, 102 for dropout, and 103 for validation
iteration. Dropout is reset before each fit. CUDA deterministic
algorithms are enabled; benchmark and TF32 are off. Epoch order hashes,
training/validation losses, initial/final checkpoint hashes and all
stream seeds are saved. A complete second Raw fit at seed 20260919 in
each cohort must reproduce every epoch's order and loss and the final
checkpoint bit for bit.

Training/validation observations and first-repeat test observations
replay the original PCG64 seed 20260919, split/operator/block order, and
Gaussian draws, including discarded bipolar draws to preserve the random
state. All retained observation hashes match the MR2 protocol.
Additional test repeats use independent SeedSequence streams
{[}20260919, 301, repeat, operator index{]}, where repeat is 1 or 2.
Each consecutive block of at most 256 trials retains the original
nominal 10 dB calibration. Augmentation arrays are fixed across training
seeds; streams {[}20260919, 201{]} and {[}20260919, 202{]} independently
generate reference choices and noise. Corresponding test observations
are shared across every front end and training seed.

CAR Pinv and complete-site interpolation/centering are algebraically
equal, as are CAR MOC and fixed smoothing after Pinv. Before any fit,
float64 relative Frobenius map differences must not exceed \(10^{-12}\).
Actual float32 train/validation inputs must have relative RMS error at
most \(2\times10^{-6}\) and maximum mixed relative error
\(|a-b|/(|a|+\mathrm{RMS}(a))\) at most 32 machine epsilons. The initial
proposed maximum-error/global-RMS threshold rejected a high-amplitude
float32 rounding difference before any training attempt. The frozen
preflight amendment documents the original threshold, observed errors
and replacement criterion; it did not change observations, optimization,
seeds, endpoints or test families. After passing, each alias uses
exactly the same training/validation tensor and checkpoint. This shares
Pinv with interpolation and MOC with fixed smoothing, requiring four
training groups per cohort/seed. At test time each non-CAR front end is
still computed from its own operator and the shared noisy array.
Numerical equivalence at CAR does not imply equality at every noisy
reference.

For fixed smoothing, the float64 map solves \((I+0.1L)S=I\) and applies
\(S M^\dagger y\). MOC retains its operator-dependent constrained
fidelity term. Both use the same CAR-trained checkpoint. The paired
zero-row condition overwrites only exactly zero measurement rows with
zero, retaining every other noise entry and its scale. No additional
model is trained. For each fixed operator, the three unordered pairs of
independent noise realizations provide a disagreement reference; these
values are averaged over six operators. They do not identify a causal
component of cross-reference disagreement.

For each subject, D averages fifteen operator pairs within a
realization; BAC and macro-F1 average six operators. Realizations and
then training seeds are averaged within subject before cohort means,
percentile bootstrap intervals or paired sign-flip tests. The
independent sample sizes remain 10 and 21. Per-seed cohort effects are
shown below, with training variability separately reported. CIs crossing
zero do not establish equivalence or a reverse effect. The same-operator
and zero-row controls are descriptive; no extra significance family is
selected from their results.

The ledger contains 26 completed training attempts: 24 distinct fits and
two complete reproducibility repeats. Both repeats passed bitwise
checkpoint and full-history equality. No failed training attempt or
additional architecture fit was required. Preflight and evaluation
records give exact observation hashes, numerical alias errors, and
fixed-checkpoint hashes.

\setlength{\LTpre}{4pt}\setlength{\LTpost}{4pt}

\begingroup\small\setlength{\tabcolsep}{4pt}\renewcommand{\arraystretch}{1.0}
\begin{longtable}{p{0.14\linewidth} p{0.2\linewidth} p{0.18\linewidth} p{0.18\linewidth} p{0.18\linewidth}}
\caption{Per-seed MOC-minus-control cohort differences for D; subjects and noise realizations are averaged within each seed.}\\
\toprule
Dataset & Control & Seed 20260919 & Seed 20260920 & Seed 20260921 \\
\midrule\endfirsthead
\toprule
Dataset & Control & Seed 20260919 & Seed 20260920 & Seed 20260921 \\
\midrule\endhead
Cho2017 & Raw & -0.165 & -0.145 & -0.121 \\
Cho2017 & Pinv & -0.008 & -0.028 & -0.019 \\
Cho2017 & Interpolation & -0.008 & -0.030 & -0.019 \\
Cho2017 & Aug. + centering & 0.007 & -0.021 & -0.011 \\
Cho2017 & Fixed smoothing & 0.000878 & 0.000154 & -0.000254 \\
PhysioNet & Raw & -0.127 & -0.088 & -0.107 \\
PhysioNet & Pinv & -0.034 & -0.007 & -0.015 \\
PhysioNet & Interpolation & -0.037 & -0.009 & -0.014 \\
PhysioNet & Aug. + centering & -0.029 & -0.035 & -0.004 \\
PhysioNet & Fixed smoothing & 0.000658 & -0.000306 & -0.000306 \\
\bottomrule\end{longtable}\endgroup

\begingroup\small\setlength{\tabcolsep}{4pt}\renewcommand{\arraystretch}{1.0}
\begin{longtable}{p{0.14\linewidth} p{0.2\linewidth} p{0.18\linewidth} p{0.18\linewidth} p{0.18\linewidth}}
\caption{Per-seed MOC-minus-control cohort differences for BAC; subjects and noise realizations are averaged within each seed.}\\
\toprule
Dataset & Control & Seed 20260919 & Seed 20260920 & Seed 20260921 \\
\midrule\endfirsthead
\toprule
Dataset & Control & Seed 20260919 & Seed 20260920 & Seed 20260921 \\
\midrule\endhead
Cho2017 & Raw & 0.051 & 0.029 & 0.051 \\
Cho2017 & Pinv & 0.002 & 0.009 & 0.009 \\
Cho2017 & Interpolation & 0.003 & 0.009 & 0.010 \\
Cho2017 & Aug. + centering & -0.002 & 0.023 & 0.009 \\
Cho2017 & Fixed smoothing & -0.000213 & 0.000079 & 0.000125 \\
PhysioNet & Raw & 0.044 & 0.067 & 0.037 \\
PhysioNet & Pinv & 0.010 & 0.009 & 0.014 \\
PhysioNet & Interpolation & 0.010 & 0.011 & 0.014 \\
PhysioNet & Aug. + centering & -0.000 & 0.008 & 0.017 \\
PhysioNet & Fixed smoothing & 0.000021 & 0.000050 & 0.000211 \\
\bottomrule\end{longtable}\endgroup

\begingroup\small\setlength{\tabcolsep}{4pt}\renewcommand{\arraystretch}{1.0}
\begin{longtable}{p{0.14\linewidth} p{0.2\linewidth} p{0.18\linewidth} p{0.18\linewidth} p{0.18\linewidth}}
\caption{Per-seed MOC-minus-control cohort differences for Macro-F1; subjects and noise realizations are averaged within each seed.}\\
\toprule
Dataset & Control & Seed 20260919 & Seed 20260920 & Seed 20260921 \\
\midrule\endfirsthead
\toprule
Dataset & Control & Seed 20260919 & Seed 20260920 & Seed 20260921 \\
\midrule\endhead
Cho2017 & Raw & 0.078 & 0.034 & 0.069 \\
Cho2017 & Pinv & 0.001 & 0.009 & 0.010 \\
Cho2017 & Interpolation & 0.002 & 0.009 & 0.010 \\
Cho2017 & Aug. + centering & 0.001 & 0.023 & 0.008 \\
Cho2017 & Fixed smoothing & -0.000203 & 0.000105 & 0.000093 \\
PhysioNet & Raw & 0.058 & 0.115 & 0.057 \\
PhysioNet & Pinv & 0.007 & 0.009 & 0.019 \\
PhysioNet & Interpolation & 0.007 & 0.011 & 0.019 \\
PhysioNet & Aug. + centering & -0.003 & 0.005 & 0.018 \\
PhysioNet & Fixed smoothing & 0.000091 & -0.000029 & 0.000227 \\
\bottomrule\end{longtable}\endgroup

\begingroup\small\setlength{\tabcolsep}{2.5pt}\renewcommand{\arraystretch}{1.0}
\begin{longtable}{p{0.12\linewidth} p{0.17\linewidth} p{0.13\linewidth} p{0.07\linewidth} p{0.11\linewidth} p{0.13\linewidth} p{0.07\linewidth} p{0.11\linewidth}}
\caption{Training variability over all three fixed seeds. SD uses divisor three; the final column for each endpoint averages within-subject seed SD.}\\
\toprule
Dataset & Front end & D range & D SD & Within D SD & BAC range & BAC SD & Within BAC SD \\
\midrule\endfirsthead
\toprule
Dataset & Front end & D range & D SD & Within D SD & BAC range & BAC SD & Within BAC SD \\
\midrule\endhead
Cho2017 & Raw & 0.220 to 0.284 & 0.026 & 0.039 & 0.650 to 0.680 & 0.012 & 0.018 \\
Cho2017 & Pinv & 0.118 to 0.130 & 0.005 & 0.014 & 0.699 to 0.706 & 0.003 & 0.016 \\
Cho2017 & Interpolation & 0.118 to 0.131 & 0.006 & 0.014 & 0.698 to 0.706 & 0.003 & 0.016 \\
Cho2017 & Aug. + centering & 0.110 to 0.122 & 0.005 & 0.011 & 0.687 to 0.707 & 0.009 & 0.017 \\
Cho2017 & MOC & 0.099 to 0.119 & 0.009 & 0.014 & 0.701 to 0.716 & 0.006 & 0.019 \\
Cho2017 & Fixed smoothing & 0.100 to 0.118 & 0.008 & 0.014 & 0.701 to 0.716 & 0.006 & 0.018 \\
PhysioNet & Raw & 0.251 to 0.304 & 0.022 & 0.048 & 0.604 to 0.642 & 0.016 & 0.033 \\
PhysioNet & Pinv & 0.170 to 0.212 & 0.017 & 0.031 & 0.657 to 0.676 & 0.008 & 0.035 \\
PhysioNet & Interpolation & 0.172 to 0.211 & 0.016 & 0.030 & 0.657 to 0.676 & 0.008 & 0.035 \\
PhysioNet & Aug. + centering & 0.179 to 0.201 & 0.010 & 0.025 & 0.654 to 0.686 & 0.013 & 0.034 \\
PhysioNet & MOC & 0.150 to 0.197 & 0.020 & 0.031 & 0.671 to 0.686 & 0.007 & 0.033 \\
PhysioNet & Fixed smoothing & 0.149 to 0.197 & 0.020 & 0.032 & 0.671 to 0.686 & 0.007 & 0.033 \\
\bottomrule\end{longtable}\endgroup

\begingroup\small\setlength{\tabcolsep}{4pt}\renewcommand{\arraystretch}{1.0}
\begin{longtable}{p{0.13\linewidth} p{0.19\linewidth} p{0.3\linewidth} p{0.3\linewidth}}
\caption{Fixed-checkpoint cross-reference and same-operator independent-noise disagreement. Values aggregate repeats within subject; intervals are descriptive.}\\
\toprule
Dataset & Front end & Cross-reference D [95\% CI] & Same-operator noise D [95\% CI] \\
\midrule\endfirsthead
\toprule
Dataset & Front end & Cross-reference D [95\% CI] & Same-operator noise D [95\% CI] \\
\midrule\endhead
Cho2017 & Raw & 0.250 [0.225, 0.275] & 0.099 [0.086, 0.112] \\
Cho2017 & Pinv & 0.125 [0.102, 0.151] & 0.123 [0.101, 0.149] \\
Cho2017 & Interpolation & 0.126 [0.103, 0.152] & 0.124 [0.101, 0.150] \\
Cho2017 & Aug. + centering & 0.115 [0.096, 0.136] & 0.113 [0.095, 0.133] \\
Cho2017 & MOC & 0.107 [0.086, 0.130] & 0.104 [0.085, 0.128] \\
Cho2017 & Fixed smoothing & 0.106 [0.086, 0.130] & 0.104 [0.084, 0.128] \\
PhysioNet & Raw & 0.277 [0.251, 0.307] & 0.157 [0.111, 0.211] \\
PhysioNet & Pinv & 0.189 [0.142, 0.242] & 0.190 [0.142, 0.246] \\
PhysioNet & Interpolation & 0.190 [0.143, 0.244] & 0.192 [0.144, 0.248] \\
PhysioNet & Aug. + centering & 0.192 [0.149, 0.241] & 0.193 [0.148, 0.243] \\
PhysioNet & MOC & 0.170 [0.124, 0.223] & 0.170 [0.123, 0.224] \\
PhysioNet & Fixed smoothing & 0.170 [0.125, 0.223] & 0.170 [0.124, 0.224] \\
\bottomrule\end{longtable}\endgroup

\begingroup\small\setlength{\tabcolsep}{4pt}\renewcommand{\arraystretch}{1.0}
\begin{longtable}{p{0.13\linewidth} p{0.19\linewidth} p{0.3\linewidth} p{0.3\linewidth}}
\caption{Fixed-checkpoint zero-row sensitivity. Zero minus all changes preserve every nonzero-row noise entry without rescaling; intervals resample subjects.}\\
\toprule
Dataset & Front end & Delta D [95\% CI] & Delta BAC [95\% CI] \\
\midrule\endfirsthead
\toprule
Dataset & Front end & Delta D [95\% CI] & Delta BAC [95\% CI] \\
\midrule\endhead
Cho2017 & Raw & 0.000335 [-0.000376, 0.000919] & -0.000048 [-0.000574, 0.000528] \\
Cho2017 & Pinv & 0 [0, 0] & 0 [0, 0] \\
Cho2017 & Interpolation & -0.000925 [-0.001565, -0.000296] & 0.000597 [0.000356, 0.000866] \\
Cho2017 & Aug. + centering & -0.001085 [-0.001704, -0.000532] & 0.000088 [-0.000319, 0.000505] \\
Cho2017 & MOC & 0 [0, 0] & 0 [0, 0] \\
Cho2017 & Fixed smoothing & 0 [0, 0] & 0 [0, 0] \\
PhysioNet & Raw & -0.000807 [-0.001450, -0.000149] & -0.000379 [-0.001307, 0.000389] \\
PhysioNet & Pinv & 0 [0, 0] & 0 [0, 0] \\
PhysioNet & Interpolation & -0.001082 [-0.002163, -0.000055] & 0.000456 [-0.000290, 0.001208] \\
PhysioNet & Aug. + centering & -0.000345 [-0.001058, 0.000345] & -0.000171 [-0.000975, 0.000571] \\
PhysioNet & MOC & 0 [0, 0] & 0 [0, 0] \\
PhysioNet & Fixed smoothing & 0 [0, 0] & 0 [0, 0] \\
\bottomrule\end{longtable}\endgroup

\begingroup\small\setlength{\tabcolsep}{4pt}\renewcommand{\arraystretch}{1.0}
\begin{longtable}{p{0.16\linewidth} p{0.13\linewidth} p{0.35\linewidth} p{0.13\linewidth} p{0.13\linewidth}}
\caption{MOC minus fixed smoothing: separate two-cohort Holm family for each endpoint.}\\
\toprule
Dataset & Endpoint & Difference [95\% CI] & p & Holm p \\
\midrule\endfirsthead
\toprule
Dataset & Endpoint & Difference [95\% CI] & p & Holm p \\
\midrule\endhead
Cho2017 & D & 0.000259 [0.000093, 0.000430] & 0.0312 & 0.0625 \\
Cho2017 & BAC & -0.000003 [-0.000315, 0.000312] & 1.0000 & 1.0000 \\
Cho2017 & Macro-F1 & -0.000002 [-0.000327, 0.000323] & 0.9961 & 1.0000 \\
PhysioNet & D & 0.000016 [-0.000423, 0.000486] & 0.9732 & 0.9732 \\
PhysioNet & BAC & 0.000094 [-0.000195, 0.000363] & 0.5395 & 1.0000 \\
PhysioNet & Macro-F1 & 0.000096 [-0.000217, 0.000389] & 0.5485 & 1.0000 \\
\bottomrule\end{longtable}\endgroup